\documentclass[aps,prb,twocolumn, amsfonts, amssymb, amsmath, reprint, showkeys, nofootinbib, superscriptaddress]{revtex4-2}

\usepackage[normalem]{ulem}
\usepackage[english]{babel}
\usepackage[utf8]{inputenc}
\usepackage{enumitem}
\usepackage[colorinlistoftodos, color=green!40, prependcaption]{todonotes}
\usepackage[pdftex, pdftitle={Article}, pdfauthor={Author}]{hyperref}
\newcommand{\be}{\begin{equation}}
\newcommand{\ee}{\end{equation}}
\newcommand{\ba}{\begin{eqnarray}}
\newcommand{\ea}{\end{eqnarray}}

\begin{document}
\title{Non-linear Transport Phenomena and Current-induced Hydrodynamics in Ultra-high Mobility Two-dimensional Electron Gas}

\author{Z. T.~Wang}
\author{M.~Hilke}
\email[]{hilke@physics.mcgill.ca}
\affiliation{Department of Physics, McGill University, Montr\'eal, Quebec, Canada, H3A 2T8}
\author{N.~Fong}
\affiliation{Emerging Technology Division, National Research Council of Canada, Ottawa, Ontario, Canada, K1A 0R6}
\author{D.~G.~Austing}
\email[]{guy.austing@nrc-cnrc.gc.ca}
\affiliation{Department of Physics, McGill University, Montr\'eal, Quebec, Canada, H3A 2T8}
\affiliation{Emerging Technology Division, National Research Council of Canada, Ottawa, Ontario, Canada, K1A 0R6}
\author{S.~A.~Studenikin}
\affiliation{Emerging Technology Division, National Research Council of Canada, Ottawa, Ontario, Canada, K1A 0R6}
\author{K. W.~West}
\author{L. N.~Pfeiffer}
\affiliation{Department of Electrical Engineering, Princeton University, Princeton, New Jersey, USA, 08544}

\date{\today} 

\begin{abstract}
We report on non-linear transport phenomena at high filling factor and DC current-induced electronic hydrodynamics in an ultra-high mobility ($\mu$=$20\times10^6$ c$m^2$/Vs) two-dimensional electron gas in a narrow (15 $\mu m$ wide) GaAs/AlGaAs Hall bar for DC current densities reaching 0.67 A/m. The various phenomena and the boundaries between the phenomena are captured together in a two-dimensional differential resistivity map as a function of magnetic field (up to 250 mT) and DC current. This map, which resembles a phase diagram, demarcate distinct regions dominated by Shubnikov-de Haas (SdH) oscillations (and phase inversion of these oscillations) around zero DC current; negative magnetoresistance and a double-peak feature (both ballistic in origin) around zero field; and Hall field-induced resistance oscillations (HIROs) radiating out from the origin. From a detailed analysis of the data near zero field, we show that increasing the DC current suppresses the electron-electron scattering length that drives a growing hydrodynamic contribution to both the differential longitudinal and transverse (Hall) resistivities. Our approach to induce hydrodynamics with DC current differs from the more usual approach of changing the temperature. We also find a significant (factor of two to four) difference between the quantum lifetime extracted from SdH oscillations, and the quantum lifetime extracted from HIROs. In addition to observing HIRO peaks up to the seventh order, we observe an unexpected HIRO-like feature close to mid-way between the first-order and the second-order HIRO maxima at high DC current.
\end{abstract}

\maketitle

\section{Introduction}
High mobility two-dimensional electron gas (2DEG) systems exhibit a remarkable richness of phenomena. In a strong magnetic (B-) field there are various topological phases such as integer and fractional quantum Hall phases, which stem from an interplay of disorder, electron correlations and B-field. These phases are distinguished by their Hall quantization \cite{Prange1990, Chakraborty1995}. On the other hand, non-linear phenomena, such as Hall field-induced, phonon-induced, and microwave-induced resistance oscillations, in Landau levels (LLs) at high filling factor close to zero B-field are examples of phenomena that cannot be characterized by conductance quantization. These non-linear phenomena have attracted significant interest over the last two decades: see the extensive reviews in Refs. \cite{dmitriev2012nonequilibrium, vitkalov2009nonlinear} and references therein.

Regarding non-linear DC transport, the principal topic of our work, the quintessential example of a DC current-induced phenomenon at low B-field is Hall field-induced resistance oscillations (HIROs)\cite{yang2002zener}. In published experimental works on HIROs\cite{yang2002zener, Bykov2005, zhang2007num3, zhang2007effect, zhang2007magnetotransport,bykov2008, dai2009,   Hatke2009,Hatke2010, Hatke2011, bykov2012zener, wiedmann2011, shi2017, Zudov2017,Mi2019}, typically data are presented in the form of selected traces of differential resistance versus B-field at fixed DC current (or differential resistance versus DC current at fixed B-field). Such an approach may not reveal all aspects of HIROs or HIRO-like phenomena, and their relationship to other distinct linear and non-linear phenomena at low B-field may not be clear. An alternative approach to gain a more complete picture is to map out the resistance as a function of B-field and DC current. Such a technique has been applied at high magnetic field in the quantum Hall regime and demonstrated to reveal a wealth of phenomena: see Refs. \cite{phaseinversionsergei, panos2014, baer2015, Rossokhaty2016, phaseinversionYu}.

Recently viscous transport in two-dimensional (2D) systems has also drawn significant interest. Hydrodynamic phenomena are expected to be most pronounced when the (momentum-conserving) electron-electron scattering length $l_{ee}$ is much less than the device width $W$, and in turn, $W$ is much less than the classical transport mean free path $l_{mfp}$, i.e., $l_{ee}$ $\ll$ $W$ $\ll$ $l_{mfp}$, distinct to the condition $W$ $\ll$ $l_{ee}$, $l_{mfp}$ for which ballistic effects dominate. The subject of hydrodynamic effects in solids at low temperature was pioneered by Gurzhi in the 1960's\cite{gurzhi1963,gurzhi1968hydrodynamic}, and initially drew theoretical attention, see for example Refs. \cite{giuliani1982lifetime,gurzhi1989hydrodynamic, jaggi1991electron,molenkamp1994,de1995hydrodynamic}. With the advent of materials in the 1990's for which (momentum-relaxing) scattering with defects and phonons was sufficiently weak, Molenkamp and de Jong investigated experimentally hydrodynamic electron flow in high-mobility GaAs/AlGaAs hetero-structure wires, and could distinguish the Knudsen and Poiseuille (Gurzhi) transport regimes in the differential resistance \cite{de1995hydrodynamic,molenkamp1994,Molenkamp1994part2}. Two decades ago analogies with fluid dynamics were also explored to explain voltage steps in the quantum Hall effect breakdown regime including an eddy viscosity model for the disruption of laminar flow around charged impurities \cite{eaves1998,eaves1999hydrodynamic,eaves2000,eaves2001,eaves2001quantum,martin2004}. In recent years there has been renewed interest in hydrodynamic electron transport following the development of material systems with ever higher transport mean free path. Experimental works studying hydrodynamic effects in 2D systems feature graphene\cite{Zaanen2016,Bandurin2016,Crossno2016,Levitov2016,sulpizio2019visualizing,Berdyugin2019,gallagher2019quantum, ella2019simultaneous,Ku2020,Pusep2022,Jenkins2022}, high-mobility semiconductor 2DEGs\cite{shi2014,gusev2018viscous,Levin2018,Gusev2018b,Braem2018,Gusev2020a,Keser2021,Gupta2021,Afanasiev2021,HornCosfeld2021,Monch2022}, 2D metals\cite{Moll2016}, semi-metals\cite{AharonSteinberg2022}, and semi-metal micro-ribbons\cite{Gooth2018}. This effort has inspired numerous theoretical works of which Refs. \cite{alekseev2016negative,guo2017higher,scaffidi2017hydrodynamic,Alekseev2019,Alekseev2019b,Alekseev2020a,Matthaiakakis2020,raichev2020,Afanasiev2021,DasSarma2022} are examples of those focusing on semiconductor 2DEGs. Change of temperature to suppress $l_{ee}$ is the most common approach to reach the hydrodynamic regime, although channel thinning in steps to effectively change $W$ can be employed in certain instances as in Ref. \cite{Moll2016}. It was also predicted in Ref. \cite{giuliani1982lifetime} that $l_{ee}$ decreases with increasing DC current, which is the basis of current-induced viscous transport as originally investigated by Molenkamp and de Jong\cite{de1995hydrodynamic,molenkamp1994,Molenkamp1994part2}- see also Ref. \cite{Hara2004}.

Here, we examine in detail the differential resistance of an ultra-high mobility 2DEG in a narrow Hall bar fabricated from a GaAs/AlGaAs quantum well hetero-structure as a function of B-field and DC current. This approach allows us to study numerous phenomena in one global 2D diagram as they evolve with increasing DC current up to 10 $\mu$A for B-fields up to $0.25$ T. In the differential longitudinal resistivity, we observe: Shubnikov-de Haas (SdH) oscillations and phase inversion (PhI) of SdH oscillations near zero current, negative magnetoresistance (nMR) and a double-peak feature near zero field, and HIROs up to the seventh order. In addition, we find evidence for DC current-induced electron hydrodynamics effects for B-fields less than 10 mT in both the differential longitudinal and transverses resistivities. Since the global 2D diagram (which identifies distinct regions within which different phenomena dominate) resembles a phase diagram in appearance, we will henceforth refer to the diagram as a ``phase diagram''.

This paper is organized into seven sections. Section \ref{sec:exp} describes the experimental details pertaining to the 2DEG material, the Hall bar device, and the measurement configuration. Section \ref{sec:diagram} presents the differential resistivity map, and introduces the different phenomena observed and delineates the boundaries between the phenomena in the phase diagram. Sections \ref{sec:sdh} to \ref{sec:hiro} give extensive analysis of the various phenomena. In more detail, in Sec. \ref{sec:sdh}, we extract parameters of interest such as the quantum lifetime from the SdH oscillations, and simulate the PhI of the SdH oscillations with an existing model. Section \ref{sec:nmr} provides evidence that the origin of the nMR and the double-peak feature is purely ballistic in nature. Section \ref{sec:hydro} presents analysis, based on current theories for viscous transport, of the observed evolution of both the differential longitudinal and transverse (Hall) resistivities near $0$ T with respect to the DC current which we attribute to hydrodynamic flow. In Sec. \ref{sec:hiro}, we compare properties of the HIROs observed in the experiment to those expected from existing theories, and extract various parameters such as the quantum lifetime and the electronic width. We also report an unexpected additional HIRO-like feature in the phase diagram located in between the first-order and the second-order HIROs peak. We end with a conclusion in Section VIII.


\section{Experimental Details}  \label{sec:exp}

\setlength{\tabcolsep}{3pt}
\begin{table}[t] 
\renewcommand{\arraystretch}{1.4}
\centering
\begin{tabular}{  c c c c c c  } 
 \hline \hline 
 $n$ & $\mu$ & $E_F$ & $l_{mfp}$  \\ 
 (c$m^{-2}$) & (c$m^2$/V s) & (meV) & ($\mu$m)  \\  
 \hline 
  $2.0 \times 10^{11}$ & $20 \times 10^{6}$ & $7.2$ & 145  \\ 
 \hline \hline
\end{tabular}
\caption{Key parameters of the 2DEG material system. The (bulk) parameters given are determined from a large area Van der Pauw device measured in a helium-3 cryostat at base temperature after illumination.}
\label{table:1}
\end{table}

\begin{figure}[!b]
\includegraphics[width=1\columnwidth]{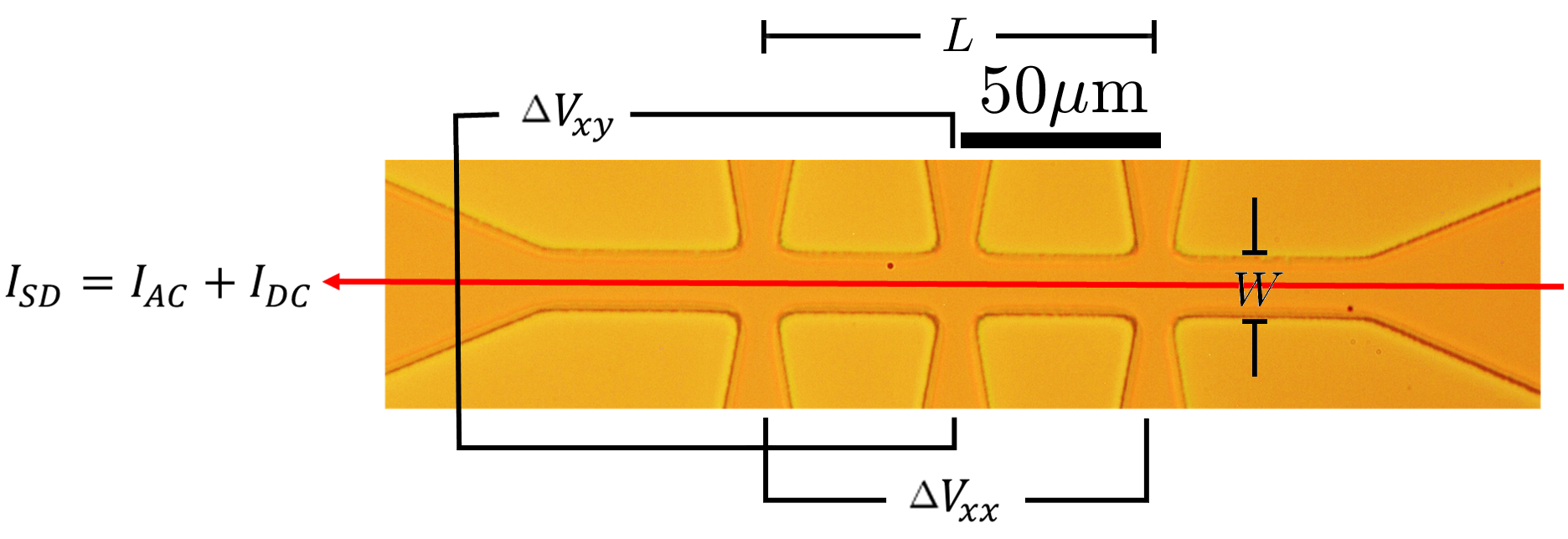}
\caption{Image of central region of the Hall bar device showing the measurement configuration. The Ohmic contacts are out of view. The lithographic width of the Hall bar is $W$=15 $\mu$m, and the two voltage probes employed to measure $\Delta V_{xx}$ are separated by distance $L$=100 $\mu$m.}
\label{fig:Set-up}
\end{figure}

The Hall bar device is made from a 2DEG confined in a GaAs/AlGaAs quantum well (QW) hetero-structure grown by molecular beam epitaxy. The 2DEG is located in a 30 nm wide GaAs QW at a depth $\sim$200 nm below the surface. The barriers on either side of the QW are composed of A$l_{0.3}$G$a_{0.7}$As and incorporate QW doping regions. From a large area Van der Pauw device measured in a helium-3 cryostat at base temperature after illumination, the 2DEG sheet density $n$ and (transport) mobility $\mu$ respectively for this material are found to be 2.0$\times10^{11}$ $cm^{-2}$ and 20$\times10^6$ c$m^2$/V s. The corresponding Fermi energy $E_F$ and transport mean free path $l_{mfp}$ respectively are determined to be 7.2 meV and 145 $\mu$m. These parameters are tabulated in Table \ref{table:1}.

\begin{figure*}
\includegraphics[width=2.0\columnwidth]{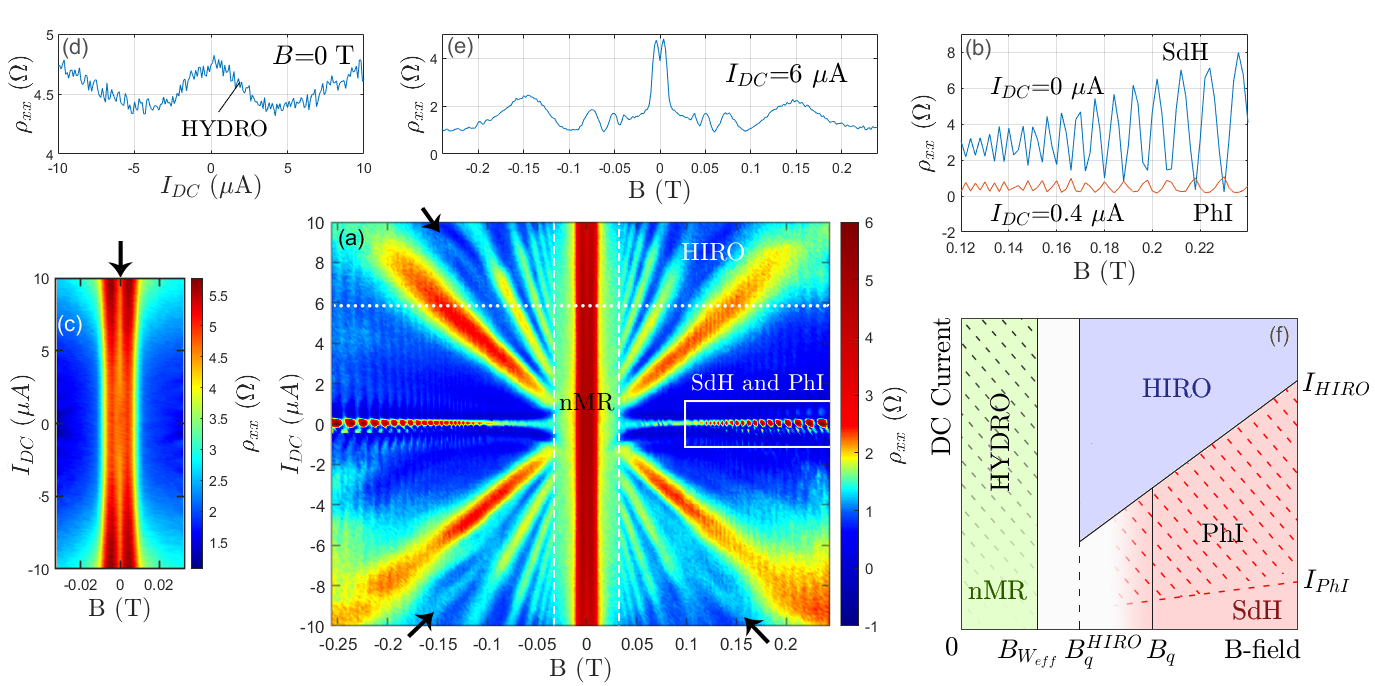}
\caption{
(a) Map identifying the various phenomena (``phase diagram''). $\rho_{xx}$ measured by sweeping the DC current and stepping the B-field. A background parabolic dependence with $I_{DC}$ and a small uniform linear dependence with B-field are subtracted from the raw data to emphasize the features of interest in the phase diagram. The background $I_{DC}$ dependence is discussed in Sec. \ref{sec:lee}, and the small linear $B$ dependence is likely due to inhomogeneity of the 2DEG material. The Shubniknov-de Haas (SdH) oscillations and the accompanying phase inversion (PhI) of the SdH oscillations are visible close to zero DC current. The vertical band near $B$=0 T identifies the negative magnetoresistance (nMR). The peaks forming a fan are the Hall-field induced resistance oscillations (HIROs). Black arrows in three of the four quadrants mark the ``1.5'' (HIRO-like) feature. (b) $\rho_{xx}$ versus B sections at $I_{DC}$=0 $\mu$A and $I_{DC}$=0.4 $\mu$A showing respectively SdH oscillations and phase inverted SdH oscillations. (c) Expanded view of phase diagram around B=0 T where nMR is observed. Note that here $\rho_{xx}$ is plotted without removing the above mentioned background parabolic dependence with $I_{DC}$ (see Sec. \ref{sec:nmr}). A double-peak feature is present on top of the nMR [see also panel (e)]. (d) $\rho_{xx}$ versus $I_{DC}$ trace at $B=0$ T [marked by the vertical arrow in panel (c)]. The initial decrease in $\rho_{xx}$ with current, i.e., negative differential resistivity, is a signature of hydrodynamics\cite{molenkamp1994} (HYDRO). (e) $\rho_{xx}$ versus $B$ section taken at $I_{DC}$=6 $\mu$A where we see the nMR and the double-peak feature around B=0 T, and HIROs at higher B-field. (f) Cartoon summarizing the regions identified in one quadrant of the phase diagram in panel (a). A more detailed description of the various phenomena, the boundaries between the phenomena, and certain marked characteristic B-fields and DC currents can be found in the text.}
\label{fig:PhaseDiagram}
\end{figure*}

The Hall bar device is made by standard fabrication techniques. The nominal (lithographic) width $W$ of the Hall bar is 15 $\mu$m. The separation between adjacent voltage probes along each side of the Hall bar is 50 $\mu$m. An image of the central region of the device is shown in Fig. \ref{fig:Set-up}. For all experiments described, a DC current $I_{DC}$ is combined with an AC excitation current $I_{AC}$ of 20 nA (unless otherwise stated) at 148 Hz, and the net current $I_{SD}$ is driven through the Hall bar from the source contact to the grounded drain contact. $\Delta V_{xx}$, the change in AC voltage along the Hall bar, and $\Delta V_{xy}$, the change in AC voltage across the width of the Hall bar, are measured with a standard lock-in technique and the differential longitudinal and transverse resistances respectively are given by $r_{xx}= \Delta V_{xx}/I_{AC}=dV_{xx}/dI$ and $r_{xy}= \Delta V_{xy}/I_{AC}=dV_{xy}/dI$: see Refs. \cite{Zhang2009,phaseinversionYu} for further details of the technique. Voltages $\Delta V_{xx}$ and $\Delta V_{xy}$ are measured between the voltage probes indicated in Fig. \ref{fig:Set-up}. Note $\Delta V_{xx}$ is dropped between voltage probes separated by a distance $L$=100 $\mu$m. Although not measured, a DC voltage drop is also discussed and estimated in section \ref{sec:hydro}. The differential longitudinal and transverse resistivities respectively are $\rho_{xx}$ = W$r_{xx}$/L and $\rho_{xy}$ = $r_{xy}$. All measurements are performed in the dark in a dilution refrigerator at base temperature where the mixing chamber temperature is $\sim$15 mK. The electron temperature $T_e$ is estimated in separate measurements to be $\sim$40 mK. The B-field is applied perpendicular to the plane of the 2DEG. Note that a small 6 mT correction has been applied to the data to account for a field offset.

\section{Phenomena Observed in the Differential Resistivity}\label{sec:diagram}

We start by looking at the global phase diagram and identify all the different phenomena therein. This is accomplished by measuring $\rho_{xx}$ on sweeping the DC current and stepping the B-field. The resulting map is presented in Fig. \ref{fig:PhaseDiagram}(a). In our Hall bar device, we can identify several phenomena of interest: SdH oscillations, phase inverted SdH oscillations, nMR, a double peak-feature on top of the nMR, DC current-induced hydrodynamic effects, HIROs, and a HIRO-like feature. In the rest of this section we provide a general introduction to these phenomena before going into details in subsequent sections. To aid this introduction, Fig. \ref{fig:PhaseDiagram}(f) provides a simple schematic showing the regions, in one quadrant of the phase diagram, where the phenomena are observed, the boundaries between the phenomena, and characteristic B-fields and DC currents.

\begin{figure*}
\includegraphics[width=2.0\columnwidth]{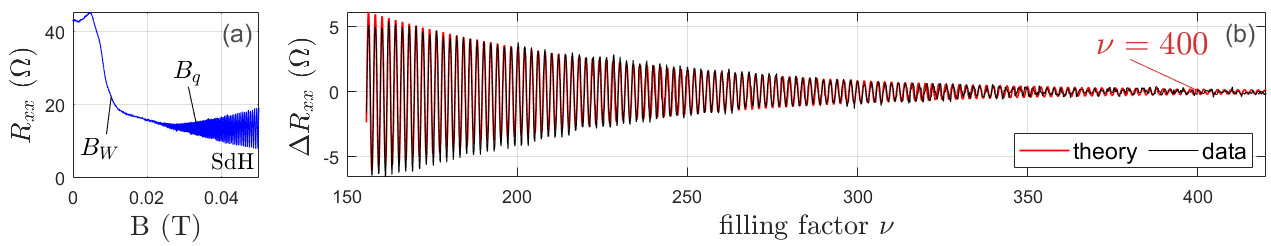}
\caption{(a) Resistance $R_{xx}$ versus $B$ trace. Here $R_{xx}$ is taken to be $r_{xx}$ at $I_{DC}$=0 $\mu$A, and for this particular measurement where the B-field is swept at 1 mT/min the AC excitation current is 40 nA. At low B-field, there is strong negative magnetoresistance up to field $B_{W}$ $\sim$10 mT as discussed in Sec. \ref{sec:diagram} and in Sec. \ref{sec:nmr}. $B_q$=33 mT is related to $\tau_q$ and is defined in Sec. \ref{sec:diagram}. (b) SdH oscillation amplitude $\Delta R_{xx}$ determined from panel (a) plotted versus filling factor ($\nu \propto 1/B$). The black line is the data, fitted to Eq. \eqref{eq:sdh}, red line, with obtained parameters $n$=$2.0 \times 10^{11} cm^{-2}$, $R_0$=$11.4 \pm 0.6$  $\Omega$, and $\tau_q$=$11.5 \pm 0.3$ ps. $\nu$=400 corresponds to B=21 mT.}
\label{fig:SdHBigTrace}
\end{figure*}

SdH oscillations at zero DC current are observed above 21 mT and their amplitude in differential resistivity is found to decay with increasing $I_{DC}$ [see additionally both Fig. \ref{fig:SdHBigTrace} and Fig. \ref{fig:inversion3d}]. In general, SdH oscillations are absent when $\omega_c\tau_q\ll$1, where $\omega_c=eB/m^{*}$ is the cyclotron frequency, $m^{*}$=$0.067m_e$ is the effective mass of the charge carrier, $m_e$ is the mass of an electron, $e$ is the electron charge and $\tau_q$ is the quantum lifetime. We can determine a characteristic B-field $B_q$ of 33 mT from the condition $\omega_c \tau_q$=1 [see also Fig. \ref{fig:SdHBigTrace}(a), and note that weak SdH oscillations are still visible below $B_q$ as represented by the red colored region with diminished shading in Fig. \ref{fig:PhaseDiagram}(f)]. With increasing $I_{DC}$, the SdH oscillation amplitude first decays before the maxima and minima of the SdH oscillations invert, an effect called phase inversion. Examples of inverted and non-inverted SdH oscillations are given in Fig. \ref{fig:PhaseDiagram}(b). See also Fig. \ref{fig:inversion3d} where inversion occurs at a DC current $I_{PhI}\sim$0.2 $\mu$A near 0.2 T.

In the vicinity of B=0 T, we observe pronounced nMR- see panels (a) and (e) in Fig. \ref{fig:PhaseDiagram}. The magnetoresistance decays rapidly with increasing B-field until a strong change of slope marks the nMR boundary. Assuming a purely ballistic regime, an abrupt change of slope is predicted\cite{scaffidi2017hydrodynamic} to occur when $W=2r_c$, where $r_c$=$m^{*}v_F/eB$=$v_F/\omega_c$ is the cyclotron radius, with a corresponding B-field $B_{W}$, and $v_F$=$1.9 \times 10^7$ cm/s is the Fermi velocity. In our case, $W$=15 $\mu$m.  Figure \ref{fig:SdHBigTrace}(a) shows a resistance versus B-field trace at zero DC current. The change in slope is most rapid at $\sim$ 10 mT. We take this field to be an estimate of $B_{W}$, and this reasonably gives $W\sim$ 15 $\mu$m (as an upper bound). However, the observed change of slope at the boundary is not as abrupt as in the calculations in Ref. \cite{scaffidi2017hydrodynamic}, which is a source of imprecision in the estimation of the location of the boundary. Furthermore, accounting for undercut during the wet etching step in the fabrication of the Hall bar, and sidewall depletion, we expect the \emph{effective} electronic width of the Hall bar $W_{eff}$ to be smaller than $W$. In Sec. \ref{sec:hiro}, from a detailed analysis of the HIROs, we determine $W_{eff}\sim$11 $\mu$m, and the corresponding B-field $B_{W_{eff}}\sim$ 14 mT. Throughout the text we will be careful in our usage of $W$ and $W_{eff}$ and make clear when it is important to distinguish the difference. Note in Fig. \ref{fig:PhaseDiagram}(f) we have marked the nMR boundary with $B_{W_{eff}}$ rather than $B_{W}$. Also accompanying the nMR is a distinctive double-peak feature, clear in panels (c) and (e) in Fig. \ref{fig:PhaseDiagram}, that we ascribe to ballistic transport.

Also close to $B=0$ T, we can identify modifications to the nMR and the double-peak feature with increasing DC current that point to a growing influence of hydrodynamics [depicted by region marked HYDRO with cross hatching of increasing weight in Fig. \ref{fig:PhaseDiagram}(f)]. In a purely hydrodynamic regime, the hydrodynamic contribution to $\rho_{xx}$ is expected to be strongest at B=0 T, and in order to reach the strong hydrodynamic regime, the electron-electron scattering length $l_{ee}$ should be smaller than the width of the Hall bar. We show in Sec. \ref{sec:lee} that $l_{ee}$ decreases with increasing $I_{DC}$, and so the signatures of viscous transport are stronger at high $I_{DC}$. See also the $\rho_{xx}$ versus $I_{DC}$ trace at $B=0$ T in Fig. \ref{fig:PhaseDiagram}(d) which displays an initial decrease in $\rho_{xx}$ with DC current. This behavior resembles that observed in a 2D wire and attributed to the Gurzhi effect, i.e., Poiseuille flow in electron transport\cite{molenkamp1994}. In section \ref{sec:hydro} we will also examine the DC-current induced hydrodynamic correction to the Hall resistivity near zero field.   

Lastly, in the phase diagram, we can identify HIRO peaks that fan out from the origin [see also the section in Fig. \ref{fig:PhaseDiagram}(e) that shows the HIROs either side of the nMR region near zero field]. The boundary above which HIROs are observed is delimited by $I_{HIRO}$ in Fig. \ref{fig:PhaseDiagram}(f), and tracks the DC current position of the first-order HIRO peak that is linear in B [see Eq. \eqref{eq:hirogamma} and Fig. \ref{fig:fanhiro}(a)]. Similar to SdH oscillations, the HIRO peak amplitude is in principal related to the quantum lifetime. However, we find in Sec. \ref{sec:hiro} that the quantum lifetime extracted from the HIROs does not match the quantum lifetime extracted from the SdH oscillations. For this reason, the B-field onset of HIROs is marked $B_q^{HIRO}$ in Fig. \ref{fig:PhaseDiagram}(f). We can discern up to seven orders of HIRO peaks for a 10 $\mu$A DC current. Additionally, we observe unexpected HIRO-like features located between the first-order and the second-order HIRO peaks at high DC current, that are marked by black arrows in Fig. \ref{fig:PhaseDiagram}(a) [see also Fig. \ref{fig:fanhiro}(a) and Fig. \ref{fig:hirohalf}], which we will refer to as the “1.5” features throughout this paper.


\section{Shubnikov-de Haas Oscillations and Phase Inversion}  \label{sec:sdh}

SdH oscillations arise from B-field induced LLs and the oscillating density-of-states (DOS) at the Fermi level \cite{ando1974theory,ando1974,ando1975,ando1982}. For an extensive review of SdH oscillations see Ref. \cite{dmitriev2012nonequilibrium}. In the small B-field regime the SdH oscillations in the resistance $R_{xx}=V_{xx}/I$ (which is equivalent to $r_{xx}$ in the limit of zero DC current) can be analytically described by the conventional Ando formula\cite{coleridge1991, isihara1986}:
\be \label{eq:sdh}
\Delta R_{xx}=4R_0 D_T \cos{\left(\frac{2\pi E_F}{\hbar \omega_c}-\pi\right)}\exp\left({-\frac{\pi }{\omega_c\tau_q}}\right),
\ee
where $\Delta R_{xx}$ is the SdH oscillation amplitude, $R_0$ is the zero-field resistance, and $D_T$ is the thermal damping factor:
\be \label{eq:sdh2}
D_T=\frac{X_T}{\sinh(X_T)},\;\;   X_T=\frac{\mathrm{2}\pi^2k_BT}{\hbar \omega_c}.
\ee
Here $k_B$ is the Boltzmann constant, $\hbar$ is the reduced Planck constant, and $T$ is the bath temperature. 
In the phase diagram in Fig. \ref{fig:PhaseDiagram}, the SdH oscillations exhibit two notable features. First, the SdH oscillation amplitude decreases with increasing DC current, and second, an inversion of the SdH oscillation extrema occurs on increasing the DC current further. By fitting the experimental SdH oscillation data to Eq. \eqref{eq:sdh} it is possible to extract the electron concentration, the quantum lifetime, the amplitude (4$R_0$), and the current dependence of the electron temperature $T_e$. Note the use of $T_e$ instead of $T$ has been shown \cite{phaseinversionsergei,phaseinversiongraphene} to be essential to describe the decrease of the SdH oscillation amplitude and the PhI with increasing DC current. We emphasize that for our narrow HB device, there is pronounced nMR around $B$=0 T [see Fig. \ref{fig:SdHBigTrace}(a)], and whether one can simply equate $R_0$ in Eq. (1) for the SdH oscillations \cite{mancoff1996} with $R$(B=0 T), the measured value of the zero-field resistance, in this case is important since this impacts the determination of the mobility. 

At $I_{DC}$=0 $\mu$A, by fitting the SdH oscillations in Fig. \ref{fig:SdHBigTrace}(a) to Eq. \eqref{eq:sdh}, we obtain parameters $n$=$2.0 \times 10^{11} cm^{-2}$, $R_0$=$11.4 \pm 0.6$ $\Omega$, and $\tau_q$=$11.5 \pm 0.3$ ps. For our experimental conditions, it is possible to detect SdH oscillations up to filling factor $\nu$=$2 \pi \hbar n/eB$ of $400$ [see Fig. \ref{fig:SdHBigTrace}(b)]. Strongly note that the value of $R_0$ as determined from SdH is significantly lower than the measured resistance at $B$=0 T  [R(B=0)=43 $\Omega$], indicating that the phenomenon of nMR near zero field is independent from the SdH oscillation effect. Consequently, without accounting for the presence of the nMR (and the double-peak feature superimposed on top of the nMR), taking the $R$(B=0 T)=43 $\Omega$ value rather than the fitted value of $R_0$=11.4 $\Omega$ relevant to the SdH oscillations, the estimated mobility would be lower than the bulk mobility by almost a factor of four \cite{sdhmobility}. We note in our study $R_0$ is close to the bulk resistance, whereas in other experiments \cite{coleridge1996} the value of $R_0$ extracted from SdH oscillations does not necessarily correspond to the resistance at zero field, i.e., for our situation the assumption implicit in the Ando formula that the B-field dependence of the DOS is sinusoidal in nature to first-order is justified.

\begin{figure}[!b]\hspace*{-0.1in}
\includegraphics[width=1\columnwidth]{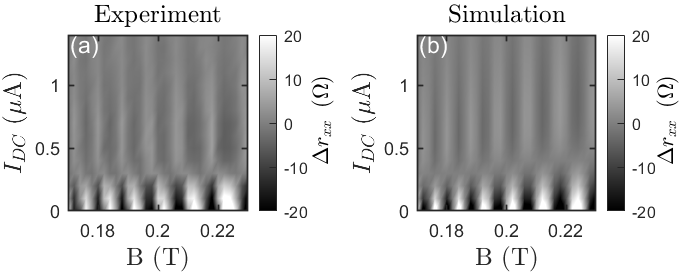}
\caption{Grayscale plots showing PhI of the SdH oscillation amplitude $\Delta r_{xx}$ in $r_{xx}$ data near 0.2 T (a), and a simulation of the data (b). The simulation is generated from Eq. \eqref{eq:phaseinversion}.}
\label{fig:inversion3d}
\end{figure}

Away from $I_{DC}$=0 $\mu$A, the SdH oscillation amplitude decreases, and eventually PhI occurs- see Figure \ref{fig:inversion3d} where this takes place at $\sim 0.2$ $\mu$A near 0.2 T. PhI whereby the extrema of the SdH oscillations invert with increasing DC current was discussed theoretically in Refs. \cite{vitkalov2009nonlinear,Dmitriev2011}, and has been reported experimentally for high mobility semiconductor 2DEGs \cite{Kalmanovitz2008, Webber2012, Dietrich2012, phaseinversionsergei,phaseinversionYu,Zhang2009}, and for graphene \cite{phaseinversiongraphene}. For $r_{xx}$, PhI with increasing DC current can be explained by an electron heating model in terms of the following relation\cite{phaseinversionsergei}:
\be \label{eq:phaseinversion}
\Delta r_{xx} = \Delta R_{xx} + I_{DC} \frac{\partial    \Delta R_{xx}}{\partial  T_e} \frac{\partial  T_e}{\partial  I},
\ee
derived from $\Delta r_{xx}$=$d(\Delta R_{xx} I)/dI$, where $\Delta r_{xx}$ is the SdH oscillation amplitude in the differential resistance, and $\Delta R_{xx}$ is the magnetoresistance described in Eq. \eqref{eq:sdh}. The term $\partial    \Delta R_{xx} / \partial  T_e$ introduces oscillations shifted by a phase of $\pi$ that dominate at larger $I_{DC}$. At higher $T_e$, the damping factor $D(T)$ from Eq. \eqref{eq:sdh2} decreases the amplitude of the SdH oscillations. Taken from the data set shown in full in Fig. \ref{fig:PhaseDiagram}, Fig. \ref{fig:inversion3d}(a) shows $\Delta r_{xx}$ near 0.2 T. Figure \ref{fig:inversion3d}(b) shows a simulation of the experimental data obtained with Eq. \eqref{eq:phaseinversion}. The simulation assumes the electron temperature is a fitting parameter of the form $T_e(I_{DC})$=$T_0 + \alpha I_{DC}^{\beta}$, where $T_0$=0.04 K is the expected electron temperature at zero current, and $\alpha$ and $\beta$ are constants. The empirical dependence of the electron temperature with DC current is found to be $T_e(I_{DC})$=$T_0+1.0I_{DC}^{0.65}$ where $I_{DC}$ is in $\mu$A. To obtain the parameters $\alpha$ and $\beta$, a Monte Carlo approach is used as both $T_e$ and $ \partial T_e/ \partial I$ need to be fitting parameters in the model. We find $\beta$ $\sim 0.65$ consistent with the value reported in previous work by Studenikin \textit{et al.} \cite{phaseinversionsergei} in an InGaAs/InP QW HB device with a much smaller carrier mobility of $1.9 \times 10^5$ $cm^2/Vs$. This demonstrates that electron heating is also an important mechanism for PhI in very high mobility HB devices in the large filling factor regime.

\section{Negative Magnetoresistance} \label{sec:nmr}

nMR is commonly observed in high-mobility 2DEG Hall bar devices, \cite{bockhorn2013magnetoresistance,hatke2012giant,mani2013size,shi2014,Wang2016,Samaraweera2017,Samaraweera2018,Samaraweera2020,HornCosfeld2021,Zudov2001,Bockhorn2011,Bockhorn2014} and narrow micron-sized quantum devices such as quantum wires and quantum point contacts \cite{Choi1986,thornton1989boundary,van1988aharonov,van1988four,Mani1993}.
nMR can arise from different effects. In one limit, nMR is often associated with weak localization (WL), see for example Refs.\cite{Paalanen1984,kramer1993localization,mani2013size}, which is due to coherent back-scattering with impurities, when the mean free path is small compared to the phase coherence length. nMR also occurs in the opposite (ballistic) regime, when the mean free path is larger than the Hall bar width and the distance between voltage probes, and coherent back-scattering is negligible, see for example Refs. \cite{bockhorn2013magnetoresistance,hatke2012giant,mani2013size,Samaraweera2018,Samaraweera2020,HornCosfeld2021,Zudov2001,Bockhorn2011}. Our Hall bar is in this regime, and so the strong nMR we observe in Fig. \ref{fig:SdHBigTrace} arises from the ballistic effect. 

For a fully ballistic conductor with lateral confinement, the nMR is a direct consequence of the non-zero resistance due to the finite number of quantum channels. In the limiting case, where there is no electron scattering, the two-terminal resistance is simply given by $R_{2T}=h/g_s e^2N$, where $h$ is the Planck constant, $g_s=2$ is the spin degeneracy, and $N$ is the number of quantum transmission channels. For a confinement potential $V(y)$ in the transverse direction, the number of channels can be found by using the Bohr-Sommerfeld quantization rule: 
\be 
N=\frac{2}{h}\int_{y_1}^{y_2}pdy.
\label{Bohr}
\ee 
Here $p=\sqrt{2m^{*}(E_F-V(y))}$ is the quasi-classical momentum, and $y_1$ and $y_2$ are the boundaries where $p$ is real. For a hard wall confinement potential of width $W$ ($y_1=-W/2$ and $y_2=W/2$), and without a B-field, this leads to the well know expression $N=Wk_F/\pi$, where the number of channels is twice the width divided by the Fermi wave length. In the presence of a B-field, $N$ is reduced due to Landau quantization, and the quasi-classical momentum is now given by $p=\sqrt{2m^{*}(E_F-V(y))-(eBy)^2}$. For a hard-wall confinement potential, Eq. \eqref{Bohr} can be evaluated directly and one obtains: 
\be \begin{split}
N=\frac{Wk_F}{\pi}f\left(\frac{W}{2r_c}\right)\mbox{ ; }  \\
f(s)=\frac{1}{2}\left(\sqrt{1-s^2}+\frac{\arcsin{s}}{s}\right).
\label{G_theory} \end{split}\ee
Equation \eqref{G_theory} is a generalization to the approximate expression obtained by Glazman and Khaetskii\cite{glazman1989quantum}, who considered the two limits $W\ll 2r_c$ and $W\gg 2r_c$ separately. In these two limits both expressions coincide. For a smooth confinement potential, $N$ can be obtained by evaluating the integral in Eq. \eqref{Bohr} numerically. 

\begin{figure}[t]
\includegraphics[width=.8\columnwidth]{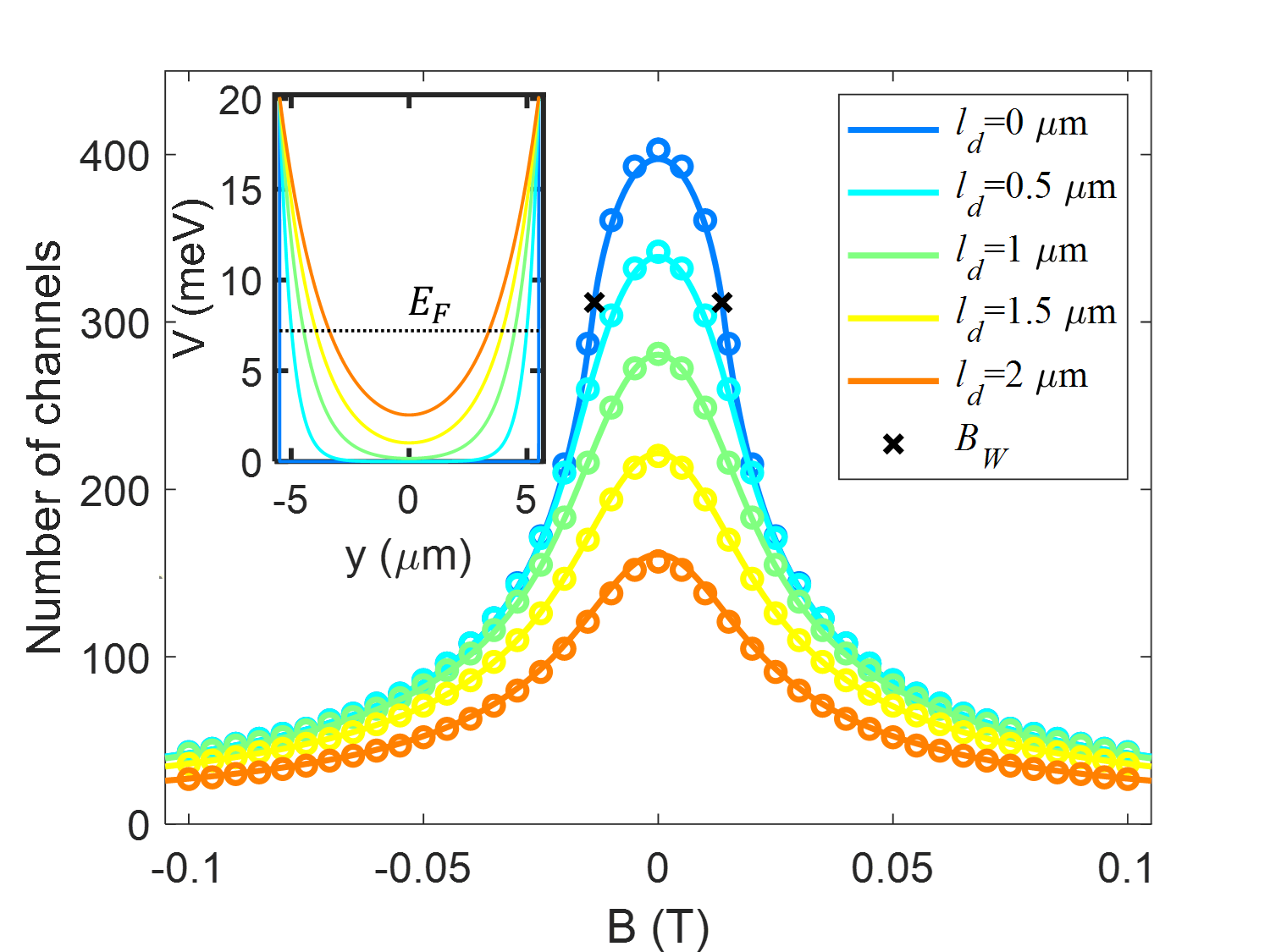}
\caption{Calculated number of transmission channels, $N$ (proportional to the conductance) as a function of B-field for different values of the sidewall depletion length, $l_d$ in $\mu$m. The lines are obtained using Eq. \eqref{Bohr}, while the open dots are obtained by computing the two-terminal transmission numerically using a discretized Schr\"odinger equation with two semi-infinite leads. The crosses indicate the B-fields $B_W$ where $W=2r_c$. Here a width of $W=11$ $\mu$m is assumed, and we have taken $U_0$ to be 20 meV. Inset: calculated depletion potential for a section through the Hall bar for different values of $l_d$. $l_d$=0 $\mu$m corresponds to a hard-wall potential.}
\label{fig:depletion}
\end{figure}

For instance, if we now consider, as depicted in the inset to Fig.  \ref{fig:depletion}, an exponential depletion potential of the form: 
\be
V(y)=U_0{(e}^{-|y-y_1|/l_d}+e^{-|y-y_2|/l_d}), 
\ee
with $l_d$ the depletion length, and $U_0$ the barrier potential, we can compare the numerically obtained conductance, assuming no diffuse scattering (in the bulk or at the edge of the Hall bar), with Eq. \eqref{Bohr}. Both are in excellent agreement as shown in Fig. \ref{fig:depletion}. The most distinctive feature is the rapid decrease of the number of transmission channels with increasing B-field. The characteristic field $B_W$ for the rapid decrease is given by $W$=$2r_c$ for which the slope of $N(B)$ is maximal. Our simple model therefore can explain the strong nMR observed in Fig. \ref{fig:SdHBigTrace}(a). However, note that $B_W$ here is closer to the maximum of $N(B)$, while in Fig. \ref{fig:SdHBigTrace}(a), $B_{W}$ is further down the side of the nMR. In the Hall bar measured, this is likely due to the presence of diffuse scattering, and the voltage probe regions, both of which are neglected in the simple model and the numerical simulation. 
Indeed, for a Hall bar in the strong ballistic regime, scattering is dominated by scattering off the openings to the voltage probes along the sides, since no scattering would lead to a zero four-terminal resistance. Scattering from the voltage probes is proportional to $N$, hence the four-terminal resistance will show a similar $B$-dependence as $N(B)$, leading to the observed nMR. Moreover, in ballistic narrow Hall bars, the two-terminal resistance at $B=0$ T is bounded by the number of conducting channels, i.e., $R>h/2e^2N$ ($R>16$ $\Omega$ in our Hall bar), regardless of the scattering potential and Hall bar aspect ratio. As a consequence, the transport mobility (or transport lifetime) cannot be extracted from the measured resistance at $B=0$ T when nMR is present but only from a fit of the SdH oscillations as discussed in Sec. \ref{sec:sdh}.

In addition to the nMR we also observe a dip at $B=0$ T leading to the distinctive double-peak feature seen in Fig. \ref{fig:PhaseDiagram}(e). The double-peak feature has been reported in recent experiments on narrow channels in 2D metals \cite{Moll2016}, and nanowires in graphene \cite{Masubuchi2012}, and was attributed to ballistic transport. The double-peak feature has also been discussed elsewhere both theoretically \cite{scaffidi2017hydrodynamic} and experimentally \cite{Ditlefsen1966,Masubuchi2012,sulpizio2019visualizing}, where the peaks were reported to occur at W$\simeq$0.55$r_c$. For our Hall bar device, we infer $W_{eff}\simeq$0.65$r_c$ from the position of the two peaks in the double-peak feature at zero DC current (B$\sim$ $\pm$4.7 mT). In the calculation displayed in Fig. \ref{fig:depletion} no double-peak feature is visible. We found that whether a double-peak feature appears in a calculation or not depends sensitively on the details of the boundary scattering assumed, and is more prominent with diffuse scattering on the edge along the Hall bar channel \cite{thornton1989boundary}. For the calculation relevant to Fig. \ref{fig:depletion} we assumed no diffuse scattering at the edge nor in the bulk. We strongly note that because we are in the regime $W$ $\ll$ $l_{mfp}$ $\sim$ 145 $\mu$m, the nMR and double-peak feature we observe are ballistic in origin and do not arise from a combination of WL and weak anti-localization (WAL)\cite{hikami1980spin}. In GaAs/AlGaAs hetero-structures spin-orbit interaction is weak, and even in the case of a moderate mobility 2DEG where WL and WAL have been observed to co-exist, the resulting double-peak feature is tightly confined within $\pm$0.1 mT of zero field \cite{Miller2003}. In the next section, we take a closer look at the evolution of $\rho_{xx}$ \emph{with DC current} in the vicinity of B=0 T and show that further corrections to $\rho_{xx}$ develop from viscous transport.


\section{Hydrodynamic Electron Transport}  \label{sec:hydro}
In this section we discuss the impact of hydrodynamics on electron transport when DC current is driven through the Hall bar. As discussed in the introduction, to reach the hydrodynamic regime, increasing the temperature is a common strategy taken to drive down $l_{ee}$ so that the condition $l_{ee}$ $\ll$ $W$ is attained. We instead employ an increasing DC current to suppress $l_{ee}$, following the predictions by Giuliani and Quinn\cite{giuliani1982lifetime} that $l_{ee}\sim 1/I_{DC}^2$. We first demonstrate the DC current induced suppression of $l_{ee}$. Subsequently, we isolate the hydrodynamic contribution (correction) to both the differential longitudinal and transverse resistivities near $B$=0 T, and compare to existing theories for hydrodynamic electron transport.

\subsection{Current-induced Suppression of Electron-Electron Scattering Mean Free Path} \label{sec:lee}

\begin{figure}[t]
\includegraphics[width=1\columnwidth]{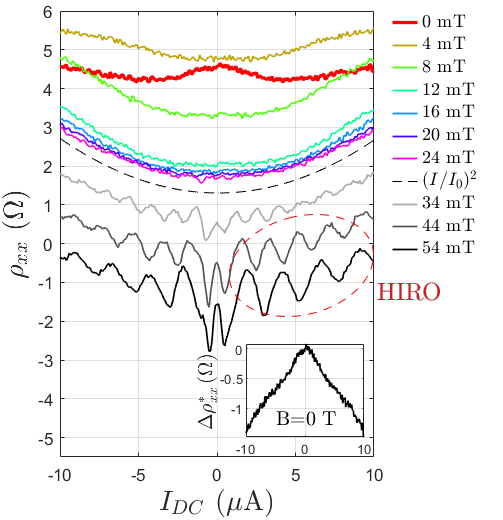}
\caption{$\rho_{xx}$ versus $I_{DC}$ for different B-fields. The traces are taken from the data set shown in Fig. \ref{fig:PhaseDiagram}. Away from $B$=0 T, $\rho_{xx}$ exhibits a background quadratic dependence with respect to $I_{DC}$. The dashed line is a quadratic fit of the average $\rho_{xx}$ from 12 mT to 24 mT, shifted down by 0.5 $\Omega$ for clarity. For this B-field range, hydrodynamic effects are suppressed, and SdH and HIRO are not yet observable. Above $B\sim 30$ mT, SdH oscillations and HIROs respectively start to appear at zero current and finite current, although the background quadratic dependence is still evident. Traces for $B>30$ mT are shifted down progressively by 1 $\Omega$ for clarity. Red dashed line ellipse: HIRO amplitude decreases with increasing $I_{DC}$ (see Sec. \ref{sec:hiro} for discussion). Inset: Hydrodynamic component $\Delta \rho_{xx}^{*}$=$\rho_{xx}-\rho_{xx,I=0}-\Delta\rho_{xx}^{bg}$ versus DC current in micro-amperes at $B$=0 T. Here $\rho_{xx,I=0}$ $\sim$ 4.6 $\Omega$. From a linear fit we find $\Delta \rho_{xx}^{*}$=$-0.155 |I_{DC}|$). This component has a negative sign. See text for further discussion.}. 
\label{fig:rhoxxcurrent}
\end{figure} 
We start by discussing the necessary pre-condition for hydrodynamics, namely $l_{ee} \lesssim W$, and how by increasing the DC current flowing through the Hall bar device $l_{ee}$ can be decreased. By careful inspection of vertical sections through the phase diagram in Fig. \ref{fig:PhaseDiagram}(a) near zero field, we can identify a general background quadratic dependence of $\rho_{xx}$ with respect to $I_{DC}$, $\Delta \rho_{xx}^{bg}$. Other than at zero field, as illustrated in Fig. \ref{fig:rhoxxcurrent}, this background quadratic dependence is clearly seen for traces when $|B|\lesssim 30$ mT. At higher B-field, the background quadratic dependence is still present but masked by the onset of SdH oscillations near zero current, and HIROs at finite current. From analysis of $\rho_{xx}$ traces between 12 mT and 24 mT in Fig. \ref{fig:rhoxxcurrent}, we find that the background quadratic dependence follows the relationship:
\be
\label{eq:rhoxxbg}
\frac{\Delta\rho_{xx}^{bg}(I_{DC})}{\rho_{xx}(0)}=\left(\frac{I_{DC}}{I_0}\right)^2, 
\ee
where $\rho_{xx}(0)$=$1.81 \pm 0.01$ $\Omega$, and $I_0$=$11.4 \pm 0.1$ $\mu$A. Strongly note that in our forthcoming discussion, $\Delta \rho_{xx}^{bg}$ is only a function of $I_{DC}$, i.e., for $|B|\lesssim 30$ mT it is assumed to have no B-field dependence, and furthermore, $\Delta \rho_{xx}^{bg}$ is defined to be zero at zero current. We attribute the quadratic dependence to a DC current induced increase of the electron-electron scattering rate $\tau_{ee}^{-1}$. Generally, the resistivity of a 2DEG is proportional to the sum of the scattering rates from different sources namely $\rho_{xx}$=$(m^{*}/e^2n)\sum_{i} \tau^{-1}_{i}$, where $\tau^{-1}_{i}$ are independent scattering rates for the different sources of scattering \cite{Yu2005}. A DC current induced increase in the electron-electron scattering rate $\tau_{ee}^{-1}$ therefore results in a correction to the resistivity on the order of $\tau_{ee}^{-1}$ at low field.

In an ideal 2DEG, the following analytical expression for the evolution of the electron-electron scattering rate with DC current, rather than with temperature, was derived by Chaplik \cite{chaplik} and Giuliani and Quinn: \cite{giuliani1982lifetime} 
\be 
\label{eq:ee}
\tau_{ee}^{-1}=\frac{E_F}{4\pi \hbar} \left(\frac{\Delta}{E_F}\right)^2 \left[\ln\left(\frac{E_F}{\Delta}\right) + \ln\left(\frac{2Q_{TF}}{k_F}\right) + \frac{1}{2} \right],
\ee
where $\Delta$ is the excitation (or excess) energy relative to $E_F$ (satisfying $\Delta \ll \hbar^2 k_F Q_{TF}/m^{*}$), $k_F$ is the Fermi wavevector, $Q_{TF}$=$2m^{*} e^2/4\pi\epsilon_r\epsilon_0\hbar^2$ is the 2D Thomas-Fermi screening wave vector, $\epsilon_r$ is the dielectric constant ($\sim$13.1 for GaAs), and $\epsilon_0$ is the vacuum permittivity. For a sufficiently small excess energy, Eq. \eqref{eq:ee} is approximately quadratic with respect to DC current. In our measurements, over the DC current range probed, the excess energy is small. The average resistance between 0 and 10 $\mu$A at zero field $R_{av}$ is approximately 40 $\Omega$, and so we estimate the maximum excess energy $\Delta$ to be of order $eV_{DC}\sim$0.4 meV, where $V_{DC}$ is the DC voltage drop between the two voltage probes along the Hall bar at 10 $\mu$A and B=0 T. This estimated value of $\Delta$ satisfies the condition $\Delta \ll \hbar^2 k_F Q_{TF}/m^{*}$ since $\hbar^2 k_F Q_{TF}/m^{*}$=24.7 meV. However, we note that the quantum interference experiment by Yacoby \textit{et al.}\cite{yacobytauee} validating this theory suggested that $\Delta$ is proportional to $eV_{DC}$ and the actual excess energy is smaller than the applied voltage\cite{yacobydelta}.

\begin{figure*}
\includegraphics[width=2\columnwidth]{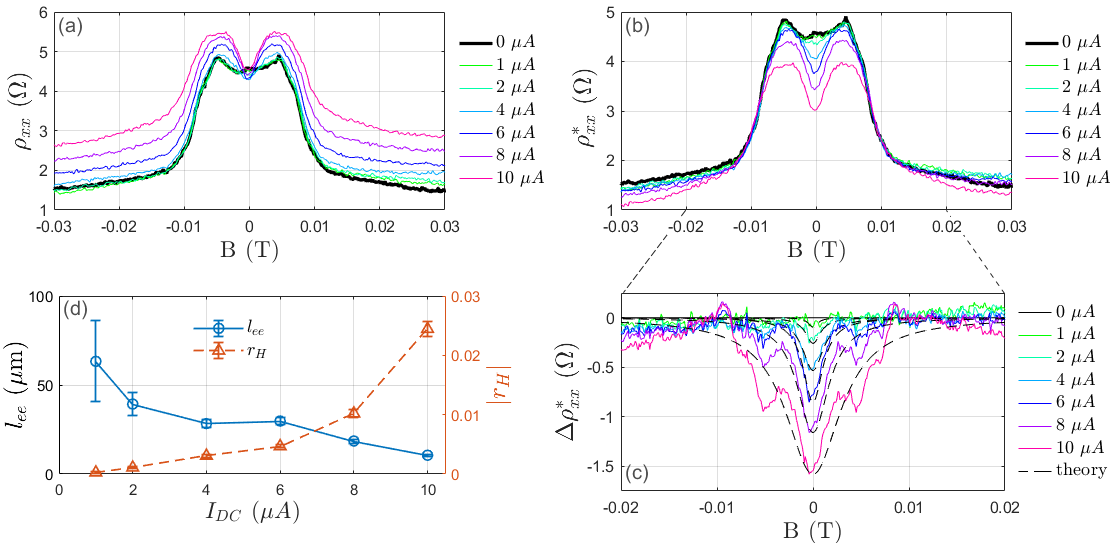}
\caption{
(a) $\rho_{xx}$ versus B traces on sweeping the B-field for different DC currents (B-field sweep rate: 10 mT/min for zero current trace and 20 mT/min for all other traces). (b) $\rho_{xx}^{*}$=$\rho_{xx}-\Delta\rho_{xx}^{bg}$ versus B-field. For each of the traces in (a) the background quadratic current dependence $\Delta\rho_{xx}^{bg}$ has been removed. For $|B| \lesssim 0.01$ T, we can see an increasingly strong decrease in $\rho_{xx}^{*}$ on raising the DC current which we attribute to hydrodynamics. (c) Solid lines:  DC current-dependent deviation $\Delta\rho_{xx}^{*}$=$\rho_{xx}^{*}-\rho_{xx, I=0}^{*}$ of $\rho_{xx}^{*}$ from the zero current residual $\rho_{xx, I=0}^{*}$, namely the hydrodynamic component. Dashed lines: $r_H \Delta \rho_{xx}^{hyd}$ fit to the data in a perturbative approach, where $r_H$ is a fitting parameter reflecting the impact of the viscous correction, and $\Delta \rho_{xx}^{hyd}$ is from the hydrodynamic theory as described by Eq. \eqref{eq:rxxhydro}. For the fits, $W_{eff}$=11 $\mu$m is used. The distinct ``dips'' at $|B|\sim$ 5 mT, and the weaker ``dips'' at $|B|\sim$ 8 mT, are artifacts of the methodology due to the subtraction of the zero current trace which has peaks at $|B|\sim$ 5 mT and weak ``shoulders'' at $|B|\sim$ 8 mT. Both constitute the double-peak feature, ballistic in origin, discussed in the nMR section, which itself has a DC current dependence. (d) $l_{ee}$ and $|r_H|$ extracted from fits of the traces in (c) plotted against DC current. As $I_{DC}$ increases, $l_{ee}$ decreases and the hydrodynamic component ($|r_H|$) becomes stronger as expected.}
\label{fig:hydrorxx}
\end{figure*}

As revealed in Fig. \ref{fig:rhoxxcurrent}, in the narrow range around zero field ($|B|\lesssim 5$ mT), $\rho_{xx}$ does not follow the $I_{dc}^2$ dependence. Specifically, at $B$=0 T, $\rho_{xx}$ decreases with $I_{DC}$ before reaching a minimum at $\sim \pm4$ $\mu$A, and then increases at higher DC current [also see Fig. \ref{fig:PhaseDiagram}(d)]. This behavior is reminiscent of that observed in an early study by Molenkamp and de Jong \cite{molenkamp1994,Molenkamp1994part2} in a wire formed from a 2DEG for which the decrease in the differential resistivity with increasing $I_{DC}$ was assigned to a hydrodynamic effect. On subtracting the background quadratic dependence $\Delta\rho_{xx}^{bg}$ from $\rho_{xx}$ at $B$=0 T, relative to the zero current value of $\rho_{xx}$, we obtain the negative residual $\Delta \rho_{xx}^{*}$=$\rho_{xx}-\Delta\rho_{xx}^{bg}$ which exhibits a linear dependence in DC current as shown in the inset to Fig. \ref{fig:rhoxxcurrent}. In other words, $\rho_{xx}$ versus DC current at $B$=0 T is the sum of two components: a positive component $\Delta \rho_{xx}^{bg}$ quadratic in DC current related to the electron-electron scattering rate, and a negative component $\Delta\rho_{xx}^{*}$ linear in DC current. We attribute the latter to a DC current-induced hydrodynamic effect, driven by the enhancement of conductance of electrons in the hydrodynamic regime as expected in the ballistic regime \cite{guo2017higher,krishna2017superballistic}, which leads to a negative hydrodynamic correction in the resistivity $\Delta\rho_{xx}^{*}$. We note that in contrast when hydrodynamics is induced by temperature in the non-ballistic regime, a positive correction to the resistivity is expected as observed \cite{gusev2018viscous}.

\subsection{Hydrodynamic Magnetoresistance}
 \label{sec:hydro2}

In the previous subsection, we established that increasing $I_{DC}$ suppresses $l_{ee}$, a necessary pre-condition to enter the hydrodynamic regime. In this subsection, we analyze the evolution of $\rho_{xx}$ with $I_{DC}$ for small B-fields ($|B|<10$ mT), and compare the change in $\rho_{xx}$ to that expected from existing hydrodynamic theory. Figure \ref{fig:hydrorxx}(a) shows $\rho_{xx}$ versus B traces for different DC currents up to 10 $\mu$A. The goal is to first identify and then isolate the various components that can contribute to the DC current dependence of $\rho_{xx}$. One component, as established in the previous subsection (see Fig. \ref{fig:rhoxxcurrent}), is the general background quadratic increase in $\rho_{xx}$ with DC current $\Delta \rho_{xx}^{bg}$, and another component near zero field, the focus of our attention, that we will argue is hydrodynamic in origin. By removing $\Delta \rho_{xx}^{bg}$ from $\rho_{xx}$ over the B range from -30 mT to +30 mT, and examining the residual $\rho_{xx}^{*}$ plotted in Fig. \ref{fig:hydrorxx}(b), we find that for $|B|\lesssim$10 mT, there is clearly a growing negative component to $\rho_{xx}$ with increasing DC current. We can isolate this negative component from the nMR and the double-peak feature by computing $\Delta\rho_{xx}^{*}$=$\rho_{xx}^{*}-\rho_{xx, I=0}^{*}$, the DC current-dependent deviation of $\rho_{xx}^{*}$ from the zero current residual $\rho_{xx, I=0}^{*}$. The deviation is plotted in Fig. \ref{fig:hydrorxx}(c). We attribute this deviation to hydrodynamics because of its growing amplitude with increasing DC current, and decreasing amplitude with increasing B-field. The latter is discussed in Ref. \cite{scaffidi2017hydrodynamic}. Next, we discuss our method to compare $\Delta\rho_{xx}^{*}$ to existing theory.

\begin{figure}[t!]
\includegraphics[width=0.85\columnwidth]{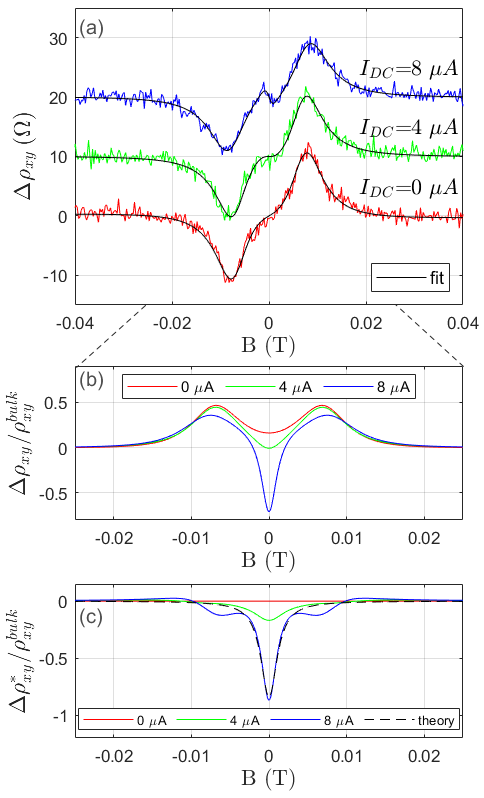}
\caption{(a) Deviation from the conventional (bulk) Hall resistivity $\Delta \rho_{xy}=\rho_{xy}-\rho_{xy}^{bulk}$ for three values of DC current. The plots are derived from $\rho_{xy}$ versus B traces on sweeping the B-field at 20 mT/min. The black lines are fits to a sum of polynomials that allow accurate division by $\rho_{xy}^{bulk}$. Plots are offset by 10 $\Omega$ for clarity. The change in the slope at zero field from positive to negative with increasing DC current is a signature of the growing influence of hydrodynamics. The extrema near $|B|\sim$ 8 mT arise from ballistic transport \cite{blaikie1995, gusev2018viscous, raichev2020}. We note that the derivative of the $\rho_{xx}$ trace for $I_{DC}$=8 $\mu A$ has a close resemblance to the $\Delta\rho_{xy}$ trace for $I_{DC}$=8 $\mu A$. (b) Calculated from the fits in panel (a), the ratio $\Delta \rho_{xy}/\rho_{xy}^{bulk}$ is plotted to emphasize the hydrodynamic component near 0 T. The negative value near 0 T in the 8 $\mu A$ curve is a clear sign of hydrodynamics. (c) Isolated hydrodynamic component $\Delta \rho_{xy}^{*}/\rho_{xy}^{bulk}=(\Delta \rho_{xy}-\Delta \rho_{xy,I=0})/\rho_{xy}^{bulk}$ determined from curves in panel (b). Dashed line: Fit to $I_{DC}$=8 $\mu A$ curve following hydrodynamic theory Eq. \eqref{eq:hallr}; see text for parameters. For the fit, $W_{eff}$=11 $\mu$m is used.}
\label{fig:hallR}
\end{figure}

We now examine in further detail the experimental data in Fig. \ref{fig:hydrorxx}(c). In the purely hydrodynamic regime, following Scaffidi \textit{et al.} \cite{scaffidi2017hydrodynamic}, the viscous correction to the magnetoresistance for a 2DEG when $l_{ee} \ll W \ll l_{mfp}$ can be expressed as:
\be \label{eq:rxxhydro}
\Delta \rho_{xx}^{hyd}=\frac{m^{*}}{e^2n}\frac{v_F l_{ee}}{W^2}\frac{3}{1+\left(\frac{{2l}_{ee}}{r_c}\right)^2}.
\ee
For our Hall bar device, $W_{eff}$ $\sim$ $W \ll l_{mfp}$ is trivially satisfied since $l_{mfp}$=145 $\mu$m. Using Eq. \eqref{eq:ee} with $\Delta$=$eV_{DC}$=$eI_{DC}R_{av}$, $l_{ee}$=$v_F \cdot \tau_{ee}$ is predicted to be smaller than $W$ for $I_{DC} \gtrsim 9.5$ $\mu$A. Therefore, for an increasing DC current up to 10 $\mu$A, the maximum applied, the 2DEG transitions from the ballistic regime ($W \ll l_{ee}, l_{mfp}$) to the hydrodynamic regime. This transitional phase has \emph{both} ballistic and hydrodynamic contributions. To describe this regime we use a perturbative approach, where we assume that (i) the 2DEG is initially in the ballistic regime, (ii) the change in $\rho_{xx}$ with increasing DC current near zero field is solely hydrodynamic in origin, and (iii) the change is proportional to the viscous correction described in  Eq. \eqref{eq:rxxhydro}. In other words, the change in $\rho_{xx}$ with increasing DC current following our model has the form:
\be \label{eq:perturbation}
\Delta \rho_{xx}=\rho_{xx}-\rho_{xx, I=0}= \Delta \rho_{xx}^{bg} + r_H \Delta \rho_{xx}^{hyd},
\ee
 where $r_H$ is a dimensionless parameter characterising the relative strength of the viscous correction at different DC currents. The term $r_H \Delta \rho_{xx}^{hyd}$ in Eq. \eqref{eq:perturbation} corresponds to $\Delta\rho_{xx}^{*}$ determined from the experimental data and plotted in Fig. \ref{fig:hydrorxx}(c). Fitting $r_H \Delta \rho_{xx}^{hyd}$ to $\Delta\rho_{xx}^{*}$, we can extract values for both $l_{ee}$ and $r_H$. The obtained DC current dependencies of $l_{ee}$ and $r_H$ are presented in Fig. \ref{fig:hydrorxx}(d). $l_{ee}$ is found to decrease with increasing DC current, and its value is comparable to that calculated from theory Eq. \eqref{eq:ee} [for example, at 10 $\mu$A, Eq. \eqref{eq:ee} predicts $l_{ee}$=14 $\mu$m and we obtain $l_{ee} \sim 11$ $\mu$m], which supports our hydrodynamic interpretation in the small B-field regime. $|r_H|$ increases faster than linear with DC current which is reflective of the growing hydrodynamic component. Lastly, we estimate the electron shear viscosity\cite{gurzhi1968hydrodynamic}, defined as $\eta$=$v_F^2\tau_{ee}/4$, to be 0.7 $m^2$/s at 10 $\mu$A. In comparison, in the work of Gusev \textit{et al.} in Ref. \cite{gusev2018viscous} for which hydrodynamics is temperature-induced and no DC current is applied, $\eta$ is found to be 0.3 $m^2$/s at T=1.4 K. 
 
 Comparing our experimental observations for DC current-induced hydrodynamics in the differential longitudinal resistivity in Fig. \ref{fig:hydrorxx}(a) with those for temperature-induced hydrodynamics in recent works of Gusev \textit{et al.} \cite{gusev2018viscous} and Raichev {\em et al.} \cite{raichev2020} there is one notable difference. In our case, the ``dip'' at zero field is seen to grow with increased DC current. In the case of Gusev \textit{et al.} and Raichev {\em et al.}, the ``dip'' is seen to weaken with increased temperature, and by 40 K the low-field double-peak feature has disappeared (a single peak showing nMR remains). In the work of Raichev {\em et al.}, a classical kinetic model was used to compute the joint ballistic and hydrodynamic contributions for temperature-induced hydrodynamics. In the theoretical analysis the authors assume that boundary scattering is independent of temperature, which is reasonable in their experiments. However, this is likely not true for our case where we increase the DC current, which will lead to a change of the effective edge potential due to the large bias along the edges of the HB. Indeed, a large current is expected to lead to less diffusivity in boundary scattering, due to the averaging over a wide energy window, and hence to an enhanced double-peak feature as observed in Fig. \ref{fig:hydrorxx}(a).

\subsection{Hydrodynamic Contribution to Hall Resistivity}
 \label{sec:hydrorxy}

In the previous subsection, we discussed the growing hydrodynamic contribution to $\rho_{xx}$ with increasing $I_{DC}$ near zero field. Similarly, hydrodynamics affects $\rho_{xy}$. In Fig. \ref{fig:hallR}(a), we show the DC current-dependent deviation of $\rho_{xy}$ from the conventional (bulk) Hall resistivity $\rho_{xy}^{bulk}$, namely $\Delta\rho_{xy}$=$\rho_{xy}$-$\rho_{xy}^{bulk}$, for $I_{DC}$ values of 0, 4, and 8 $\mu$A, where $\rho_{xy}^{bulk}$=$-B/en$. At B=0 T, the sign of the slope changes from positive to negative with increasing DC current. Other than a global sign difference, note that the form of the $I_{DC}$=8 $\mu$A trace is similar to traces reported in Ref. \cite{gusev2018viscous} measured between 1.7 K and 40 K with zero DC current supporting our hydrodynamic interpretation. Following the approach of Gusev \textit{et al.} where the induced change with temperature was tracked instead \cite{gusev2018viscous}, the induced change with DC current we see can be better visualized by examining the ratio $\Delta\rho_{xy}/\rho_{xy}^{bulk}$: see Fig. \ref{fig:hallR}(b). As with $\rho_{xx}$, we analyze the change in $\rho_{xy}$ with DC current near the 0 T and compare to theory in a perturbative method. In the purely hydrodynamic regime, incorporating earlier work by Alekseev \cite{alekseev2016negative}, the viscous Hall correction to $\rho_{xy}$ was derived by Scaffidi \textit{et al.} \cite{scaffidi2017hydrodynamic} and found to be:
\be \label{eq:hallr}
\frac{ \Delta \rho_{xy}^{hyd} }{\rho_{xy}^{bulk}} =  \left[ -\frac{6}{1+(2l_{ee}/r_c)^2}\left(\frac{l_{ee}}{W}\right)^2 \right],
\ee
where $\Delta \rho_{xy}^{hyd}$=$\rho_{xy}-\rho_{xy}^{bulk}$ is the hydrodynamic contribution to the bulk Hall resistivity. Following the same approach as for $\rho_{xx}$, the change in the deviation of $\rho_{xy}$ with increasing DC current relative to the zero current deviation $\Delta \rho_{xy}^{*}$=$\Delta \rho_{xy} - \Delta \rho_{xy,I=0}$ corresponds to $r_H \Delta \rho_{xy}^{hyd}$ in a perturbative method, where $r_H$ is a dimensionless fitting parameter reflecting the impact of the viscous correction to $\rho_{xy}$ at different DC currents. In Fig. \ref{fig:hallR}(c), $\Delta \rho_{xy}^{*}/\rho_{xy}^{bulk}=(\Delta \rho_{xy}-\Delta \rho_{xy,I=0})/\rho_{xy}^{bulk}$ is plotted, and we find that increasing the DC current ``amplifies'' the minimum at zero field. The 8 $\mu$A curve is fitted to $r_H \Delta \rho_{xy}^{hyd}/\rho_{xy}^{bulk}$. We find the minimum value at 0 T equates to a $l_{ee}$ of $29$ $\mu$m, with a $r_H$ factor of 0.019. These values are consistent with those from our earlier analysis of $\rho_{xx}$ and theory Eq. \eqref{eq:ee} for $l_{ee}$ [see also Fig. \ref{fig:hydrorxx}(d)].



Echoing our commentary at the end of the previous subsection, comparing our experimental observations for DC current-induced hydrodynamics in the differential transverse resistivity in Fig. \ref{fig:hallR}(a) with those for temperature-induced hydrodynamics in recent works of Gusev \textit{et al.} \cite{gusev2018viscous} and Raichev {\em et al.} \cite{raichev2020} there is again a notable difference. In our case, the change in sign of the slope at zero field occurs on increasing the DC current (which increases the electron temperature but not the bath temperature). In the case of of Gusev \textit{et al.} and Raichev {\em et al.}, the distinctive line shape is seen to weaken with increased temperature, and by 40 K the deviation from the bulk Hall resistivity is small. High temperature also leads to both an increase in the phonon population and decrease decrease of the mobility.   

\section{Hall-field Induced Resistance Oscillations}  \label{sec:hiro}

\begin{figure}
\includegraphics[width=1\columnwidth]{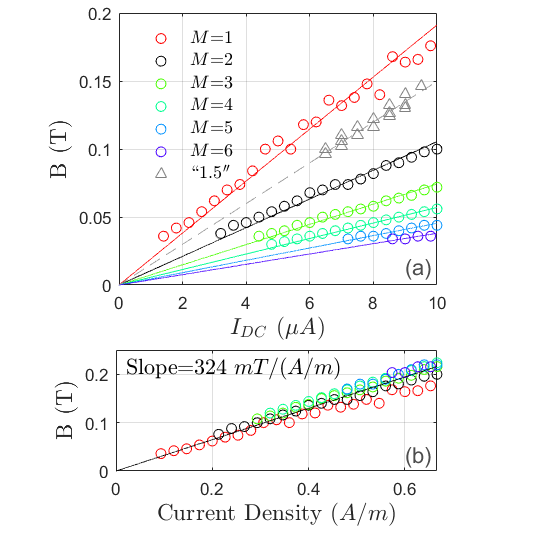}
\caption{(a) Fan diagram of HIRO maxima in $\partial \rho_{xx}/\partial |B|$ for each peak up to the sixth order. Points are extracted from the Fig. \ref{fig:PhaseDiagram}(a) data set for HIROs in the +I and +B quadrant. Triangles: Maxima in $\partial \rho_{xx}/\partial |B|$ of the ``1.5'' features visible in three of the four quadrants in Fig. \ref{fig:PhaseDiagram}(a) (the ``1.5'' feature in the +I and +B quadrant is not visible). Dashed line corresponds to (effective) M=1.44. (b) All HIRO maxima collapse on to a single line following the relation $B.M \propto I_{DC}/W$. The slope is used to extract the parameter $\gamma$. In panel (b) we have used $W$=15 $\mu$m rather than $W_{eff}$ to calculate the nominal current density so the value of the slope can be compared to those in the literature.
}
\label{fig:fanhiro}
\end{figure}

So far, in the phase diagram, we have described in detail the SdH oscillations and the accompanying PhI of the SdH oscillations located near zero current, and the nMR and the effects of DC current-induced hydrodynamics near zero field. Away from both $B$=0 T and $I_{DC}$=0 $\mu$A we observe HIROs. In Fig. \ref{fig:PhaseDiagram}(a), HIRO peaks up to the seventh order that fan out from the origin are identifiable. HIROs were first reported by Yang \textit{et al.} \cite{yang2002zener}, and have since been studied extensively both experimentally \cite{Bykov2005, zhang2007num3, zhang2007effect, zhang2007magnetotransport,bykov2008, dai2009,   Hatke2009,Hatke2010, Hatke2011, bykov2012zener, wiedmann2011, shi2017, Zudov2017,Mi2019} and theoretically \cite{vavilov2004,Dmitriev2005,vavilov2007nonlinear, khodas2008,vitkalov2009nonlinear, dmitriev2012nonequilibrium}. HIROs emerge in the weak field magnetoresistance in high mobility 2DEGs due to resonant electron transitions between LLs that are spatially tilted along the direction of the transverse electric (Hall) field when DC current is passed along the Hall bar. In the original work by Yang \textit{et al.} \cite{yang2002zener}, it was explained that Zener tunneling can occur when the electron hopping distance $\Delta Y_M$ between different Laudau levels is equal to $\gamma r_c$, where $\Delta Y_M=M\hbar \omega_c/eE_H$, $M$ is an integer, $E_H$ is the Hall field, and $\gamma$ $\approx 2$ according to theory. This leads to HIRO maxima in $\partial \rho_{xx}/\partial |B|$ with position in the $I_{DC}-B$ plane obeying the condition:
\be \label{eq:hirogamma}
    B = \gamma \frac{\sqrt{2\pi} m^{*}}{e^2 \sqrt{n}} \frac{I_{DC}}{WM}.
\ee
For the regime $k_BT\gg \hbar\omega_c$, where SdH oscillations are smeared out by temperature, Vavilov \textit{et al.} \cite{vavilov2007nonlinear}, found that the HIROs in the differential resistivity can be approximately described by the expression:
\be \label{eq:hirofit}
\rho_{xx,HIRO} \approx \frac{16m^{*}}{\pi e^2n\tau_{\pi}}\exp(-2\frac{\pi}{\omega_c\tau_{q}})\cos(2\pi\frac{2 I_{DC} k_F}{e n W\omega_c}),
\ee
where $\tau_{\pi}$ is the back-scattering lifetime. In the derivation\cite{vavilov2007nonlinear}, the oscillatory dependence of the differential resistivity stems from the product of the oscillatory DOS and the oscillatory non-equilibrium distribution function. When integrated over the non-equilibrium energy range, the remaining leading oscillatory term is squared, which explains the additional factor of two in the exponential damping factor $\exp(-2\pi/\omega_c\tau_{q})$ in Eq. \eqref{eq:hirofit} as compared to the SdH oscillation damping factor in Ando's expression in Eq. \eqref{eq:sdh}. The amplitude of the induced oscillation according to Eq. \eqref{eq:hirofit} is proportional to the back-scattering rate $\sim\tau_\pi^{-1}$, which is related to sharp disorder enabling a ``kick'' from one cyclotron orbit into another. The full theory by Vavilov \textit{et al.} also takes into account effects at relatively low current ($2\pi I_{DC}k_F/enW\omega_c<1$) due to variation of the occupation factors for the electronic states. We now compare the experimental data to the theories in Refs. \cite{yang2002zener,vavilov2007nonlinear}, report the extracted fitting parameters, and discuss other relevant observations and implications.

We first analyze our experimental HIRO data with the model of Yang \textit{et al.} \cite{yang2002zener}. We extract the HIRO maxima in the derivative of the differential resistivity $\partial \rho_{xx}/\partial |B|$ and compare with the expression in Eq. \eqref{eq:hirogamma}. The positions of the HIRO maxima for peaks up to the sixth order are plotted in Fig. \ref{fig:fanhiro}(a). The position of the maxima can be collapsed onto a single line, essentially the line tracking the first-order HIRO peak, following the relation $B\cdot M \propto I_{DC}/W$. With $W$=15 $\mu$m, the slope of the single line in Fig. \ref{fig:fanhiro}(b) is found to be 324 mT/(A/m), and from Eq. \eqref{eq:hirogamma}, we obtain $\gamma$=$2.4$. This result is consistent with the theoretical value of $\gamma\approx2.0$ reported in the work of Yang \textit{et al.} \cite{yang2002zener}, and their empirically determined $\gamma$ values (1.7-2.1) for Hall bars with a 2DEG density close to that for our Hall bar device. Note that had we used $W_{eff}\sim$ 11 $\mu$m to calculate the nominal current density, we would obtain $\gamma$=$2.1$.

\begin{figure*} 
\includegraphics[width=2\columnwidth]{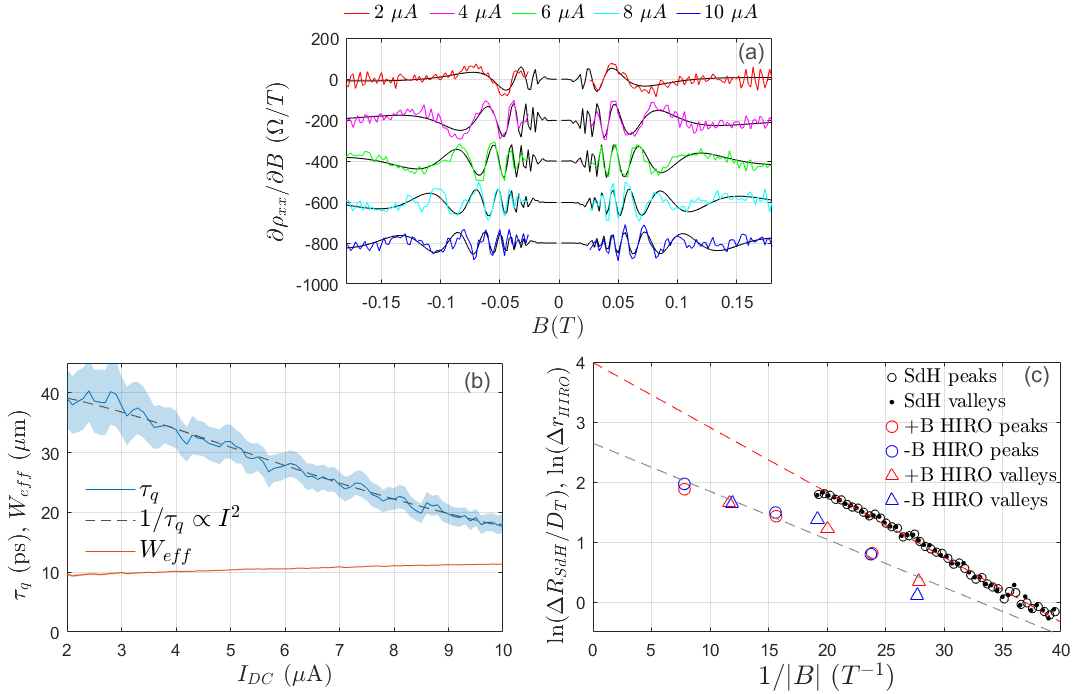}
\caption{(a) $\partial \rho_{xx}/\partial B$ versus B for selected DC currents. $\rho_{xx}$ data are from sections through the map in Fig. \ref{fig:PhaseDiagram} (a). Black lines: fits to data with derivative of Eq. \eqref{eq:hirofit}. The plots are offset by 200 $\Omega/T$ for clarity. (b) HIRO parameters $\tau_q$ and $W_{eff}$ obtained from fitting $\partial \rho_{xx}/\partial B$ versus B sections for positive $I_{DC}$. Dashed line: $1/\tau_q \propto I_{DC}^2$ fit to $\tau_q$ values. (c) Log plot versus 1/B of SdH oscillation peak and valley amplitudes in $r_{xx}$ at $I_{DC}$=0 $\mu$A, $\Delta R_{SdH}$, and HIRO peak and valley amplitudes in $r_{xx}$ at $I_{DC}$=5 $\mu A$, $\Delta r_{HIRO}$. We use both positive and negative B-field extrema for the HIRO data points. For HIROs, the 5 $\mu$A section is selected for analysis because for this current there are a sufficient number of resolved extrema to analyze, and the ``1.5'' feature  that could influence the minimum between the first- and second-order peaks has not yet developed. Red dashed line for SdH oscillations: fit with Eq. \eqref{eq:sdh} where $D_T$ is the thermal damping factor, and $\tau_q^{SdH}$ is determined from the slope ($\tau_q^{SdH}$=11 ps). Grey dashed line for HIROs: fit with Eq. \eqref{eq:hirofit} where $\tau_q^{HIRO}$ at 5 $\mu A$ is determined from the slope ($\tau_q^{HIRO}$=29 ps).}
\label{fig:HIROfit}
\end{figure*}

In the model of Vavilov \textit{et al.} \cite{vavilov2007nonlinear}, the amplitude of the HIROs depends on $\tau_{\pi}$ and $\tau_q$, respectively the back-scattering and quantum lifetimes. We fit $\rho_{xx}$ versus $B$ sections with Eq. \eqref{eq:hirofit} to obtain $\tau_{\pi}$, $\tau_q$, and additionally $W_{eff}$. Note that rather than assume $W$ is fixed and equal to the nominal lithographic width of the Hall bar, $W$ is treated as a fitting parameter which we call $W_{eff}$. Furthermore, we choose to fit $\partial \rho_{xx}/\partial B$ [see Fig. \ref{fig:HIROfit}(a)], essentially the partial derivative of Eq. \eqref{eq:hirofit} with respect to B, instead of $\rho_{xx}$, to remove the aforementioned background linear B dependence in the data, and to reduce fitting errors, although fits to either are equivalent and give nearly identical fitting parameters. Parameters $\tau_q$ and $W_{eff}$ obtained are presented in Fig. \ref{fig:HIROfit}(b). $\tau_q$ is discussed more extensively in the following paragraph. The effective electronic width, $W_{eff}$, is found to be $\sim$ 11 $\mu$m over the full 10 $\mu$A current range. Parameter $W_{eff}$ is obtained from the HIRO frequency and is extremely accurate as in Eq. \eqref{eq:hirofit} it is independent of the amplitude of the HIROs. The effective electronic width is smaller than the lithographic width of the Hall bar W=15 $\mu$m. As commented on earlier, we attribute the difference to a combination of undercut during the wet etching step in the fabrication of the Hall bar, and sidewall depletion. For the back-scattering lifetime, we obtained $\tau_{\pi}\approx$ 5 ns with no significant current dependence. We note that the corresponding scattering length associated with the back-scattering process is $l_\pi=v_F\tau_\pi\simeq 1$ mm. This length scale is much larger than our Hall bar device size and therefore cannot be interpreted as the typical distance between back-scattering impurities.

\begin{figure}[t]
\hspace*{-0.5cm}
\includegraphics[width=1\columnwidth]{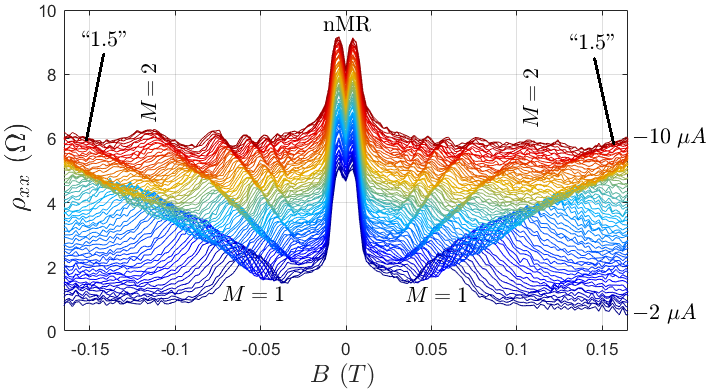}
\caption{$\rho_{xx}$ versus B sections through the map in Fig. \ref{fig:PhaseDiagram} (a) for negative DC currents between -2 $\mu$A and -10 $\mu$A. Plots are offset from each other by +0.06 $\Omega$ for clarity. 
The $I_{DC}$=-2 $\mu$A plot has no offset. The plots are arranged to emphasize the emergence of the ``1.5'' features at high current.}
\label{fig:hirohalf}
\end{figure}

The values of the quantum lifetime extracted from HIROs in Fig. \ref{fig:HIROfit}(b) are notable for two reasons. First, $\tau_q$ decreases with increasing DC current, and second, $\tau_q$ here far exceeds the value of $\tau_q$=11.5 ps extracted earlier from the SdH oscillations [see Fig. \ref{fig:SdHBigTrace}]. Concerning the first point, in the model of Vavilov \textit{et al.} \cite{vavilov2007nonlinear}, the HIROs amplitude does not depend on DC current, whereas our data shows that the amplitude of the HIROs decreases with increasing DC current [see red dashed line ellipse in Fig. \ref{fig:rhoxxcurrent}]. This decrease was also observed in another experiment featuring a 2DEG in a MgZnO/ZnO hetero-structure \cite{shi2017}, and was attributed to enhanced electron-electron scattering with increasing electron temperature. The relationship $1/\tau_q\propto I_{DC}^2$ was found. Using the same analysis for our Hall bar, we obtain $\tau_q(0)/\tau_q(I_{DC})=1+(I_{DC}/I_0)^2$ where (the extrapolated zero current quantum lifetime) $\tau_q(0)$=$41.2 \pm 0.7$ ps, and $I_0$=$8.66 \pm 0.04$ $\mu$A [see fit in Fig. \ref{fig:HIROfit}(b)]. The value of $I_0$ here is comparable to that obtained from the background quadratic dependence to $\rho_{xx}$ discussed in Sec. \ref{sec:lee} relating to Eq. \eqref{eq:rhoxxbg}. Concerning the second point, the extracted value of $\tau_q$ from fitting HIROs, which we now explicitly identify as $\tau_q^{HIRO}$, varies from $\sim$40 ps at 2 $\mu$A to $\sim$18 ps at 10 $\mu$A, whereas the extracted value of $\tau_q$ from fitting the SdH oscillations \emph{at zero DC current}, also now explicitly identified as $\tau_q^{SdH}$, is $\sim$11 ps. An alternative method presented in Refs. \cite{zhang2007magnetotransport,Zudov2017,Studenikin2005} to extract the exponential damping of SdH oscillations and HIROs is to plot the logarithm of the SdH oscillation and HIRO extrema amplitudes as a function of inverse B-field as shown in Fig. \ref{fig:HIROfit}(c). The slope of the plots corresponds to $-\zeta\pi/\mu_q$, where $\mu_q=e\tau_q/m^{*}$ is the quantum mobility. From Eqs. \eqref{eq:sdh} and \eqref{eq:hirofit} we expect $\zeta=1$ for the SdH oscillations and $\zeta=2$ for the HIROs, i.e., based on these equations the slope for HIROs should be twice the slope for the SdH oscillations. However, this is clearly not the case as can be seen in Fig. \ref{fig:HIROfit}(c). For example, for the 5 $\mu$A data, the slope for the HIROs is in fact even less than the slope for the SdH oscillations (consistent with the observation that the value for $\tau_q^{HIRO}$ is a factor of two to four times larger than the value for $\tau_q^{SdH}$). Note that the values obtained for the quantum lifetime do not depend significantly on the methods used [when we compare the value of $\tau_q^{SdH}$ obtained from the full fit method in Fig. \ref{fig:SdHBigTrace}(b) with that from the fit for the alternative method in Fig. \ref{fig:HIROfit}(c), 11.5 ps versus 11 ps, and likewise the value of $\tau_q^{HIRO}$ obtained from the full fit method in Fig. \ref{fig:HIROfit}(a) with that from the fit for the alternative method in Fig. \ref{fig:HIROfit}(c), 31 ps versus 29 ps both for 5 $\mu$A current], i.e., both methods lead to very similar values. Examining closely the theory in Ref. \cite{vavilov2007nonlinear}, it is assumed that the quantum lifetime for SdH oscillations and HIROs are the same. Furthermore, in the derivation of  Eq. \eqref{eq:hirofit} it is also assumed that $k_BT\gg\hbar\omega_c$. This equates to a temperature exceeding 0.5 K at 25 mT, which clearly does not hold for our experimental situation. This may explain the observed discrepancy between $\tau_q^{SdH}$ and $\tau_q^{HIRO}$, and suggests that a theory for HIROs extended to the low temperature regime is needed in order to correctly explain the observed amplitude of the HIROs. It is likely that at low temperature, when temperature smearing is small, the amplitude of the HIROs is proportional to the oscillatory DOS and not its square as in the derivation of Eq. \eqref{eq:hirofit} that assumes $k_BT\gg\hbar\omega_c$. We stress that the quantum lifetime is normally determined from the SdH oscillations ($\tau_q^{SdH}$) \emph{at zero DC current}. Inferring this quantum lifetime is the same as the quantum lifetime determined solely from measurement of HIROs ($\tau_q^{HIRO}$), a non-linear transport phenomenon observed at finite DC current, therefore has to be done with caution\cite{Zudov2017}. 

Lastly, we observe unexpected HIRO-like features located between the first-order and the second-order HIRO peaks at high DC current. These features are marked by black arrows in Fig. \ref{fig:PhaseDiagram}(a) in three of the four quadrants, and appear for a DC current exceeding 6 $\mu$A and a B-field above 0.1 T [see Fig. \ref{fig:fanhiro}(a), and also Fig. \ref{fig:hirohalf}]. Although not expected in the standard picture of HIROs, these features are HIRO-like in the sense that they lie on a line following the relation $B \propto I_{DC}$ and appear as part of the fan formed from the expected HIRO maxima (M=1, 2, 3, ...). We have referred to them as ``1.5'' features since their \emph{effective} index $M \sim 1.5$. We do not claim $M$ is quantized at 3/2: specifically from the fit in Fig. \ref{fig:fanhiro}(a) we determine $M$=1.44$\pm$0.04. Currently we have no clear understanding of the origin of the ``1.5'' feature, but speculate that it may be related to the lifting of the spin degeneracy which is observed in the Hall bar measured in the SdH oscillations at a similar $B$-field ($\sim$ 0.15 T: data not shown). We note that Hatke \textit{et al} \cite{Hatke2008} interestingly reported a HIRO-like feature at $M$=0.5, also referred to as fractional HIRO in Ref. \cite{dmitriev2012nonequilibrium}, however this was observed in a measurement performed under microwave illumination. We stress that we have observed the ``1.5'' features with no microwaves applied.  

\section{Conclusion}

\setlength{\tabcolsep}{3pt}
\begin{table}[t]
\renewcommand{\arraystretch}{1.4}
\centering
\begin{tabular}{  l c c c c c c  } 
 \hline \hline 
 $W_{eff}$ & $l_{ee}$ & $\eta$ & $\tau_q^{SdH}$ & $\tau_q^{HIRO}$ & $\tau_{\pi}$ \\ 
 ($\mu$m) & ($\mu$m) & ($m^2$/s) & (ps) & (ps) & (ns) \\  
 \hline 
 11 & 11 & 0.7 & 11.5 & 18 - 40 & 5 \\ 
 \hline \hline
\end{tabular}
\caption{Summary of key extracted parameters from Secs. \ref{sec:sdh} to \ref{sec:hiro}. The values of $l_{ee}$ and $\eta$ are for $I_{DC}$=10 $\mu$A determined in Sec. \ref{sec:hydro}. $\tau_q^{HIRO}$ depends on $I_{DC}$ hence a range of values is given (see Sec. \ref{sec:hiro}).}
\label{table:2}
\end{table}

In conclusion, for an ultra-high mobility 2DEG in a narrow Hall bar we measured $\rho_{xx}$ as a function of B-field and DC current at the base temperature of a dilution refrigerator. This measurement approach provides in a single 2D map a ``global'' view of different low-magnetic-field non-linear phenomena in a 2DEG. We identified and analyzed several different phenomena (and the boundaries between these phenomena): SdH oscillations, PhI of the SdH oscillations, nMR, a double-peak feature, and HIROs. By analyzing the data with existing theories, we extracted relevant parameters as summarized in Table \ref{table:2}. For the SdH oscillations, we found that the Ando formula in Eq. \eqref{eq:sdh} combined with an electron heating model can explain both the initial decrease in the SdH oscillation amplitude and then the subsequent PhI of the SdH oscillations with increasing DC current. As the zero-field resistance is largely determined by the nMR in our narrow Hall bar, we found that a fit of the SdH oscillations was essential to determine the appropriate resistance ($R_{0}$) needed to estimate the mobility. We confirmed that the nMR can be explained purely as a ballistic effect with a numerical simulation. From fits of the HIROs, we extracted the effective electronic width of the Hall bar $W_{eff}$ $\sim$ 11 $\mu$m. We also found that the quantum lifetime determined from analysis of the SdH oscillations at zero DC current, $\tau_q^{SdH}$, does not match the quantum lifetime determined from analysis of the HIROs at finite DC current, $\tau_q^{HIRO}$. The factor of two to four difference deserves further attention and would suggest a more careful theoretical treatment of HIROs at low temperature is needed. This also implies that HIROs, which is a non-linear effect, cannot be used in a direct way to infer the quantum lifetime in the linear regime (from SdH oscillations). Unexpectedly, a HIRO-like feature nestled between the first- and second-order HIRO maxima was observed with an effective index $M$=1.44$\pm$0.04. The origin of this feature is not understood at present.

A major part of our work here has demonstrated the growing influence of hydrodynamics brought about by applying a DC current, rather than changing the temperature as more commonly employed, to reduce the electron-electron scattering length $l_{ee}$ to the point that this length becomes comparable to or even smaller than $W_{eff}$. For a DC current of $\sim$ 10 $\mu$A, we determined $l_{ee} \sim W_{eff}$ $\sim$ 11 $\mu$m. DC current-induced hydrodynamics complements temperature-induced hydrodynamics. The current-induced case is a non-linear process, where the electron distribution is pushed strongly out of equilibrium, while in the temperature-induced case the electron distribution is thermal (and transport is linear), but it also increases the population of phonons, which reduces the carrier mobility and eventually suppresses hydrodynamic behavior. From a detailed analysis of the data within a -10 mT to +10 mT window, we isolated the growing hydrodynamic contribution to both the differential longitudinal and transverse (Hall) resistivities with increasing DC current. For the former (latter) the signature was found to be a growing ``dip'' (a change in sign of the slope) at zero field. We quantified the hydrodynamic corrections using a perturbative method. In theoretical works such as those in Refs. \cite{scaffidi2017hydrodynamic, raichev2020}, the magnetoresistance in the hydrodynamic and ballistic regimes is modelled for linear transport, whereas a full model that incorporates non-linear transport relevant to our experimental situation merits investigation.     

In the measurements described, the maximum DC current applied was limited to 10 $\mu$A and the temperature was fixed at the base temperature of the dilution refrigerator. Higher DC current and higher temperature could enhance further hydrodynamic corrections and bring other phenomena in to play. Measurement of differential resistivity maps as we have demonstrated here at low magnetic field is a powerful and convenient approach to probe numerous phenomena.   

We would like to thank Aviv Padawer-Blatt for his help with experiments. The authors acknowledge support from INTRIQ and RQMP.
This research is funded in part by FRQNT, NSERC and the Gordon and Betty Moore Foundation’s EPiQS Initiative, Grant GBMF9615 to L. N. Pfeiffer, and by the National Science Foundation MRSEC grant DMR 2011750 to Princeton University. 


%


\begin{thebibliography}{122}%
\makeatletter
\providecommand \@ifxundefined [1]{%
 \@ifx{#1\undefined}
}%
\providecommand \@ifnum [1]{%
 \ifnum #1\expandafter \@firstoftwo
 \else \expandafter \@secondoftwo
 \fi
}%
\providecommand \@ifx [1]{%
 \ifx #1\expandafter \@firstoftwo
 \else \expandafter \@secondoftwo
 \fi
}%
\providecommand \natexlab [1]{#1}%
\providecommand \enquote  [1]{``#1''}%
\providecommand \bibnamefont  [1]{#1}%
\providecommand \bibfnamefont [1]{#1}%
\providecommand \citenamefont [1]{#1}%
\providecommand \href@noop [0]{\@secondoftwo}%
\providecommand \href [0]{\begingroup \@sanitize@url \@href}%
\providecommand \@href[1]{\@@startlink{#1}\@@href}%
\providecommand \@@href[1]{\endgroup#1\@@endlink}%
\providecommand \@sanitize@url [0]{\catcode `\\12\catcode `\$12\catcode
  `\&12\catcode `\#12\catcode `\^12\catcode `\_12\catcode `\%12\relax}%
\providecommand \@@startlink[1]{}%
\providecommand \@@endlink[0]{}%
\providecommand \url  [0]{\begingroup\@sanitize@url \@url }%
\providecommand \@url [1]{\endgroup\@href {#1}{\urlprefix }}%
\providecommand \urlprefix  [0]{URL }%
\providecommand \Eprint [0]{\href }%
\providecommand \doibase [0]{https://doi.org/}%
\providecommand \selectlanguage [0]{\@gobble}%
\providecommand \bibinfo  [0]{\@secondoftwo}%
\providecommand \bibfield  [0]{\@secondoftwo}%
\providecommand \translation [1]{[#1]}%
\providecommand \BibitemOpen [0]{}%
\providecommand \bibitemStop [0]{}%
\providecommand \bibitemNoStop [0]{.\EOS\space}%
\providecommand \EOS [0]{\spacefactor3000\relax}%
\providecommand \BibitemShut  [1]{\csname bibitem#1\endcsname}%
\let\auto@bib@innerbib\@empty
\bibitem [{\citenamefont {Prange}\ and\ \citenamefont
  {Girvin}(1990)}]{Prange1990}%
  \BibitemOpen
  \bibinfo {editor} {\bibfnamefont {R.~E.}\ \bibnamefont {Prange}}\ and\
  \bibinfo {editor} {\bibfnamefont {S.~M.}\ \bibnamefont {Girvin}},\ eds.,\
  \href {https://doi.org/10.1007/978-1-4612-3350-3} {\emph {\bibinfo {title}
  {The Quantum Hall Effect}}}\ (\bibinfo  {publisher} {Springer New York},\
  \bibinfo {year} {1990})\BibitemShut {NoStop}%
\bibitem [{\citenamefont {Chakraborty}\ and\ \citenamefont
  {Pietil\"{a}inen}(1995)}]{Chakraborty1995}%
  \BibitemOpen
  \bibfield  {author} {\bibinfo {author} {\bibfnamefont {T.}~\bibnamefont
  {Chakraborty}}\ and\ \bibinfo {author} {\bibfnamefont {P.}~\bibnamefont
  {Pietil\"{a}inen}},\ }\href {https://doi.org/10.1007/978-3-642-79319-6}
  {\emph {\bibinfo {title} {The Quantum Hall Effects}}}\ (\bibinfo  {publisher}
  {Springer Berlin Heidelberg},\ \bibinfo {year} {1995})\BibitemShut {NoStop}%
\bibitem [{\citenamefont {Dmitriev}\ \emph {et~al.}(2012)\citenamefont
  {Dmitriev}, \citenamefont {Mirlin}, \citenamefont {Polyakov},\ and\
  \citenamefont {Zudov}}]{dmitriev2012nonequilibrium}%
  \BibitemOpen
  \bibfield  {author} {\bibinfo {author} {\bibfnamefont {I.}~\bibnamefont
  {Dmitriev}}, \bibinfo {author} {\bibfnamefont {A.}~\bibnamefont {Mirlin}},
  \bibinfo {author} {\bibfnamefont {D.}~\bibnamefont {Polyakov}},\ and\
  \bibinfo {author} {\bibfnamefont {M.}~\bibnamefont {Zudov}},\ }\href
  {https://doi.org/10.1103/RevModPhys.84.1709} {\bibfield  {journal} {\bibinfo
  {journal} {Rev. Mod. Phys.}\ }\textbf {\bibinfo {volume} {84}},\ \bibinfo
  {pages} {1709} (\bibinfo {year} {2012})}\BibitemShut {NoStop}%
\bibitem [{\citenamefont {Vitkalov}(2009)}]{vitkalov2009nonlinear}%
  \BibitemOpen
  \bibfield  {author} {\bibinfo {author} {\bibfnamefont {S.}~\bibnamefont
  {Vitkalov}},\ }\href
  {https://doi.org/https://doi.org/10.1142/S0217979209054090} {\bibfield
  {journal} {\bibinfo  {journal} {Int. J. Mod. Phys. B}\ }\textbf {\bibinfo
  {volume} {23}},\ \bibinfo {pages} {4727} (\bibinfo {year}
  {2009})}\BibitemShut {NoStop}%
\bibitem [{\citenamefont {Yang}\ \emph {et~al.}(2002)\citenamefont {Yang},
  \citenamefont {Zhang}, \citenamefont {Du}, \citenamefont {Simmons},\ and\
  \citenamefont {Reno}}]{yang2002zener}%
  \BibitemOpen
  \bibfield  {author} {\bibinfo {author} {\bibfnamefont {C.}~\bibnamefont
  {Yang}}, \bibinfo {author} {\bibfnamefont {J.}~\bibnamefont {Zhang}},
  \bibinfo {author} {\bibfnamefont {R.}~\bibnamefont {Du}}, \bibinfo {author}
  {\bibfnamefont {J.}~\bibnamefont {Simmons}},\ and\ \bibinfo {author}
  {\bibfnamefont {J.}~\bibnamefont {Reno}},\ }\href
  {https://doi.org/10.1103/PhysRevLett.89.076801} {\bibfield  {journal}
  {\bibinfo  {journal} {Phys. Rev. Lett.}\ }\textbf {\bibinfo {volume} {89}},\
  \bibinfo {pages} {076801} (\bibinfo {year} {2002})}\BibitemShut {NoStop}%
\bibitem [{\citenamefont {Bykov}\ \emph {et~al.}(2005)\citenamefont {Bykov},
  \citenamefont {Zhang}, \citenamefont {Vitkalov}, \citenamefont {Kalagin},\
  and\ \citenamefont {Bakarov}}]{Bykov2005}%
  \BibitemOpen
  \bibfield  {author} {\bibinfo {author} {\bibfnamefont {A.~A.}\ \bibnamefont
  {Bykov}}, \bibinfo {author} {\bibfnamefont {J.-q.}\ \bibnamefont {Zhang}},
  \bibinfo {author} {\bibfnamefont {S.}~\bibnamefont {Vitkalov}}, \bibinfo
  {author} {\bibfnamefont {A.~K.}\ \bibnamefont {Kalagin}},\ and\ \bibinfo
  {author} {\bibfnamefont {A.~K.}\ \bibnamefont {Bakarov}},\ }\href
  {https://doi.org/10.1103/PhysRevB.72.245307} {\bibfield  {journal} {\bibinfo
  {journal} {Phys. Rev. B}\ }\textbf {\bibinfo {volume} {72}},\ \bibinfo
  {pages} {245307} (\bibinfo {year} {2005})}\BibitemShut {NoStop}%
\bibitem [{\citenamefont {Zhang}\ \emph
  {et~al.}(2007{\natexlab{a}})\citenamefont {Zhang}, \citenamefont {Zudov},
  \citenamefont {Pfeiffer},\ and\ \citenamefont {West}}]{zhang2007num3}%
  \BibitemOpen
  \bibfield  {author} {\bibinfo {author} {\bibfnamefont {W.}~\bibnamefont
  {Zhang}}, \bibinfo {author} {\bibfnamefont {M.~A.}\ \bibnamefont {Zudov}},
  \bibinfo {author} {\bibfnamefont {L.~N.}\ \bibnamefont {Pfeiffer}},\ and\
  \bibinfo {author} {\bibfnamefont {K.~W.}\ \bibnamefont {West}},\ }\href
  {https://doi.org/10.1103/PhysRevLett.98.106804} {\bibfield  {journal}
  {\bibinfo  {journal} {Phys. Rev. Lett.}\ }\textbf {\bibinfo {volume} {98}},\
  \bibinfo {pages} {106804} (\bibinfo {year} {2007}{\natexlab{a}})}\BibitemShut
  {NoStop}%
\bibitem [{\citenamefont {Zhang}\ \emph
  {et~al.}(2007{\natexlab{b}})\citenamefont {Zhang}, \citenamefont {Vitkalov},
  \citenamefont {Bykov}, \citenamefont {Kalagin},\ and\ \citenamefont
  {Bakarov}}]{zhang2007effect}%
  \BibitemOpen
  \bibfield  {author} {\bibinfo {author} {\bibfnamefont {J.-q.}\ \bibnamefont
  {Zhang}}, \bibinfo {author} {\bibfnamefont {S.}~\bibnamefont {Vitkalov}},
  \bibinfo {author} {\bibfnamefont {A.}~\bibnamefont {Bykov}}, \bibinfo
  {author} {\bibfnamefont {A.}~\bibnamefont {Kalagin}},\ and\ \bibinfo {author}
  {\bibfnamefont {A.}~\bibnamefont {Bakarov}},\ }\href
  {https://doi.org/10.1103/PhysRevB.75.081305} {\bibfield  {journal} {\bibinfo
  {journal} {Phys. Rev. B}\ }\textbf {\bibinfo {volume} {75}},\ \bibinfo
  {pages} {081305} (\bibinfo {year} {2007}{\natexlab{b}})}\BibitemShut
  {NoStop}%
\bibitem [{\citenamefont {Zhang}\ \emph
  {et~al.}(2007{\natexlab{c}})\citenamefont {Zhang}, \citenamefont {Chiang},
  \citenamefont {Zudov}, \citenamefont {Pfeiffer},\ and\ \citenamefont
  {West}}]{zhang2007magnetotransport}%
  \BibitemOpen
  \bibfield  {author} {\bibinfo {author} {\bibfnamefont {W.}~\bibnamefont
  {Zhang}}, \bibinfo {author} {\bibfnamefont {H.-S.}\ \bibnamefont {Chiang}},
  \bibinfo {author} {\bibfnamefont {M.}~\bibnamefont {Zudov}}, \bibinfo
  {author} {\bibfnamefont {L.}~\bibnamefont {Pfeiffer}},\ and\ \bibinfo
  {author} {\bibfnamefont {K.}~\bibnamefont {West}},\ }\href
  {https://doi.org/10.1103/PhysRevB.75.041304} {\bibfield  {journal} {\bibinfo
  {journal} {Phys. Rev. B}\ }\textbf {\bibinfo {volume} {75}},\ \bibinfo
  {pages} {041304} (\bibinfo {year} {2007}{\natexlab{c}})}\BibitemShut
  {NoStop}%
\bibitem [{\citenamefont {Bykov}(2008)}]{bykov2008}%
  \BibitemOpen
  \bibfield  {author} {\bibinfo {author} {\bibfnamefont {A.~A.}\ \bibnamefont
  {Bykov}},\ }\href {https://doi.org/10.1134/S0021364008180112} {\bibfield
  {journal} {\bibinfo  {journal} {Sov. Phys. JETP}\ }\textbf {\bibinfo {volume}
  {88}},\ \bibinfo {pages} {394} (\bibinfo {year} {2008})}\BibitemShut
  {NoStop}%
\bibitem [{\citenamefont {Dai}\ \emph {et~al.}(2009)\citenamefont {Dai},
  \citenamefont {Yuan}, \citenamefont {Yang}, \citenamefont {Du}, \citenamefont
  {Manfra}, \citenamefont {Pfeiffer},\ and\ \citenamefont {West}}]{dai2009}%
  \BibitemOpen
  \bibfield  {author} {\bibinfo {author} {\bibfnamefont {Y.}~\bibnamefont
  {Dai}}, \bibinfo {author} {\bibfnamefont {Z.}~\bibnamefont {Yuan}}, \bibinfo
  {author} {\bibfnamefont {C.}~\bibnamefont {Yang}}, \bibinfo {author}
  {\bibfnamefont {R.}~\bibnamefont {Du}}, \bibinfo {author} {\bibfnamefont
  {M.}~\bibnamefont {Manfra}}, \bibinfo {author} {\bibfnamefont
  {L.}~\bibnamefont {Pfeiffer}},\ and\ \bibinfo {author} {\bibfnamefont
  {K.}~\bibnamefont {West}},\ }\href
  {https://doi.org/10.1103/PhysRevB.80.041310} {\bibfield  {journal} {\bibinfo
  {journal} {Phys. Rev. B}\ }\textbf {\bibinfo {volume} {80}},\ \bibinfo
  {pages} {041310} (\bibinfo {year} {2009})}\BibitemShut {NoStop}%
\bibitem [{\citenamefont {Hatke}\ \emph {et~al.}(2009)\citenamefont {Hatke},
  \citenamefont {Zudov}, \citenamefont {Pfeiffer},\ and\ \citenamefont
  {West}}]{Hatke2009}%
  \BibitemOpen
  \bibfield  {author} {\bibinfo {author} {\bibfnamefont {A.~T.}\ \bibnamefont
  {Hatke}}, \bibinfo {author} {\bibfnamefont {M.~A.}\ \bibnamefont {Zudov}},
  \bibinfo {author} {\bibfnamefont {L.~N.}\ \bibnamefont {Pfeiffer}},\ and\
  \bibinfo {author} {\bibfnamefont {K.~W.}\ \bibnamefont {West}},\ }\href
  {https://doi.org/10.1103/PhysRevB.79.161308} {\bibfield  {journal} {\bibinfo
  {journal} {Phys. Rev. B}\ }\textbf {\bibinfo {volume} {79}},\ \bibinfo
  {pages} {161308} (\bibinfo {year} {2009})}\BibitemShut {NoStop}%
\bibitem [{\citenamefont {Hatke}\ \emph {et~al.}(2010)\citenamefont {Hatke},
  \citenamefont {Chiang}, \citenamefont {Zudov}, \citenamefont {Pfeiffer},\
  and\ \citenamefont {West}}]{Hatke2010}%
  \BibitemOpen
  \bibfield  {author} {\bibinfo {author} {\bibfnamefont {A.~T.}\ \bibnamefont
  {Hatke}}, \bibinfo {author} {\bibfnamefont {H.-S.}\ \bibnamefont {Chiang}},
  \bibinfo {author} {\bibfnamefont {M.~A.}\ \bibnamefont {Zudov}}, \bibinfo
  {author} {\bibfnamefont {L.~N.}\ \bibnamefont {Pfeiffer}},\ and\ \bibinfo
  {author} {\bibfnamefont {K.~W.}\ \bibnamefont {West}},\ }\href
  {https://doi.org/10.1103/PhysRevB.82.041304} {\bibfield  {journal} {\bibinfo
  {journal} {Phys. Rev. B}\ }\textbf {\bibinfo {volume} {82}},\ \bibinfo
  {pages} {041304} (\bibinfo {year} {2010})}\BibitemShut {NoStop}%
\bibitem [{\citenamefont {Hatke}\ \emph {et~al.}(2011)\citenamefont {Hatke},
  \citenamefont {Zudov}, \citenamefont {Pfeiffer},\ and\ \citenamefont
  {West}}]{Hatke2011}%
  \BibitemOpen
  \bibfield  {author} {\bibinfo {author} {\bibfnamefont {A.~T.}\ \bibnamefont
  {Hatke}}, \bibinfo {author} {\bibfnamefont {M.~A.}\ \bibnamefont {Zudov}},
  \bibinfo {author} {\bibfnamefont {L.~N.}\ \bibnamefont {Pfeiffer}},\ and\
  \bibinfo {author} {\bibfnamefont {K.~W.}\ \bibnamefont {West}},\ }\href
  {https://doi.org/10.1103/PhysRevB.83.081301} {\bibfield  {journal} {\bibinfo
  {journal} {Phys. Rev. B}\ }\textbf {\bibinfo {volume} {83}},\ \bibinfo
  {pages} {081301} (\bibinfo {year} {2011})}\BibitemShut {NoStop}%
\bibitem [{\citenamefont {Bykov}\ \emph {et~al.}(2012)\citenamefont {Bykov},
  \citenamefont {Dmitriev}, \citenamefont {Marchishin}, \citenamefont
  {Byrnes},\ and\ \citenamefont {Vitkalov}}]{bykov2012zener}%
  \BibitemOpen
  \bibfield  {author} {\bibinfo {author} {\bibfnamefont {A.}~\bibnamefont
  {Bykov}}, \bibinfo {author} {\bibfnamefont {D.}~\bibnamefont {Dmitriev}},
  \bibinfo {author} {\bibfnamefont {I.}~\bibnamefont {Marchishin}}, \bibinfo
  {author} {\bibfnamefont {S.}~\bibnamefont {Byrnes}},\ and\ \bibinfo {author}
  {\bibfnamefont {S.}~\bibnamefont {Vitkalov}},\ }\href
  {https://doi.org/https://doi.org/10.1063/1.4729590} {\bibfield  {journal}
  {\bibinfo  {journal} {Appl. Phys. Lett.}\ }\textbf {\bibinfo {volume}
  {100}},\ \bibinfo {pages} {251602} (\bibinfo {year} {2012})}\BibitemShut
  {NoStop}%
\bibitem [{\citenamefont {Wiedmann}\ \emph {et~al.}(2011)\citenamefont
  {Wiedmann}, \citenamefont {Gusev}, \citenamefont {Raichev}, \citenamefont
  {Bakarov},\ and\ \citenamefont {Portal}}]{wiedmann2011}%
  \BibitemOpen
  \bibfield  {author} {\bibinfo {author} {\bibfnamefont {S.}~\bibnamefont
  {Wiedmann}}, \bibinfo {author} {\bibfnamefont {G.}~\bibnamefont {Gusev}},
  \bibinfo {author} {\bibfnamefont {O.}~\bibnamefont {Raichev}}, \bibinfo
  {author} {\bibfnamefont {A.}~\bibnamefont {Bakarov}},\ and\ \bibinfo {author}
  {\bibfnamefont {J.}~\bibnamefont {Portal}},\ }\href
  {https://doi.org/10.1103/PhysRevB.84.165303} {\bibfield  {journal} {\bibinfo
  {journal} {Phys. Rev. B}\ }\textbf {\bibinfo {volume} {84}},\ \bibinfo
  {pages} {165303} (\bibinfo {year} {2011})}\BibitemShut {NoStop}%
\bibitem [{\citenamefont {Shi}\ \emph {et~al.}(2017)\citenamefont {Shi},
  \citenamefont {Zudov}, \citenamefont {Falson}, \citenamefont {Kozuka},
  \citenamefont {Tsukazaki}, \citenamefont {Kawasaki}, \citenamefont {von
  Klitzing},\ and\ \citenamefont {Smet}}]{shi2017}%
  \BibitemOpen
  \bibfield  {author} {\bibinfo {author} {\bibfnamefont {Q.}~\bibnamefont
  {Shi}}, \bibinfo {author} {\bibfnamefont {M.~A.}\ \bibnamefont {Zudov}},
  \bibinfo {author} {\bibfnamefont {J.}~\bibnamefont {Falson}}, \bibinfo
  {author} {\bibfnamefont {Y.}~\bibnamefont {Kozuka}}, \bibinfo {author}
  {\bibfnamefont {A.}~\bibnamefont {Tsukazaki}}, \bibinfo {author}
  {\bibfnamefont {M.}~\bibnamefont {Kawasaki}}, \bibinfo {author}
  {\bibfnamefont {K.}~\bibnamefont {von Klitzing}},\ and\ \bibinfo {author}
  {\bibfnamefont {J.}~\bibnamefont {Smet}},\ }\href
  {https://doi.org/10.1103/PhysRevB.95.041411} {\bibfield  {journal} {\bibinfo
  {journal} {Phys. Rev. B}\ }\textbf {\bibinfo {volume} {95}},\ \bibinfo
  {pages} {041411} (\bibinfo {year} {2017})}\BibitemShut {NoStop}%
\bibitem [{\citenamefont {Zudov}\ \emph {et~al.}(2017)\citenamefont {Zudov},
  \citenamefont {Dmitriev}, \citenamefont {Friess}, \citenamefont {Shi},
  \citenamefont {Umansky}, \citenamefont {von Klitzing},\ and\ \citenamefont
  {Smet}}]{Zudov2017}%
  \BibitemOpen
  \bibfield  {author} {\bibinfo {author} {\bibfnamefont {M.~A.}\ \bibnamefont
  {Zudov}}, \bibinfo {author} {\bibfnamefont {I.~A.}\ \bibnamefont {Dmitriev}},
  \bibinfo {author} {\bibfnamefont {B.}~\bibnamefont {Friess}}, \bibinfo
  {author} {\bibfnamefont {Q.}~\bibnamefont {Shi}}, \bibinfo {author}
  {\bibfnamefont {V.}~\bibnamefont {Umansky}}, \bibinfo {author} {\bibfnamefont
  {K.}~\bibnamefont {von Klitzing}},\ and\ \bibinfo {author} {\bibfnamefont
  {J.}~\bibnamefont {Smet}},\ }\href
  {https://doi.org/10.1103/PhysRevB.96.121301} {\bibfield  {journal} {\bibinfo
  {journal} {Phys. Rev. B}\ }\textbf {\bibinfo {volume} {96}},\ \bibinfo
  {pages} {121301} (\bibinfo {year} {2017})}\BibitemShut {NoStop}%
\bibitem [{\citenamefont {Mi}\ \emph {et~al.}(2019)\citenamefont {Mi},
  \citenamefont {Liu}, \citenamefont {Shi}, \citenamefont {Pfeiffer},
  \citenamefont {West}, \citenamefont {Baldwin},\ and\ \citenamefont
  {Zhang}}]{Mi2019}%
  \BibitemOpen
  \bibfield  {author} {\bibinfo {author} {\bibfnamefont {J.}~\bibnamefont
  {Mi}}, \bibinfo {author} {\bibfnamefont {H.}~\bibnamefont {Liu}}, \bibinfo
  {author} {\bibfnamefont {J.}~\bibnamefont {Shi}}, \bibinfo {author}
  {\bibfnamefont {L.~N.}\ \bibnamefont {Pfeiffer}}, \bibinfo {author}
  {\bibfnamefont {K.~W.}\ \bibnamefont {West}}, \bibinfo {author}
  {\bibfnamefont {K.~W.}\ \bibnamefont {Baldwin}},\ and\ \bibinfo {author}
  {\bibfnamefont {C.}~\bibnamefont {Zhang}},\ }\href
  {https://doi.org/10.1103/PhysRevB.100.235437} {\bibfield  {journal} {\bibinfo
   {journal} {Phys. Rev. B}\ }\textbf {\bibinfo {volume} {100}},\ \bibinfo
  {pages} {235437} (\bibinfo {year} {2019})}\BibitemShut {NoStop}%
\bibitem [{\citenamefont {Studenikin}\ \emph {et~al.}(2012)\citenamefont
  {Studenikin}, \citenamefont {Granger}, \citenamefont {Kam}, \citenamefont
  {Sachrajda}, \citenamefont {Wasilewski},\ and\ \citenamefont
  {Poole}}]{phaseinversionsergei}%
  \BibitemOpen
  \bibfield  {author} {\bibinfo {author} {\bibfnamefont {S.~A.}\ \bibnamefont
  {Studenikin}}, \bibinfo {author} {\bibfnamefont {G.}~\bibnamefont {Granger}},
  \bibinfo {author} {\bibfnamefont {A.}~\bibnamefont {Kam}}, \bibinfo {author}
  {\bibfnamefont {A.~S.}\ \bibnamefont {Sachrajda}}, \bibinfo {author}
  {\bibfnamefont {Z.~R.}\ \bibnamefont {Wasilewski}},\ and\ \bibinfo {author}
  {\bibfnamefont {P.~J.}\ \bibnamefont {Poole}},\ }\href
  {https://doi.org/10.1103/PhysRevB.86.115309} {\bibfield  {journal} {\bibinfo
  {journal} {Phys. Rev. B}\ }\textbf {\bibinfo {volume} {86}},\ \bibinfo
  {pages} {115309} (\bibinfo {year} {2012})}\BibitemShut {NoStop}%
\bibitem [{\citenamefont {Panos}\ \emph {et~al.}(2014)\citenamefont {Panos},
  \citenamefont {Gerhardts}, \citenamefont {Weis},\ and\ \citenamefont
  {Von~Klitzing}}]{panos2014}%
  \BibitemOpen
  \bibfield  {author} {\bibinfo {author} {\bibfnamefont {K.}~\bibnamefont
  {Panos}}, \bibinfo {author} {\bibfnamefont {R.}~\bibnamefont {Gerhardts}},
  \bibinfo {author} {\bibfnamefont {J.}~\bibnamefont {Weis}},\ and\ \bibinfo
  {author} {\bibfnamefont {K.}~\bibnamefont {Von~Klitzing}},\ }\href
  {https://doi.org/https://doi.org/10.1088/1367-2630/16/11/113071} {\bibfield
  {journal} {\bibinfo  {journal} {New J. Phys.}\ }\textbf {\bibinfo {volume}
  {16}},\ \bibinfo {pages} {113071} (\bibinfo {year} {2014})}\BibitemShut
  {NoStop}%
\bibitem [{\citenamefont {Baer}\ \emph {et~al.}(2015)\citenamefont {Baer},
  \citenamefont {R\"ossler}, \citenamefont {Hennel}, \citenamefont {Overweg},
  \citenamefont {Ihn}, \citenamefont {Ensslin}, \citenamefont {Reichl},\ and\
  \citenamefont {Wegscheider}}]{baer2015}%
  \BibitemOpen
  \bibfield  {author} {\bibinfo {author} {\bibfnamefont {S.}~\bibnamefont
  {Baer}}, \bibinfo {author} {\bibfnamefont {C.}~\bibnamefont {R\"ossler}},
  \bibinfo {author} {\bibfnamefont {S.}~\bibnamefont {Hennel}}, \bibinfo
  {author} {\bibfnamefont {H.~C.}\ \bibnamefont {Overweg}}, \bibinfo {author}
  {\bibfnamefont {T.}~\bibnamefont {Ihn}}, \bibinfo {author} {\bibfnamefont
  {K.}~\bibnamefont {Ensslin}}, \bibinfo {author} {\bibfnamefont
  {C.}~\bibnamefont {Reichl}},\ and\ \bibinfo {author} {\bibfnamefont
  {W.}~\bibnamefont {Wegscheider}},\ }\href
  {https://doi.org/10.1103/PhysRevB.91.195414} {\bibfield  {journal} {\bibinfo
  {journal} {Phys. Rev. B}\ }\textbf {\bibinfo {volume} {91}},\ \bibinfo
  {pages} {195414} (\bibinfo {year} {2015})}\BibitemShut {NoStop}%
\bibitem [{\citenamefont {Rossokhaty}\ \emph {et~al.}(2016)\citenamefont
  {Rossokhaty}, \citenamefont {Baum}, \citenamefont {Folk}, \citenamefont
  {Watson}, \citenamefont {Gardner},\ and\ \citenamefont
  {Manfra}}]{Rossokhaty2016}%
  \BibitemOpen
  \bibfield  {author} {\bibinfo {author} {\bibfnamefont {A.~V.}\ \bibnamefont
  {Rossokhaty}}, \bibinfo {author} {\bibfnamefont {Y.}~\bibnamefont {Baum}},
  \bibinfo {author} {\bibfnamefont {J.~A.}\ \bibnamefont {Folk}}, \bibinfo
  {author} {\bibfnamefont {J.~D.}\ \bibnamefont {Watson}}, \bibinfo {author}
  {\bibfnamefont {G.~C.}\ \bibnamefont {Gardner}},\ and\ \bibinfo {author}
  {\bibfnamefont {M.~J.}\ \bibnamefont {Manfra}},\ }\href
  {https://doi.org/10.1103/PhysRevLett.117.166805} {\bibfield  {journal}
  {\bibinfo  {journal} {Phys. Rev. Lett.}\ }\textbf {\bibinfo {volume} {117}},\
  \bibinfo {pages} {166805} (\bibinfo {year} {2016})}\BibitemShut {NoStop}%
\bibitem [{\citenamefont {Yu}\ \emph {et~al.}(2018)\citenamefont {Yu},
  \citenamefont {Hilke}, \citenamefont {Poole}, \citenamefont {Studenikin},\
  and\ \citenamefont {Austing}}]{phaseinversionYu}%
  \BibitemOpen
  \bibfield  {author} {\bibinfo {author} {\bibfnamefont {V.}~\bibnamefont
  {Yu}}, \bibinfo {author} {\bibfnamefont {M.}~\bibnamefont {Hilke}}, \bibinfo
  {author} {\bibfnamefont {P.~J.}\ \bibnamefont {Poole}}, \bibinfo {author}
  {\bibfnamefont {S.}~\bibnamefont {Studenikin}},\ and\ \bibinfo {author}
  {\bibfnamefont {D.~G.}\ \bibnamefont {Austing}},\ }\href
  {https://doi.org/10.1103/PhysRevB.98.165434} {\bibfield  {journal} {\bibinfo
  {journal} {Phys. Rev. B}\ }\textbf {\bibinfo {volume} {98}},\ \bibinfo
  {pages} {165434} (\bibinfo {year} {2018})}\BibitemShut {NoStop}%
\bibitem [{\citenamefont {Gurzhi}(1963)}]{gurzhi1963}%
  \BibitemOpen
  \bibfield  {author} {\bibinfo {author} {\bibfnamefont {R.}~\bibnamefont
  {Gurzhi}},\ }\href@noop {} {\bibfield  {journal} {\bibinfo  {journal} {Sov.
  Phys. JETP}\ }\textbf {\bibinfo {volume} {44}},\ \bibinfo {pages} {771}
  (\bibinfo {year} {1963})}\BibitemShut {NoStop}%
\bibitem [{\citenamefont {Gurzhi}(1968)}]{gurzhi1968hydrodynamic}%
  \BibitemOpen
  \bibfield  {author} {\bibinfo {author} {\bibfnamefont {R.}~\bibnamefont
  {Gurzhi}},\ }\href@noop {} {\bibfield  {journal} {\bibinfo  {journal} {Sov.
  Phys. Usp}\ }\textbf {\bibinfo {volume} {11}},\ \bibinfo {pages} {255}
  (\bibinfo {year} {1968})}\BibitemShut {NoStop}%
\bibitem [{\citenamefont {Giuliani}\ and\ \citenamefont
  {Quinn}(1982)}]{giuliani1982lifetime}%
  \BibitemOpen
  \bibfield  {author} {\bibinfo {author} {\bibfnamefont {G.~F.}\ \bibnamefont
  {Giuliani}}\ and\ \bibinfo {author} {\bibfnamefont {J.~J.}\ \bibnamefont
  {Quinn}},\ }\href {https://doi.org/10.1103/PhysRevB.26.4421} {\bibfield
  {journal} {\bibinfo  {journal} {Phys. Rev. B}\ }\textbf {\bibinfo {volume}
  {26}},\ \bibinfo {pages} {4421} (\bibinfo {year} {1982})}\BibitemShut
  {NoStop}%
\bibitem [{\citenamefont {Gurzhi}\ \emph {et~al.}(1989)\citenamefont {Gurzhi},
  \citenamefont {Kalinenko},\ and\ \citenamefont
  {Kopeliovich}}]{gurzhi1989hydrodynamic}%
  \BibitemOpen
  \bibfield  {author} {\bibinfo {author} {\bibfnamefont {R.}~\bibnamefont
  {Gurzhi}}, \bibinfo {author} {\bibfnamefont {A.}~\bibnamefont {Kalinenko}},\
  and\ \bibinfo {author} {\bibfnamefont {A.}~\bibnamefont {Kopeliovich}},\
  }\href@noop {} {\bibfield  {journal} {\bibinfo  {journal} {Sov. Phys. JETP}\
  }\textbf {\bibinfo {volume} {69}},\ \bibinfo {pages} {863} (\bibinfo {year}
  {1989})}\BibitemShut {NoStop}%
\bibitem [{\citenamefont {Jaggi}(1991)}]{jaggi1991electron}%
  \BibitemOpen
  \bibfield  {author} {\bibinfo {author} {\bibfnamefont {R.}~\bibnamefont
  {Jaggi}},\ }\href {https://doi.org/10.1063/1.347315} {\bibfield  {journal}
  {\bibinfo  {journal} {J. Appl. Phys.}\ }\textbf {\bibinfo {volume} {69}},\
  \bibinfo {pages} {816} (\bibinfo {year} {1991})}\BibitemShut {NoStop}%
\bibitem [{\citenamefont {Molenkamp}\ and\ \citenamefont
  {de~Jong}(1994)}]{molenkamp1994}%
  \BibitemOpen
  \bibfield  {author} {\bibinfo {author} {\bibfnamefont {L.~W.}\ \bibnamefont
  {Molenkamp}}\ and\ \bibinfo {author} {\bibfnamefont {M.~J.~M.}\ \bibnamefont
  {de~Jong}},\ }\href {https://doi.org/10.1103/PhysRevB.49.5038} {\bibfield
  {journal} {\bibinfo  {journal} {Phys. Rev. B}\ }\textbf {\bibinfo {volume}
  {49}},\ \bibinfo {pages} {5038} (\bibinfo {year} {1994})}\BibitemShut
  {NoStop}%
\bibitem [{\citenamefont {de~Jong}\ and\ \citenamefont
  {Molenkamp}(1995)}]{de1995hydrodynamic}%
  \BibitemOpen
  \bibfield  {author} {\bibinfo {author} {\bibfnamefont {M.}~\bibnamefont
  {de~Jong}}\ and\ \bibinfo {author} {\bibfnamefont {L.}~\bibnamefont
  {Molenkamp}},\ }\href {https://doi.org/10.1103/PhysRevB.51.13389} {\bibfield
  {journal} {\bibinfo  {journal} {Phys. Rev. B}\ }\textbf {\bibinfo {volume}
  {51}},\ \bibinfo {pages} {13389} (\bibinfo {year} {1995})}\BibitemShut
  {NoStop}%
\bibitem [{\citenamefont {Molenkamp}\ and\ \citenamefont {{de
  Jong}}(1994)}]{Molenkamp1994part2}%
  \BibitemOpen
  \bibfield  {author} {\bibinfo {author} {\bibfnamefont {L.}~\bibnamefont
  {Molenkamp}}\ and\ \bibinfo {author} {\bibfnamefont {M.}~\bibnamefont {{de
  Jong}}},\ }\href
  {https://doi.org/https://doi.org/10.1016/0038-1101(94)90244-5} {\bibfield
  {journal} {\bibinfo  {journal} {Solid State Electron.}\ }\textbf {\bibinfo
  {volume} {37}},\ \bibinfo {pages} {551} (\bibinfo {year} {1994})}\BibitemShut
  {NoStop}%
\bibitem [{\citenamefont {Eaves}(1998)}]{eaves1998}%
  \BibitemOpen
  \bibfield  {author} {\bibinfo {author} {\bibfnamefont {L.}~\bibnamefont
  {Eaves}},\ }\href
  {https://doi.org/https://doi.org/10.1016/S0921-4526(98)00753-4} {\bibfield
  {journal} {\bibinfo  {journal} {Physica B}\ }\textbf {\bibinfo {volume}
  {256-258}},\ \bibinfo {pages} {47} (\bibinfo {year} {1998})}\BibitemShut
  {NoStop}%
\bibitem [{\citenamefont {Eaves}(1999)}]{eaves1999hydrodynamic}%
  \BibitemOpen
  \bibfield  {author} {\bibinfo {author} {\bibfnamefont {L.}~\bibnamefont
  {Eaves}},\ }\href {https://doi.org/10.1016/S0921-4526(99)00255-0} {\bibfield
  {journal} {\bibinfo  {journal} {Physica B}\ }\textbf {\bibinfo {volume}
  {272}},\ \bibinfo {pages} {130} (\bibinfo {year} {1999})}\BibitemShut
  {NoStop}%
\bibitem [{\citenamefont {Eaves}\ \emph {et~al.}(2000)\citenamefont {Eaves},
  \citenamefont {Stoddart}, \citenamefont {Wirtz}, \citenamefont {Neumann},
  \citenamefont {Gallagher}, \citenamefont {Main},\ and\ \citenamefont
  {Henini}}]{eaves2000}%
  \BibitemOpen
  \bibfield  {author} {\bibinfo {author} {\bibfnamefont {L.}~\bibnamefont
  {Eaves}}, \bibinfo {author} {\bibfnamefont {S.}~\bibnamefont {Stoddart}},
  \bibinfo {author} {\bibfnamefont {R.}~\bibnamefont {Wirtz}}, \bibinfo
  {author} {\bibfnamefont {A.}~\bibnamefont {Neumann}}, \bibinfo {author}
  {\bibfnamefont {B.}~\bibnamefont {Gallagher}}, \bibinfo {author}
  {\bibfnamefont {P.}~\bibnamefont {Main}},\ and\ \bibinfo {author}
  {\bibfnamefont {M.}~\bibnamefont {Henini}},\ }\href
  {https://doi.org/https://doi.org/10.1016/S1386-9477(99)00080-6} {\bibfield
  {journal} {\bibinfo  {journal} {Physica E}\ }\textbf {\bibinfo {volume}
  {6}},\ \bibinfo {pages} {136} (\bibinfo {year} {2000})}\BibitemShut {NoStop}%
\bibitem [{\citenamefont {Eaves}(2001{\natexlab{a}})}]{eaves2001}%
  \BibitemOpen
  \bibfield  {author} {\bibinfo {author} {\bibfnamefont {L.}~\bibnamefont
  {Eaves}},\ }\href
  {https://doi.org/https://doi.org/10.1016/S1386-9477(00)00176-4} {\bibfield
  {journal} {\bibinfo  {journal} {Physica E}\ }\textbf {\bibinfo {volume}
  {9}},\ \bibinfo {pages} {45} (\bibinfo {year}
  {2001}{\natexlab{a}})}\BibitemShut {NoStop}%
\bibitem [{\citenamefont {Eaves}(2001{\natexlab{b}})}]{eaves2001quantum}%
  \BibitemOpen
  \bibfield  {author} {\bibinfo {author} {\bibfnamefont {L.}~\bibnamefont
  {Eaves}},\ }\href
  {https://doi.org/https://doi.org/10.1016/S0921-4526(01)00244-7} {\bibfield
  {journal} {\bibinfo  {journal} {Physica B}\ }\textbf {\bibinfo {volume}
  {298}},\ \bibinfo {pages} {1} (\bibinfo {year}
  {2001}{\natexlab{b}})}\BibitemShut {NoStop}%
\bibitem [{\citenamefont {Martin}\ \emph {et~al.}(2004)\citenamefont {Martin},
  \citenamefont {Benedict}, \citenamefont {Sheard},\ and\ \citenamefont
  {Eaves}}]{martin2004}%
  \BibitemOpen
  \bibfield  {author} {\bibinfo {author} {\bibfnamefont {A.}~\bibnamefont
  {Martin}}, \bibinfo {author} {\bibfnamefont {K.}~\bibnamefont {Benedict}},
  \bibinfo {author} {\bibfnamefont {F.}~\bibnamefont {Sheard}},\ and\ \bibinfo
  {author} {\bibfnamefont {L.}~\bibnamefont {Eaves}},\ }\href
  {https://doi.org/https://doi.org/10.1016/j.physe.2003.11.250} {\bibfield
  {journal} {\bibinfo  {journal} {Physica E}\ }\textbf {\bibinfo {volume}
  {22}},\ \bibinfo {pages} {205} (\bibinfo {year} {2004})}\BibitemShut
  {NoStop}%
\bibitem [{\citenamefont {Zaanen}(2016)}]{Zaanen2016}%
  \BibitemOpen
  \bibfield  {author} {\bibinfo {author} {\bibfnamefont {J.}~\bibnamefont
  {Zaanen}},\ }\href {https://doi.org/10.1126/science.aaf2487} {\bibfield
  {journal} {\bibinfo  {journal} {Science}\ }\textbf {\bibinfo {volume}
  {351}},\ \bibinfo {pages} {1026} (\bibinfo {year} {2016})}\BibitemShut
  {NoStop}%
\bibitem [{\citenamefont {Bandurin}\ \emph {et~al.}(2016)\citenamefont
  {Bandurin}, \citenamefont {Torre}, \citenamefont {Kumar}, \citenamefont
  {Shalom}, \citenamefont {Tomadin}, \citenamefont {Principi}, \citenamefont
  {Auton}, \citenamefont {Khestanova}, \citenamefont {Novoselov}, \citenamefont
  {Grigorieva}, \citenamefont {Ponomarenko}, \citenamefont {Geim},\ and\
  \citenamefont {Polini}}]{Bandurin2016}%
  \BibitemOpen
  \bibfield  {author} {\bibinfo {author} {\bibfnamefont {D.~A.}\ \bibnamefont
  {Bandurin}}, \bibinfo {author} {\bibfnamefont {I.}~\bibnamefont {Torre}},
  \bibinfo {author} {\bibfnamefont {R.~K.}\ \bibnamefont {Kumar}}, \bibinfo
  {author} {\bibfnamefont {M.~B.}\ \bibnamefont {Shalom}}, \bibinfo {author}
  {\bibfnamefont {A.}~\bibnamefont {Tomadin}}, \bibinfo {author} {\bibfnamefont
  {A.}~\bibnamefont {Principi}}, \bibinfo {author} {\bibfnamefont {G.~H.}\
  \bibnamefont {Auton}}, \bibinfo {author} {\bibfnamefont {E.}~\bibnamefont
  {Khestanova}}, \bibinfo {author} {\bibfnamefont {K.~S.}\ \bibnamefont
  {Novoselov}}, \bibinfo {author} {\bibfnamefont {I.~V.}\ \bibnamefont
  {Grigorieva}}, \bibinfo {author} {\bibfnamefont {L.~A.}\ \bibnamefont
  {Ponomarenko}}, \bibinfo {author} {\bibfnamefont {A.~K.}\ \bibnamefont
  {Geim}},\ and\ \bibinfo {author} {\bibfnamefont {M.}~\bibnamefont {Polini}},\
  }\href {https://doi.org/10.1126/science.aad0201} {\bibfield  {journal}
  {\bibinfo  {journal} {Science}\ }\textbf {\bibinfo {volume} {351}},\ \bibinfo
  {pages} {1055} (\bibinfo {year} {2016})}\BibitemShut {NoStop}%
\bibitem [{\citenamefont {Crossno}\ \emph {et~al.}(2016)\citenamefont
  {Crossno}, \citenamefont {Shi}, \citenamefont {Wang}, \citenamefont {Liu},
  \citenamefont {Harzheim}, \citenamefont {Lucas}, \citenamefont {Sachdev},
  \citenamefont {Kim}, \citenamefont {Taniguchi}, \citenamefont {Watanabe},
  \citenamefont {Ohki},\ and\ \citenamefont {Fong}}]{Crossno2016}%
  \BibitemOpen
  \bibfield  {author} {\bibinfo {author} {\bibfnamefont {J.}~\bibnamefont
  {Crossno}}, \bibinfo {author} {\bibfnamefont {J.~K.}\ \bibnamefont {Shi}},
  \bibinfo {author} {\bibfnamefont {K.}~\bibnamefont {Wang}}, \bibinfo {author}
  {\bibfnamefont {X.}~\bibnamefont {Liu}}, \bibinfo {author} {\bibfnamefont
  {A.}~\bibnamefont {Harzheim}}, \bibinfo {author} {\bibfnamefont
  {A.}~\bibnamefont {Lucas}}, \bibinfo {author} {\bibfnamefont
  {S.}~\bibnamefont {Sachdev}}, \bibinfo {author} {\bibfnamefont
  {P.}~\bibnamefont {Kim}}, \bibinfo {author} {\bibfnamefont {T.}~\bibnamefont
  {Taniguchi}}, \bibinfo {author} {\bibfnamefont {K.}~\bibnamefont {Watanabe}},
  \bibinfo {author} {\bibfnamefont {T.~A.}\ \bibnamefont {Ohki}},\ and\
  \bibinfo {author} {\bibfnamefont {K.~C.}\ \bibnamefont {Fong}},\ }\href
  {https://doi.org/10.1126/science.aad0343} {\bibfield  {journal} {\bibinfo
  {journal} {Science}\ }\textbf {\bibinfo {volume} {351}},\ \bibinfo {pages}
  {1058} (\bibinfo {year} {2016})}\BibitemShut {NoStop}%
\bibitem [{\citenamefont {Levitov}\ and\ \citenamefont
  {Falkovich}(2016)}]{Levitov2016}%
  \BibitemOpen
  \bibfield  {author} {\bibinfo {author} {\bibfnamefont {L.}~\bibnamefont
  {Levitov}}\ and\ \bibinfo {author} {\bibfnamefont {G.}~\bibnamefont
  {Falkovich}},\ }\href {https://doi.org/10.1038/nphys3667} {\bibfield
  {journal} {\bibinfo  {journal} {Nat. Phys.}\ }\textbf {\bibinfo {volume}
  {12}},\ \bibinfo {pages} {672} (\bibinfo {year} {2016})}\BibitemShut
  {NoStop}%
\bibitem [{\citenamefont {Sulpizio}\ \emph {et~al.}(2019)\citenamefont
  {Sulpizio}, \citenamefont {Ella}, \citenamefont {Rozen}, \citenamefont
  {Birkbeck}, \citenamefont {Perello}, \citenamefont {Dutta}, \citenamefont
  {Ben-Shalom}, \citenamefont {Taniguchi}, \citenamefont {Watanabe},
  \citenamefont {Holder}, \citenamefont {Queiroz}, \citenamefont {Principi},
  \citenamefont {Stern}, \citenamefont {Scaffidi}, \citenamefont {Geim},\ and\
  \citenamefont {Ilani}}]{sulpizio2019visualizing}%
  \BibitemOpen
  \bibfield  {author} {\bibinfo {author} {\bibfnamefont {J.~A.}\ \bibnamefont
  {Sulpizio}}, \bibinfo {author} {\bibfnamefont {L.}~\bibnamefont {Ella}},
  \bibinfo {author} {\bibfnamefont {A.}~\bibnamefont {Rozen}}, \bibinfo
  {author} {\bibfnamefont {J.}~\bibnamefont {Birkbeck}}, \bibinfo {author}
  {\bibfnamefont {D.~J.}\ \bibnamefont {Perello}}, \bibinfo {author}
  {\bibfnamefont {D.}~\bibnamefont {Dutta}}, \bibinfo {author} {\bibfnamefont
  {M.}~\bibnamefont {Ben-Shalom}}, \bibinfo {author} {\bibfnamefont
  {T.}~\bibnamefont {Taniguchi}}, \bibinfo {author} {\bibfnamefont
  {K.}~\bibnamefont {Watanabe}}, \bibinfo {author} {\bibfnamefont
  {T.}~\bibnamefont {Holder}}, \bibinfo {author} {\bibfnamefont
  {R.}~\bibnamefont {Queiroz}}, \bibinfo {author} {\bibfnamefont
  {A.}~\bibnamefont {Principi}}, \bibinfo {author} {\bibfnamefont
  {A.}~\bibnamefont {Stern}}, \bibinfo {author} {\bibfnamefont
  {T.}~\bibnamefont {Scaffidi}}, \bibinfo {author} {\bibfnamefont {A.~K.}\
  \bibnamefont {Geim}},\ and\ \bibinfo {author} {\bibfnamefont
  {S.}~\bibnamefont {Ilani}},\ }\href
  {https://doi.org/10.1038/s41586-019-1788-9} {\bibfield  {journal} {\bibinfo
  {journal} {Nature}\ }\textbf {\bibinfo {volume} {576}},\ \bibinfo {pages}
  {75} (\bibinfo {year} {2019})}\BibitemShut {NoStop}%
\bibitem [{\citenamefont {Berdyugin}\ \emph {et~al.}(2019)\citenamefont
  {Berdyugin}, \citenamefont {Xu}, \citenamefont {Pellegrino}, \citenamefont
  {Kumar}, \citenamefont {Principi}, \citenamefont {Torre}, \citenamefont
  {Shalom}, \citenamefont {Taniguchi}, \citenamefont {Watanabe}, \citenamefont
  {Grigorieva}, \citenamefont {Polini}, \citenamefont {Geim},\ and\
  \citenamefont {Bandurin}}]{Berdyugin2019}%
  \BibitemOpen
  \bibfield  {author} {\bibinfo {author} {\bibfnamefont {A.~I.}\ \bibnamefont
  {Berdyugin}}, \bibinfo {author} {\bibfnamefont {S.~G.}\ \bibnamefont {Xu}},
  \bibinfo {author} {\bibfnamefont {F.~M.~D.}\ \bibnamefont {Pellegrino}},
  \bibinfo {author} {\bibfnamefont {R.~K.}\ \bibnamefont {Kumar}}, \bibinfo
  {author} {\bibfnamefont {A.}~\bibnamefont {Principi}}, \bibinfo {author}
  {\bibfnamefont {I.}~\bibnamefont {Torre}}, \bibinfo {author} {\bibfnamefont
  {M.~B.}\ \bibnamefont {Shalom}}, \bibinfo {author} {\bibfnamefont
  {T.}~\bibnamefont {Taniguchi}}, \bibinfo {author} {\bibfnamefont
  {K.}~\bibnamefont {Watanabe}}, \bibinfo {author} {\bibfnamefont {I.~V.}\
  \bibnamefont {Grigorieva}}, \bibinfo {author} {\bibfnamefont
  {M.}~\bibnamefont {Polini}}, \bibinfo {author} {\bibfnamefont {A.~K.}\
  \bibnamefont {Geim}},\ and\ \bibinfo {author} {\bibfnamefont {D.~A.}\
  \bibnamefont {Bandurin}},\ }\href {https://doi.org/10.1126/science.aau0685}
  {\bibfield  {journal} {\bibinfo  {journal} {Science}\ }\textbf {\bibinfo
  {volume} {364}},\ \bibinfo {pages} {162} (\bibinfo {year}
  {2019})}\BibitemShut {NoStop}%
\bibitem [{\citenamefont {Gallagher}\ \emph {et~al.}(2019)\citenamefont
  {Gallagher}, \citenamefont {Yang}, \citenamefont {Lyu}, \citenamefont {Tian},
  \citenamefont {Kou}, \citenamefont {Zhang}, \citenamefont {Watanabe},
  \citenamefont {Taniguchi},\ and\ \citenamefont
  {Wang}}]{gallagher2019quantum}%
  \BibitemOpen
  \bibfield  {author} {\bibinfo {author} {\bibfnamefont {P.}~\bibnamefont
  {Gallagher}}, \bibinfo {author} {\bibfnamefont {C.-S.}\ \bibnamefont {Yang}},
  \bibinfo {author} {\bibfnamefont {T.}~\bibnamefont {Lyu}}, \bibinfo {author}
  {\bibfnamefont {F.}~\bibnamefont {Tian}}, \bibinfo {author} {\bibfnamefont
  {R.}~\bibnamefont {Kou}}, \bibinfo {author} {\bibfnamefont {H.}~\bibnamefont
  {Zhang}}, \bibinfo {author} {\bibfnamefont {K.}~\bibnamefont {Watanabe}},
  \bibinfo {author} {\bibfnamefont {T.}~\bibnamefont {Taniguchi}},\ and\
  \bibinfo {author} {\bibfnamefont {F.}~\bibnamefont {Wang}},\ }\href
  {https://doi.org/https://doi.org/10.1126/science.aat8687} {\bibfield
  {journal} {\bibinfo  {journal} {Science}\ }\textbf {\bibinfo {volume}
  {364}},\ \bibinfo {pages} {158} (\bibinfo {year} {2019})}\BibitemShut
  {NoStop}%
\bibitem [{\citenamefont {Ella}\ \emph {et~al.}(2019)\citenamefont {Ella},
  \citenamefont {Rozen}, \citenamefont {Birkbeck}, \citenamefont {Ben-Shalom},
  \citenamefont {Perello}, \citenamefont {Zultak}, \citenamefont {Taniguchi},
  \citenamefont {Watanabe}, \citenamefont {Geim}, \citenamefont {Ilani},\ and\
  \citenamefont {Sulpizio}}]{ella2019simultaneous}%
  \BibitemOpen
  \bibfield  {author} {\bibinfo {author} {\bibfnamefont {L.}~\bibnamefont
  {Ella}}, \bibinfo {author} {\bibfnamefont {A.}~\bibnamefont {Rozen}},
  \bibinfo {author} {\bibfnamefont {J.}~\bibnamefont {Birkbeck}}, \bibinfo
  {author} {\bibfnamefont {M.}~\bibnamefont {Ben-Shalom}}, \bibinfo {author}
  {\bibfnamefont {D.}~\bibnamefont {Perello}}, \bibinfo {author} {\bibfnamefont
  {J.}~\bibnamefont {Zultak}}, \bibinfo {author} {\bibfnamefont
  {T.}~\bibnamefont {Taniguchi}}, \bibinfo {author} {\bibfnamefont
  {K.}~\bibnamefont {Watanabe}}, \bibinfo {author} {\bibfnamefont {A.~K.}\
  \bibnamefont {Geim}}, \bibinfo {author} {\bibfnamefont {S.}~\bibnamefont
  {Ilani}},\ and\ \bibinfo {author} {\bibfnamefont {J.~A.}\ \bibnamefont
  {Sulpizio}},\ }\href
  {https://doi.org/https://doi.org/10.1038/s41565-019-0398-x} {\bibfield
  {journal} {\bibinfo  {journal} {Nat. Nanotechnol.}\ }\textbf {\bibinfo
  {volume} {14}},\ \bibinfo {pages} {480} (\bibinfo {year} {2019})}\BibitemShut
  {NoStop}%
\bibitem [{\citenamefont {Ku}\ \emph {et~al.}(2020)\citenamefont {Ku},
  \citenamefont {Zhou}, \citenamefont {Li}, \citenamefont {Shin}, \citenamefont
  {Shi}, \citenamefont {Burch}, \citenamefont {Anderson}, \citenamefont
  {Pierce}, \citenamefont {Xie}, \citenamefont {Hamo}, \citenamefont {Vool},
  \citenamefont {Zhang}, \citenamefont {Casola}, \citenamefont {Taniguchi},
  \citenamefont {Watanabe}, \citenamefont {Fogler}, \citenamefont {Kim},
  \citenamefont {Yacoby},\ and\ \citenamefont {Walsworth}}]{Ku2020}%
  \BibitemOpen
  \bibfield  {author} {\bibinfo {author} {\bibfnamefont {M.~J.~H.}\
  \bibnamefont {Ku}}, \bibinfo {author} {\bibfnamefont {T.~X.}\ \bibnamefont
  {Zhou}}, \bibinfo {author} {\bibfnamefont {Q.}~\bibnamefont {Li}}, \bibinfo
  {author} {\bibfnamefont {Y.~J.}\ \bibnamefont {Shin}}, \bibinfo {author}
  {\bibfnamefont {J.~K.}\ \bibnamefont {Shi}}, \bibinfo {author} {\bibfnamefont
  {C.}~\bibnamefont {Burch}}, \bibinfo {author} {\bibfnamefont {L.~E.}\
  \bibnamefont {Anderson}}, \bibinfo {author} {\bibfnamefont {A.~T.}\
  \bibnamefont {Pierce}}, \bibinfo {author} {\bibfnamefont {Y.}~\bibnamefont
  {Xie}}, \bibinfo {author} {\bibfnamefont {A.}~\bibnamefont {Hamo}}, \bibinfo
  {author} {\bibfnamefont {U.}~\bibnamefont {Vool}}, \bibinfo {author}
  {\bibfnamefont {H.}~\bibnamefont {Zhang}}, \bibinfo {author} {\bibfnamefont
  {F.}~\bibnamefont {Casola}}, \bibinfo {author} {\bibfnamefont
  {T.}~\bibnamefont {Taniguchi}}, \bibinfo {author} {\bibfnamefont
  {K.}~\bibnamefont {Watanabe}}, \bibinfo {author} {\bibfnamefont {M.~M.}\
  \bibnamefont {Fogler}}, \bibinfo {author} {\bibfnamefont {P.}~\bibnamefont
  {Kim}}, \bibinfo {author} {\bibfnamefont {A.}~\bibnamefont {Yacoby}},\ and\
  \bibinfo {author} {\bibfnamefont {R.~L.}\ \bibnamefont {Walsworth}},\ }\href
  {https://doi.org/10.1038/s41586-020-2507-2} {\bibfield  {journal} {\bibinfo
  {journal} {Nature}\ }\textbf {\bibinfo {volume} {583}},\ \bibinfo {pages}
  {537} (\bibinfo {year} {2020})}\BibitemShut {NoStop}%
\bibitem [{\citenamefont {Pusep}\ \emph {et~al.}(2022)\citenamefont {Pusep},
  \citenamefont {Teodoro}, \citenamefont {Laurindo}, \citenamefont {Cardozo~de
  Oliveira}, \citenamefont {Gusev},\ and\ \citenamefont {Bakarov}}]{Pusep2022}%
  \BibitemOpen
  \bibfield  {author} {\bibinfo {author} {\bibfnamefont {Y.~A.}\ \bibnamefont
  {Pusep}}, \bibinfo {author} {\bibfnamefont {M.~D.}\ \bibnamefont {Teodoro}},
  \bibinfo {author} {\bibfnamefont {V.}~\bibnamefont {Laurindo}}, \bibinfo
  {author} {\bibfnamefont {E.~R.}\ \bibnamefont {Cardozo~de Oliveira}},
  \bibinfo {author} {\bibfnamefont {G.~M.}\ \bibnamefont {Gusev}},\ and\
  \bibinfo {author} {\bibfnamefont {A.~K.}\ \bibnamefont {Bakarov}},\ }\href
  {https://doi.org/10.1103/PhysRevLett.128.136801} {\bibfield  {journal}
  {\bibinfo  {journal} {Phys. Rev. Lett.}\ }\textbf {\bibinfo {volume} {128}},\
  \bibinfo {pages} {136801} (\bibinfo {year} {2022})}\BibitemShut {NoStop}%
\bibitem [{\citenamefont {Jenkins}\ \emph {et~al.}(2022)\citenamefont
  {Jenkins}, \citenamefont {Baumann}, \citenamefont {Zhou}, \citenamefont
  {Meynell}, \citenamefont {Daipeng}, \citenamefont {Watanabe}, \citenamefont
  {Taniguchi}, \citenamefont {Lucas}, \citenamefont {Young},\ and\
  \citenamefont {Bleszynski~Jayich}}]{Jenkins2022}%
  \BibitemOpen
  \bibfield  {author} {\bibinfo {author} {\bibfnamefont {A.}~\bibnamefont
  {Jenkins}}, \bibinfo {author} {\bibfnamefont {S.}~\bibnamefont {Baumann}},
  \bibinfo {author} {\bibfnamefont {H.}~\bibnamefont {Zhou}}, \bibinfo {author}
  {\bibfnamefont {S.~A.}\ \bibnamefont {Meynell}}, \bibinfo {author}
  {\bibfnamefont {Y.}~\bibnamefont {Daipeng}}, \bibinfo {author} {\bibfnamefont
  {K.}~\bibnamefont {Watanabe}}, \bibinfo {author} {\bibfnamefont
  {T.}~\bibnamefont {Taniguchi}}, \bibinfo {author} {\bibfnamefont
  {A.}~\bibnamefont {Lucas}}, \bibinfo {author} {\bibfnamefont {A.~F.}\
  \bibnamefont {Young}},\ and\ \bibinfo {author} {\bibfnamefont {A.~C.}\
  \bibnamefont {Bleszynski~Jayich}},\ }\href
  {https://doi.org/10.1103/PhysRevLett.129.087701} {\bibfield  {journal}
  {\bibinfo  {journal} {Phys. Rev. Lett.}\ }\textbf {\bibinfo {volume} {129}},\
  \bibinfo {pages} {087701} (\bibinfo {year} {2022})}\BibitemShut {NoStop}%
\bibitem [{\citenamefont {Shi}\ \emph {et~al.}(2014)\citenamefont {Shi},
  \citenamefont {Martin}, \citenamefont {Ebner}, \citenamefont {Zudov},
  \citenamefont {Pfeiffer},\ and\ \citenamefont {West}}]{shi2014}%
  \BibitemOpen
  \bibfield  {author} {\bibinfo {author} {\bibfnamefont {Q.}~\bibnamefont
  {Shi}}, \bibinfo {author} {\bibfnamefont {P.~D.}\ \bibnamefont {Martin}},
  \bibinfo {author} {\bibfnamefont {Q.~A.}\ \bibnamefont {Ebner}}, \bibinfo
  {author} {\bibfnamefont {M.~A.}\ \bibnamefont {Zudov}}, \bibinfo {author}
  {\bibfnamefont {L.~N.}\ \bibnamefont {Pfeiffer}},\ and\ \bibinfo {author}
  {\bibfnamefont {K.~W.}\ \bibnamefont {West}},\ }\href
  {https://doi.org/10.1103/PhysRevB.89.201301} {\bibfield  {journal} {\bibinfo
  {journal} {Phys. Rev. B}\ }\textbf {\bibinfo {volume} {89}},\ \bibinfo
  {pages} {201301} (\bibinfo {year} {2014})}\BibitemShut {NoStop}%
\bibitem [{\citenamefont {Gusev}\ \emph
  {et~al.}(2018{\natexlab{a}})\citenamefont {Gusev}, \citenamefont {Levin},
  \citenamefont {Levinson},\ and\ \citenamefont {Bakarov}}]{gusev2018viscous}%
  \BibitemOpen
  \bibfield  {author} {\bibinfo {author} {\bibfnamefont {G.}~\bibnamefont
  {Gusev}}, \bibinfo {author} {\bibfnamefont {A.}~\bibnamefont {Levin}},
  \bibinfo {author} {\bibfnamefont {E.}~\bibnamefont {Levinson}},\ and\
  \bibinfo {author} {\bibfnamefont {A.}~\bibnamefont {Bakarov}},\ }\href
  {https://doi.org/10.1103/PhysRevB.98.161303} {\bibfield  {journal} {\bibinfo
  {journal} {Phys. Rev. B}\ }\textbf {\bibinfo {volume} {98}},\ \bibinfo
  {pages} {161303} (\bibinfo {year} {2018}{\natexlab{a}})}\BibitemShut
  {NoStop}%
\bibitem [{\citenamefont {Levin}\ \emph {et~al.}(2018)\citenamefont {Levin},
  \citenamefont {Gusev}, \citenamefont {Levinson}, \citenamefont {Kvon},\ and\
  \citenamefont {Bakarov}}]{Levin2018}%
  \BibitemOpen
  \bibfield  {author} {\bibinfo {author} {\bibfnamefont {A.~D.}\ \bibnamefont
  {Levin}}, \bibinfo {author} {\bibfnamefont {G.~M.}\ \bibnamefont {Gusev}},
  \bibinfo {author} {\bibfnamefont {E.~V.}\ \bibnamefont {Levinson}}, \bibinfo
  {author} {\bibfnamefont {Z.~D.}\ \bibnamefont {Kvon}},\ and\ \bibinfo
  {author} {\bibfnamefont {A.~K.}\ \bibnamefont {Bakarov}},\ }\href
  {https://doi.org/10.1103/PhysRevB.97.245308} {\bibfield  {journal} {\bibinfo
  {journal} {Phys. Rev. B}\ }\textbf {\bibinfo {volume} {97}},\ \bibinfo
  {pages} {245308} (\bibinfo {year} {2018})}\BibitemShut {NoStop}%
\bibitem [{\citenamefont {Gusev}\ \emph
  {et~al.}(2018{\natexlab{b}})\citenamefont {Gusev}, \citenamefont {Levin},
  \citenamefont {Levinson},\ and\ \citenamefont {Bakarov}}]{Gusev2018b}%
  \BibitemOpen
  \bibfield  {author} {\bibinfo {author} {\bibfnamefont {G.~M.}\ \bibnamefont
  {Gusev}}, \bibinfo {author} {\bibfnamefont {A.~D.}\ \bibnamefont {Levin}},
  \bibinfo {author} {\bibfnamefont {E.~V.}\ \bibnamefont {Levinson}},\ and\
  \bibinfo {author} {\bibfnamefont {A.~K.}\ \bibnamefont {Bakarov}},\ }\href
  {https://doi.org/10.1063/1.5020763} {\bibfield  {journal} {\bibinfo
  {journal} {AIP Adv.}\ }\textbf {\bibinfo {volume} {8}},\ \bibinfo {pages}
  {025318} (\bibinfo {year} {2018}{\natexlab{b}})}\BibitemShut {NoStop}%
\bibitem [{\citenamefont {Braem}\ \emph {et~al.}(2018)\citenamefont {Braem},
  \citenamefont {Pellegrino}, \citenamefont {Principi}, \citenamefont
  {R\"o\"osli}, \citenamefont {Gold}, \citenamefont {Hennel}, \citenamefont
  {Koski}, \citenamefont {Berl}, \citenamefont {Dietsche}, \citenamefont
  {Wegscheider}, \citenamefont {Polini}, \citenamefont {Ihn},\ and\
  \citenamefont {Ensslin}}]{Braem2018}%
  \BibitemOpen
  \bibfield  {author} {\bibinfo {author} {\bibfnamefont {B.~A.}\ \bibnamefont
  {Braem}}, \bibinfo {author} {\bibfnamefont {F.~M.~D.}\ \bibnamefont
  {Pellegrino}}, \bibinfo {author} {\bibfnamefont {A.}~\bibnamefont
  {Principi}}, \bibinfo {author} {\bibfnamefont {M.}~\bibnamefont
  {R\"o\"osli}}, \bibinfo {author} {\bibfnamefont {C.}~\bibnamefont {Gold}},
  \bibinfo {author} {\bibfnamefont {S.}~\bibnamefont {Hennel}}, \bibinfo
  {author} {\bibfnamefont {J.~V.}\ \bibnamefont {Koski}}, \bibinfo {author}
  {\bibfnamefont {M.}~\bibnamefont {Berl}}, \bibinfo {author} {\bibfnamefont
  {W.}~\bibnamefont {Dietsche}}, \bibinfo {author} {\bibfnamefont
  {W.}~\bibnamefont {Wegscheider}}, \bibinfo {author} {\bibfnamefont
  {M.}~\bibnamefont {Polini}}, \bibinfo {author} {\bibfnamefont
  {T.}~\bibnamefont {Ihn}},\ and\ \bibinfo {author} {\bibfnamefont
  {K.}~\bibnamefont {Ensslin}},\ }\href
  {https://doi.org/10.1103/PhysRevB.98.241304} {\bibfield  {journal} {\bibinfo
  {journal} {Phys. Rev. B}\ }\textbf {\bibinfo {volume} {98}},\ \bibinfo
  {pages} {241304} (\bibinfo {year} {2018})}\BibitemShut {NoStop}%
\bibitem [{\citenamefont {Gusev}\ \emph {et~al.}(2020)\citenamefont {Gusev},
  \citenamefont {Jaroshevich}, \citenamefont {Levin}, \citenamefont {Kvon},\
  and\ \citenamefont {Bakarov}}]{Gusev2020a}%
  \BibitemOpen
  \bibfield  {author} {\bibinfo {author} {\bibfnamefont {G.~M.}\ \bibnamefont
  {Gusev}}, \bibinfo {author} {\bibfnamefont {A.~S.}\ \bibnamefont
  {Jaroshevich}}, \bibinfo {author} {\bibfnamefont {A.~D.}\ \bibnamefont
  {Levin}}, \bibinfo {author} {\bibfnamefont {Z.~D.}\ \bibnamefont {Kvon}},\
  and\ \bibinfo {author} {\bibfnamefont {A.~K.}\ \bibnamefont {Bakarov}},\
  }\href {https://doi.org/10.1038/s41598-020-64807-6} {\bibfield  {journal}
  {\bibinfo  {journal} {Sci. Rep.}\ }\textbf {\bibinfo {volume} {10}},\
  \bibinfo {pages} {7860} (\bibinfo {year} {2020})}\BibitemShut {NoStop}%
\bibitem [{\citenamefont {Keser}\ \emph {et~al.}(2021)\citenamefont {Keser},
  \citenamefont {Wang}, \citenamefont {Klochan}, \citenamefont {Ho},
  \citenamefont {Tkachenko}, \citenamefont {Tkachenko}, \citenamefont {Culcer},
  \citenamefont {Adam}, \citenamefont {Farrer}, \citenamefont {Ritchie},
  \citenamefont {Sushkov},\ and\ \citenamefont {Hamilton}}]{Keser2021}%
  \BibitemOpen
  \bibfield  {author} {\bibinfo {author} {\bibfnamefont {A.~C.}\ \bibnamefont
  {Keser}}, \bibinfo {author} {\bibfnamefont {D.~Q.}\ \bibnamefont {Wang}},
  \bibinfo {author} {\bibfnamefont {O.}~\bibnamefont {Klochan}}, \bibinfo
  {author} {\bibfnamefont {D.~Y.~H.}\ \bibnamefont {Ho}}, \bibinfo {author}
  {\bibfnamefont {O.~A.}\ \bibnamefont {Tkachenko}}, \bibinfo {author}
  {\bibfnamefont {V.~A.}\ \bibnamefont {Tkachenko}}, \bibinfo {author}
  {\bibfnamefont {D.}~\bibnamefont {Culcer}}, \bibinfo {author} {\bibfnamefont
  {S.}~\bibnamefont {Adam}}, \bibinfo {author} {\bibfnamefont {I.}~\bibnamefont
  {Farrer}}, \bibinfo {author} {\bibfnamefont {D.~A.}\ \bibnamefont {Ritchie}},
  \bibinfo {author} {\bibfnamefont {O.~P.}\ \bibnamefont {Sushkov}},\ and\
  \bibinfo {author} {\bibfnamefont {A.~R.}\ \bibnamefont {Hamilton}},\ }\href
  {https://doi.org/10.1103/PhysRevX.11.031030} {\bibfield  {journal} {\bibinfo
  {journal} {Phys. Rev. X}\ }\textbf {\bibinfo {volume} {11}},\ \bibinfo
  {pages} {031030} (\bibinfo {year} {2021})}\BibitemShut {NoStop}%
\bibitem [{\citenamefont {Gupta}\ \emph {et~al.}(2021)\citenamefont {Gupta},
  \citenamefont {Heremans}, \citenamefont {Kataria}, \citenamefont {Chandra},
  \citenamefont {Fallahi}, \citenamefont {Gardner},\ and\ \citenamefont
  {Manfra}}]{Gupta2021}%
  \BibitemOpen
  \bibfield  {author} {\bibinfo {author} {\bibfnamefont {A.}~\bibnamefont
  {Gupta}}, \bibinfo {author} {\bibfnamefont {J.~J.}\ \bibnamefont {Heremans}},
  \bibinfo {author} {\bibfnamefont {G.}~\bibnamefont {Kataria}}, \bibinfo
  {author} {\bibfnamefont {M.}~\bibnamefont {Chandra}}, \bibinfo {author}
  {\bibfnamefont {S.}~\bibnamefont {Fallahi}}, \bibinfo {author} {\bibfnamefont
  {G.~C.}\ \bibnamefont {Gardner}},\ and\ \bibinfo {author} {\bibfnamefont
  {M.~J.}\ \bibnamefont {Manfra}},\ }\href
  {https://doi.org/10.1103/PhysRevLett.126.076803} {\bibfield  {journal}
  {\bibinfo  {journal} {Phys. Rev. Lett.}\ }\textbf {\bibinfo {volume} {126}},\
  \bibinfo {pages} {076803} (\bibinfo {year} {2021})}\BibitemShut {NoStop}%
\bibitem [{\citenamefont {Afanasiev}\ \emph {et~al.}(2021)\citenamefont
  {Afanasiev}, \citenamefont {Alekseev}, \citenamefont {Greshnov},\ and\
  \citenamefont {Semina}}]{Afanasiev2021}%
  \BibitemOpen
  \bibfield  {author} {\bibinfo {author} {\bibfnamefont {A.~N.}\ \bibnamefont
  {Afanasiev}}, \bibinfo {author} {\bibfnamefont {P.~S.}\ \bibnamefont
  {Alekseev}}, \bibinfo {author} {\bibfnamefont {A.~A.}\ \bibnamefont
  {Greshnov}},\ and\ \bibinfo {author} {\bibfnamefont {M.~A.}\ \bibnamefont
  {Semina}},\ }\href {https://doi.org/10.1103/PhysRevB.104.195415} {\bibfield
  {journal} {\bibinfo  {journal} {Phys. Rev. B}\ }\textbf {\bibinfo {volume}
  {104}},\ \bibinfo {pages} {195415} (\bibinfo {year} {2021})}\BibitemShut
  {NoStop}%
\bibitem [{\citenamefont {Horn-Cosfeld}\ \emph {et~al.}(2021)\citenamefont
  {Horn-Cosfeld}, \citenamefont {Schluck}, \citenamefont {Lammert},
  \citenamefont {Cerchez}, \citenamefont {Heinzel}, \citenamefont {Pierz},
  \citenamefont {Schumacher},\ and\ \citenamefont {Mailly}}]{HornCosfeld2021}%
  \BibitemOpen
  \bibfield  {author} {\bibinfo {author} {\bibfnamefont {B.}~\bibnamefont
  {Horn-Cosfeld}}, \bibinfo {author} {\bibfnamefont {J.}~\bibnamefont
  {Schluck}}, \bibinfo {author} {\bibfnamefont {J.}~\bibnamefont {Lammert}},
  \bibinfo {author} {\bibfnamefont {M.}~\bibnamefont {Cerchez}}, \bibinfo
  {author} {\bibfnamefont {T.}~\bibnamefont {Heinzel}}, \bibinfo {author}
  {\bibfnamefont {K.}~\bibnamefont {Pierz}}, \bibinfo {author} {\bibfnamefont
  {H.~W.}\ \bibnamefont {Schumacher}},\ and\ \bibinfo {author} {\bibfnamefont
  {D.}~\bibnamefont {Mailly}},\ }\href
  {https://doi.org/10.1103/PhysRevB.104.045306} {\bibfield  {journal} {\bibinfo
   {journal} {Phys. Rev. B}\ }\textbf {\bibinfo {volume} {104}},\ \bibinfo
  {pages} {045306} (\bibinfo {year} {2021})}\BibitemShut {NoStop}%
\bibitem [{\citenamefont {M\"onch}\ \emph {et~al.}(2022)\citenamefont
  {M\"onch}, \citenamefont {Potashin}, \citenamefont {Lindner}, \citenamefont
  {Yahniuk}, \citenamefont {Golub}, \citenamefont {Kachorovskii}, \citenamefont
  {Bel'kov}, \citenamefont {Huber}, \citenamefont {Watanabe}, \citenamefont
  {Taniguchi}, \citenamefont {Eroms}, \citenamefont {Weiss},\ and\
  \citenamefont {Ganichev}}]{Monch2022}%
  \BibitemOpen
  \bibfield  {author} {\bibinfo {author} {\bibfnamefont {E.}~\bibnamefont
  {M\"onch}}, \bibinfo {author} {\bibfnamefont {S.~O.}\ \bibnamefont
  {Potashin}}, \bibinfo {author} {\bibfnamefont {K.}~\bibnamefont {Lindner}},
  \bibinfo {author} {\bibfnamefont {I.}~\bibnamefont {Yahniuk}}, \bibinfo
  {author} {\bibfnamefont {L.~E.}\ \bibnamefont {Golub}}, \bibinfo {author}
  {\bibfnamefont {V.~Y.}\ \bibnamefont {Kachorovskii}}, \bibinfo {author}
  {\bibfnamefont {V.~V.}\ \bibnamefont {Bel'kov}}, \bibinfo {author}
  {\bibfnamefont {R.}~\bibnamefont {Huber}}, \bibinfo {author} {\bibfnamefont
  {K.}~\bibnamefont {Watanabe}}, \bibinfo {author} {\bibfnamefont
  {T.}~\bibnamefont {Taniguchi}}, \bibinfo {author} {\bibfnamefont
  {J.}~\bibnamefont {Eroms}}, \bibinfo {author} {\bibfnamefont
  {D.}~\bibnamefont {Weiss}},\ and\ \bibinfo {author} {\bibfnamefont {S.~D.}\
  \bibnamefont {Ganichev}},\ }\href
  {https://doi.org/10.1103/PhysRevB.105.045404} {\bibfield  {journal} {\bibinfo
   {journal} {Phys. Rev. B}\ }\textbf {\bibinfo {volume} {105}},\ \bibinfo
  {pages} {045404} (\bibinfo {year} {2022})}\BibitemShut {NoStop}%
\bibitem [{\citenamefont {Moll}\ \emph {et~al.}(2016)\citenamefont {Moll},
  \citenamefont {Kushwaha}, \citenamefont {Nandi}, \citenamefont {Schmidt},\
  and\ \citenamefont {Mackenzie}}]{Moll2016}%
  \BibitemOpen
  \bibfield  {author} {\bibinfo {author} {\bibfnamefont {P.~J.~W.}\
  \bibnamefont {Moll}}, \bibinfo {author} {\bibfnamefont {P.}~\bibnamefont
  {Kushwaha}}, \bibinfo {author} {\bibfnamefont {N.}~\bibnamefont {Nandi}},
  \bibinfo {author} {\bibfnamefont {B.}~\bibnamefont {Schmidt}},\ and\ \bibinfo
  {author} {\bibfnamefont {A.~P.}\ \bibnamefont {Mackenzie}},\ }\href
  {https://doi.org/10.1126/science.aac8385} {\bibfield  {journal} {\bibinfo
  {journal} {Science}\ }\textbf {\bibinfo {volume} {351}},\ \bibinfo {pages}
  {1061} (\bibinfo {year} {2016})}\BibitemShut {NoStop}%
\bibitem [{\citenamefont {Aharon-Steinberg}\ \emph {et~al.}(2022)\citenamefont
  {Aharon-Steinberg}, \citenamefont {Völkl}, \citenamefont {Kaplan},
  \citenamefont {Pariari}, \citenamefont {Roy}, \citenamefont {Holder},
  \citenamefont {Wolf}, \citenamefont {Meltzer}, \citenamefont {Myasoedov},
  \citenamefont {Huber}, \citenamefont {Yan}, \citenamefont {Falkovich},
  \citenamefont {Levitov}, \citenamefont {Hücker},\ and\ \citenamefont
  {Zeldov}}]{AharonSteinberg2022}%
  \BibitemOpen
  \bibfield  {author} {\bibinfo {author} {\bibfnamefont {A.}~\bibnamefont
  {Aharon-Steinberg}}, \bibinfo {author} {\bibfnamefont {T.}~\bibnamefont
  {Völkl}}, \bibinfo {author} {\bibfnamefont {A.}~\bibnamefont {Kaplan}},
  \bibinfo {author} {\bibfnamefont {A.~K.}\ \bibnamefont {Pariari}}, \bibinfo
  {author} {\bibfnamefont {I.}~\bibnamefont {Roy}}, \bibinfo {author}
  {\bibfnamefont {T.}~\bibnamefont {Holder}}, \bibinfo {author} {\bibfnamefont
  {Y.}~\bibnamefont {Wolf}}, \bibinfo {author} {\bibfnamefont {A.~Y.}\
  \bibnamefont {Meltzer}}, \bibinfo {author} {\bibfnamefont {Y.}~\bibnamefont
  {Myasoedov}}, \bibinfo {author} {\bibfnamefont {M.~E.}\ \bibnamefont
  {Huber}}, \bibinfo {author} {\bibfnamefont {B.}~\bibnamefont {Yan}}, \bibinfo
  {author} {\bibfnamefont {G.}~\bibnamefont {Falkovich}}, \bibinfo {author}
  {\bibfnamefont {L.~S.}\ \bibnamefont {Levitov}}, \bibinfo {author}
  {\bibfnamefont {M.}~\bibnamefont {Hücker}},\ and\ \bibinfo {author}
  {\bibfnamefont {E.}~\bibnamefont {Zeldov}},\ }\href
  {https://doi.org/10.1038/s41586-022-04794-y} {\bibfield  {journal} {\bibinfo
  {journal} {Nature}\ }\textbf {\bibinfo {volume} {607}},\ \bibinfo {pages}
  {74} (\bibinfo {year} {2022})}\BibitemShut {NoStop}%
\bibitem [{\citenamefont {Gooth}\ \emph {et~al.}(2018)\citenamefont {Gooth},
  \citenamefont {Menges}, \citenamefont {Kumar}, \citenamefont {S\"{u}ss},
  \citenamefont {Shekhar}, \citenamefont {Sun}, \citenamefont {Drechsler},
  \citenamefont {Zierold}, \citenamefont {Felser},\ and\ \citenamefont
  {Gotsmann}}]{Gooth2018}%
  \BibitemOpen
  \bibfield  {author} {\bibinfo {author} {\bibfnamefont {J.}~\bibnamefont
  {Gooth}}, \bibinfo {author} {\bibfnamefont {F.}~\bibnamefont {Menges}},
  \bibinfo {author} {\bibfnamefont {N.}~\bibnamefont {Kumar}}, \bibinfo
  {author} {\bibfnamefont {V.}~\bibnamefont {S\"{u}ss}}, \bibinfo {author}
  {\bibfnamefont {C.}~\bibnamefont {Shekhar}}, \bibinfo {author} {\bibfnamefont
  {Y.}~\bibnamefont {Sun}}, \bibinfo {author} {\bibfnamefont {U.}~\bibnamefont
  {Drechsler}}, \bibinfo {author} {\bibfnamefont {R.}~\bibnamefont {Zierold}},
  \bibinfo {author} {\bibfnamefont {C.}~\bibnamefont {Felser}},\ and\ \bibinfo
  {author} {\bibfnamefont {B.}~\bibnamefont {Gotsmann}},\ }\href
  {https://doi.org/10.1038/s41467-018-06688-y} {\bibfield  {journal} {\bibinfo
  {journal} {Nat. Commun.}\ }\textbf {\bibinfo {volume} {9}},\ \bibinfo {pages}
  {4093} (\bibinfo {year} {2018})}\BibitemShut {NoStop}%
\bibitem [{\citenamefont {Alekseev}(2016)}]{alekseev2016negative}%
  \BibitemOpen
  \bibfield  {author} {\bibinfo {author} {\bibfnamefont {P.}~\bibnamefont
  {Alekseev}},\ }\href {https://doi.org/10.1103/PhysRevLett.117.166601}
  {\bibfield  {journal} {\bibinfo  {journal} {Phys. Rev. Lett.}\ }\textbf
  {\bibinfo {volume} {117}},\ \bibinfo {pages} {166601} (\bibinfo {year}
  {2016})}\BibitemShut {NoStop}%
\bibitem [{\citenamefont {Guo}\ \emph {et~al.}(2017)\citenamefont {Guo},
  \citenamefont {Ilseven}, \citenamefont {Falkovich},\ and\ \citenamefont
  {Levitov}}]{guo2017higher}%
  \BibitemOpen
  \bibfield  {author} {\bibinfo {author} {\bibfnamefont {H.}~\bibnamefont
  {Guo}}, \bibinfo {author} {\bibfnamefont {E.}~\bibnamefont {Ilseven}},
  \bibinfo {author} {\bibfnamefont {G.}~\bibnamefont {Falkovich}},\ and\
  \bibinfo {author} {\bibfnamefont {L.~S.}\ \bibnamefont {Levitov}},\ }\href
  {https://doi.org/10.1073/pnas.1612181114} {\bibfield  {journal} {\bibinfo
  {journal} {Proc. Natl. Acad. Sci. U.S.A.}\ }\textbf {\bibinfo {volume}
  {114}},\ \bibinfo {pages} {3068} (\bibinfo {year} {2017})}\BibitemShut
  {NoStop}%
\bibitem [{\citenamefont {Scaffidi}\ \emph {et~al.}(2017)\citenamefont
  {Scaffidi}, \citenamefont {Nandi}, \citenamefont {Schmidt}, \citenamefont
  {Mackenzie},\ and\ \citenamefont {Moore}}]{scaffidi2017hydrodynamic}%
  \BibitemOpen
  \bibfield  {author} {\bibinfo {author} {\bibfnamefont {T.}~\bibnamefont
  {Scaffidi}}, \bibinfo {author} {\bibfnamefont {N.}~\bibnamefont {Nandi}},
  \bibinfo {author} {\bibfnamefont {B.}~\bibnamefont {Schmidt}}, \bibinfo
  {author} {\bibfnamefont {A.~P.}\ \bibnamefont {Mackenzie}},\ and\ \bibinfo
  {author} {\bibfnamefont {J.~E.}\ \bibnamefont {Moore}},\ }\href
  {https://doi.org/10.1103/PhysRevLett.118.226601} {\bibfield  {journal}
  {\bibinfo  {journal} {Phys. Rev. Lett.}\ }\textbf {\bibinfo {volume} {118}},\
  \bibinfo {pages} {226601} (\bibinfo {year} {2017})}\BibitemShut {NoStop}%
\bibitem [{\citenamefont {Alekseev}\ and\ \citenamefont
  {Semina}(2019)}]{Alekseev2019}%
  \BibitemOpen
  \bibfield  {author} {\bibinfo {author} {\bibfnamefont {P.~S.}\ \bibnamefont
  {Alekseev}}\ and\ \bibinfo {author} {\bibfnamefont {M.~A.}\ \bibnamefont
  {Semina}},\ }\href {https://doi.org/10.1103/PhysRevB.100.125419} {\bibfield
  {journal} {\bibinfo  {journal} {Phys. Rev. B}\ }\textbf {\bibinfo {volume}
  {100}},\ \bibinfo {pages} {125419} (\bibinfo {year} {2019})}\BibitemShut
  {NoStop}%
\bibitem [{\citenamefont {Alekseev}\ and\ \citenamefont
  {Alekseeva}(2019)}]{Alekseev2019b}%
  \BibitemOpen
  \bibfield  {author} {\bibinfo {author} {\bibfnamefont {P.~S.}\ \bibnamefont
  {Alekseev}}\ and\ \bibinfo {author} {\bibfnamefont {A.~P.}\ \bibnamefont
  {Alekseeva}},\ }\href {https://doi.org/10.1103/PhysRevLett.123.236801}
  {\bibfield  {journal} {\bibinfo  {journal} {Phys. Rev. Lett.}\ }\textbf
  {\bibinfo {volume} {123}},\ \bibinfo {pages} {236801} (\bibinfo {year}
  {2019})}\BibitemShut {NoStop}%
\bibitem [{\citenamefont {Alekseev}\ and\ \citenamefont
  {Dmitriev}(2020)}]{Alekseev2020a}%
  \BibitemOpen
  \bibfield  {author} {\bibinfo {author} {\bibfnamefont {P.~S.}\ \bibnamefont
  {Alekseev}}\ and\ \bibinfo {author} {\bibfnamefont {A.~P.}\ \bibnamefont
  {Dmitriev}},\ }\href {https://doi.org/10.1103/PhysRevB.102.241409} {\bibfield
   {journal} {\bibinfo  {journal} {Phys. Rev. B}\ }\textbf {\bibinfo {volume}
  {102}},\ \bibinfo {pages} {241409} (\bibinfo {year} {2020})}\BibitemShut
  {NoStop}%
\bibitem [{\citenamefont {Matthaiakakis}\ \emph {et~al.}(2020)\citenamefont
  {Matthaiakakis}, \citenamefont {Rodr\'{\i}guez~Fern\'andez}, \citenamefont
  {Tutschku}, \citenamefont {Hankiewicz}, \citenamefont {Erdmenger},\ and\
  \citenamefont {Meyer}}]{Matthaiakakis2020}%
  \BibitemOpen
  \bibfield  {author} {\bibinfo {author} {\bibfnamefont {I.}~\bibnamefont
  {Matthaiakakis}}, \bibinfo {author} {\bibfnamefont {D.}~\bibnamefont
  {Rodr\'{\i}guez~Fern\'andez}}, \bibinfo {author} {\bibfnamefont
  {C.}~\bibnamefont {Tutschku}}, \bibinfo {author} {\bibfnamefont {E.~M.}\
  \bibnamefont {Hankiewicz}}, \bibinfo {author} {\bibfnamefont
  {J.}~\bibnamefont {Erdmenger}},\ and\ \bibinfo {author} {\bibfnamefont
  {R.}~\bibnamefont {Meyer}},\ }\href
  {https://doi.org/10.1103/PhysRevB.101.045423} {\bibfield  {journal} {\bibinfo
   {journal} {Phys. Rev. B}\ }\textbf {\bibinfo {volume} {101}},\ \bibinfo
  {pages} {045423} (\bibinfo {year} {2020})}\BibitemShut {NoStop}%
\bibitem [{\citenamefont {Raichev}\ \emph {et~al.}(2020)\citenamefont
  {Raichev}, \citenamefont {Gusev}, \citenamefont {Levin},\ and\ \citenamefont
  {Bakarov}}]{raichev2020}%
  \BibitemOpen
  \bibfield  {author} {\bibinfo {author} {\bibfnamefont {O.~E.}\ \bibnamefont
  {Raichev}}, \bibinfo {author} {\bibfnamefont {G.~M.}\ \bibnamefont {Gusev}},
  \bibinfo {author} {\bibfnamefont {A.~D.}\ \bibnamefont {Levin}},\ and\
  \bibinfo {author} {\bibfnamefont {A.~K.}\ \bibnamefont {Bakarov}},\ }\href
  {https://doi.org/10.1103/PhysRevB.101.235314} {\bibfield  {journal} {\bibinfo
   {journal} {Phys. Rev. B}\ }\textbf {\bibinfo {volume} {101}},\ \bibinfo
  {pages} {235314} (\bibinfo {year} {2020})}\BibitemShut {NoStop}%
\bibitem [{\citenamefont {Ahn}\ and\ \citenamefont
  {Das~Sarma}(2022)}]{DasSarma2022}%
  \BibitemOpen
  \bibfield  {author} {\bibinfo {author} {\bibfnamefont {S.}~\bibnamefont
  {Ahn}}\ and\ \bibinfo {author} {\bibfnamefont {S.}~\bibnamefont
  {Das~Sarma}},\ }\href {https://doi.org/10.1103/PhysRevB.106.L081303}
  {\bibfield  {journal} {\bibinfo  {journal} {Phys. Rev. B}\ }\textbf {\bibinfo
  {volume} {106}},\ \bibinfo {pages} {L081303} (\bibinfo {year}
  {2022})}\BibitemShut {NoStop}%
\bibitem [{\citenamefont {Hara}\ \emph {et~al.}(2004)\citenamefont {Hara},
  \citenamefont {Endo}, \citenamefont {Katsumoto},\ and\ \citenamefont
  {Iye}}]{Hara2004}%
  \BibitemOpen
  \bibfield  {author} {\bibinfo {author} {\bibfnamefont {M.}~\bibnamefont
  {Hara}}, \bibinfo {author} {\bibfnamefont {A.}~\bibnamefont {Endo}}, \bibinfo
  {author} {\bibfnamefont {S.}~\bibnamefont {Katsumoto}},\ and\ \bibinfo
  {author} {\bibfnamefont {Y.}~\bibnamefont {Iye}},\ }\href
  {https://doi.org/10.1103/PhysRevB.69.153304} {\bibfield  {journal} {\bibinfo
  {journal} {Phys. Rev. B}\ }\textbf {\bibinfo {volume} {69}},\ \bibinfo
  {pages} {153304} (\bibinfo {year} {2004})}\BibitemShut {NoStop}%
\bibitem [{\citenamefont {Zhang}\ \emph {et~al.}(2009)\citenamefont {Zhang},
  \citenamefont {Vitkalov},\ and\ \citenamefont {Bykov}}]{Zhang2009}%
  \BibitemOpen
  \bibfield  {author} {\bibinfo {author} {\bibfnamefont {J.~Q.}\ \bibnamefont
  {Zhang}}, \bibinfo {author} {\bibfnamefont {S.}~\bibnamefont {Vitkalov}},\
  and\ \bibinfo {author} {\bibfnamefont {A.~A.}\ \bibnamefont {Bykov}},\ }\href
  {https://doi.org/10.1103/PhysRevB.80.045310} {\bibfield  {journal} {\bibinfo
  {journal} {Phys. Rev. B}\ }\textbf {\bibinfo {volume} {80}},\ \bibinfo
  {pages} {045310} (\bibinfo {year} {2009})}\BibitemShut {NoStop}%
\bibitem [{\citenamefont {Ando}(1974{\natexlab{a}})}]{ando1974theory}%
  \BibitemOpen
  \bibfield  {author} {\bibinfo {author} {\bibfnamefont {T.}~\bibnamefont
  {Ando}},\ }\href {https://doi.org/10.1143/JPSJ.36.1521} {\bibfield  {journal}
  {\bibinfo  {journal} {J. Phys. Soc. Jpn.}\ }\textbf {\bibinfo {volume}
  {36}},\ \bibinfo {pages} {1521} (\bibinfo {year}
  {1974}{\natexlab{a}})}\BibitemShut {NoStop}%
\bibitem [{\citenamefont {Ando}(1974{\natexlab{b}})}]{ando1974}%
  \BibitemOpen
  \bibfield  {author} {\bibinfo {author} {\bibfnamefont {T.}~\bibnamefont
  {Ando}},\ }\href {https://doi.org/10.1143/JPSJ.37.1233} {\bibfield  {journal}
  {\bibinfo  {journal} {J. Phys. Soc. Jpn.}\ }\textbf {\bibinfo {volume}
  {37}},\ \bibinfo {pages} {1233} (\bibinfo {year}
  {1974}{\natexlab{b}})}\BibitemShut {NoStop}%
\bibitem [{\citenamefont {Ando}(1975)}]{ando1975}%
  \BibitemOpen
  \bibfield  {author} {\bibinfo {author} {\bibfnamefont {T.}~\bibnamefont
  {Ando}},\ }\href {https://doi.org/10.1143/JPSJ.38.989} {\bibfield  {journal}
  {\bibinfo  {journal} {J. Phys. Soc. Jpn.}\ }\textbf {\bibinfo {volume}
  {38}},\ \bibinfo {pages} {989} (\bibinfo {year} {1975})}\BibitemShut
  {NoStop}%
\bibitem [{\citenamefont {Ando}(1982)}]{ando1982}%
  \BibitemOpen
  \bibfield  {author} {\bibinfo {author} {\bibfnamefont {T.}~\bibnamefont
  {Ando}},\ }\href {https://doi.org/10.1143/JPSJ.51.3900} {\bibfield  {journal}
  {\bibinfo  {journal} {J. Phys. Soc. Jpn.}\ }\textbf {\bibinfo {volume}
  {51}},\ \bibinfo {pages} {3900} (\bibinfo {year} {1982})}\BibitemShut
  {NoStop}%
\bibitem [{\citenamefont {Coleridge}(1991)}]{coleridge1991}%
  \BibitemOpen
  \bibfield  {author} {\bibinfo {author} {\bibfnamefont {P.~T.}\ \bibnamefont
  {Coleridge}},\ }\href {https://doi.org/10.1103/PhysRevB.44.3793} {\bibfield
  {journal} {\bibinfo  {journal} {Phys. Rev. B}\ }\textbf {\bibinfo {volume}
  {44}},\ \bibinfo {pages} {3793} (\bibinfo {year} {1991})}\BibitemShut
  {NoStop}%
\bibitem [{\citenamefont {Isihara}\ and\ \citenamefont
  {Smrcka}(1986)}]{isihara1986}%
  \BibitemOpen
  \bibfield  {author} {\bibinfo {author} {\bibfnamefont {A.}~\bibnamefont
  {Isihara}}\ and\ \bibinfo {author} {\bibfnamefont {L.}~\bibnamefont
  {Smrcka}},\ }\href {https://doi.org/10.1088/0022-3719/19/34/015} {\bibfield
  {journal} {\bibinfo  {journal} {J. Phys. C: Solid State Phys.}\ }\textbf
  {\bibinfo {volume} {19}},\ \bibinfo {pages} {6777} (\bibinfo {year}
  {1986})}\BibitemShut {NoStop}%
\bibitem [{\citenamefont {Tan}\ \emph {et~al.}(2011)\citenamefont {Tan},
  \citenamefont {Tan}, \citenamefont {Ma}, \citenamefont {Liu}, \citenamefont
  {Lu},\ and\ \citenamefont {Yang}}]{phaseinversiongraphene}%
  \BibitemOpen
  \bibfield  {author} {\bibinfo {author} {\bibfnamefont {Z.}~\bibnamefont
  {Tan}}, \bibinfo {author} {\bibfnamefont {C.}~\bibnamefont {Tan}}, \bibinfo
  {author} {\bibfnamefont {L.}~\bibnamefont {Ma}}, \bibinfo {author}
  {\bibfnamefont {G.~T.}\ \bibnamefont {Liu}}, \bibinfo {author} {\bibfnamefont
  {L.}~\bibnamefont {Lu}},\ and\ \bibinfo {author} {\bibfnamefont {C.~L.}\
  \bibnamefont {Yang}},\ }\href {https://doi.org/10.1103/PhysRevB.84.115429}
  {\bibfield  {journal} {\bibinfo  {journal} {Phys. Rev. B}\ }\textbf {\bibinfo
  {volume} {84}},\ \bibinfo {pages} {115429} (\bibinfo {year}
  {2011})}\BibitemShut {NoStop}%
\bibitem [{\citenamefont {Mancoff}\ \emph {et~al.}(1996)\citenamefont
  {Mancoff}, \citenamefont {Zielinski}, \citenamefont {Marcus}, \citenamefont
  {Campman},\ and\ \citenamefont {Gossard}}]{mancoff1996}%
  \BibitemOpen
  \bibfield  {author} {\bibinfo {author} {\bibfnamefont {F.~B.}\ \bibnamefont
  {Mancoff}}, \bibinfo {author} {\bibfnamefont {L.~J.}\ \bibnamefont
  {Zielinski}}, \bibinfo {author} {\bibfnamefont {C.~M.}\ \bibnamefont
  {Marcus}}, \bibinfo {author} {\bibfnamefont {K.}~\bibnamefont {Campman}},\
  and\ \bibinfo {author} {\bibfnamefont {A.~C.}\ \bibnamefont {Gossard}},\
  }\href {https://doi.org/10.1103/PhysRevB.53.R7599} {\bibfield  {journal}
  {\bibinfo  {journal} {Phys. Rev. B}\ }\textbf {\bibinfo {volume} {53}},\
  \bibinfo {pages} {R7599} (\bibinfo {year} {1996})}\BibitemShut {NoStop}%
\bibitem [{sdh()}]{sdhmobility}%
  \BibitemOpen
  \href@noop {} {}\bibinfo {note} {The electron mobility measured at 0.28 K for
  a large area Van der Pauw device in which we expect nMR arsing from ballistic
  transport to be negligible is $20 \times 10^{6}$ c$m^2$/V s. For the narrow
  Hall bar measured at the dilution refrigerator base temperature, from fitting
  the SdH oscillations in Fig. \ref{fig:SdHBigTrace}(b), we extracted
  $R_0$=11.4 $\Omega$ which corresponds to an electron mobility of $25 \times
  10^{6}$ c$m^2$/V s. Independently, from the plot of the SdH oscillation
  extrema amplitudes in Fig. \ref{fig:HIROfit}(c), we obtain $R_0$=13.5
  $\Omega$ which corresponds to an electron mobility of $22 \times 10^{6}$
  c$m^2$/V s.}\BibitemShut {Stop}%
\bibitem [{\citenamefont {Coleridge}\ \emph {et~al.}(1996)\citenamefont
  {Coleridge}, \citenamefont {Hayne}, \citenamefont {Zawadzki},\ and\
  \citenamefont {Sachrajda}}]{coleridge1996}%
  \BibitemOpen
  \bibfield  {author} {\bibinfo {author} {\bibfnamefont {P.~T.}\ \bibnamefont
  {Coleridge}}, \bibinfo {author} {\bibfnamefont {M.}~\bibnamefont {Hayne}},
  \bibinfo {author} {\bibfnamefont {P.}~\bibnamefont {Zawadzki}},\ and\
  \bibinfo {author} {\bibfnamefont {A.~S.}\ \bibnamefont {Sachrajda}},\ }\href
  {https://doi.org/https://doi.org/10.1016/0039-6028(96)00469-4} {\bibfield
  {journal} {\bibinfo  {journal} {Appl. Surf. Sci.}\ }\textbf {\bibinfo
  {volume} {361}},\ \bibinfo {pages} {560} (\bibinfo {year}
  {1996})}\BibitemShut {NoStop}%
\bibitem [{\citenamefont {Dmitriev}(2011)}]{Dmitriev2011}%
  \BibitemOpen
  \bibfield  {author} {\bibinfo {author} {\bibfnamefont {I.~A.}\ \bibnamefont
  {Dmitriev}},\ }\href {https://doi.org/10.1088/1742-6596/334/1/012015}
  {\bibfield  {journal} {\bibinfo  {journal} {J. Phys. Conf. Ser.}\ }\textbf
  {\bibinfo {volume} {334}},\ \bibinfo {pages} {012015} (\bibinfo {year}
  {2011})}\BibitemShut {NoStop}%
\bibitem [{\citenamefont {Kalmanovitz}\ \emph {et~al.}(2008)\citenamefont
  {Kalmanovitz}, \citenamefont {Bykov}, \citenamefont {Vitkalov},\ and\
  \citenamefont {Toropov}}]{Kalmanovitz2008}%
  \BibitemOpen
  \bibfield  {author} {\bibinfo {author} {\bibfnamefont {N.~R.}\ \bibnamefont
  {Kalmanovitz}}, \bibinfo {author} {\bibfnamefont {A.~A.}\ \bibnamefont
  {Bykov}}, \bibinfo {author} {\bibfnamefont {S.}~\bibnamefont {Vitkalov}},\
  and\ \bibinfo {author} {\bibfnamefont {A.~I.}\ \bibnamefont {Toropov}},\
  }\href {https://doi.org/10.1103/PhysRevB.78.085306} {\bibfield  {journal}
  {\bibinfo  {journal} {Phys. Rev. B}\ }\textbf {\bibinfo {volume} {78}},\
  \bibinfo {pages} {085306} (\bibinfo {year} {2008})}\BibitemShut {NoStop}%
\bibitem [{\citenamefont {Alexander-Webber}\ \emph {et~al.}(2012)\citenamefont
  {Alexander-Webber}, \citenamefont {Baker}, \citenamefont {Buckle},
  \citenamefont {Ashley},\ and\ \citenamefont {Nicholas}}]{Webber2012}%
  \BibitemOpen
  \bibfield  {author} {\bibinfo {author} {\bibfnamefont {J.~A.}\ \bibnamefont
  {Alexander-Webber}}, \bibinfo {author} {\bibfnamefont {A.~M.~R.}\
  \bibnamefont {Baker}}, \bibinfo {author} {\bibfnamefont {P.~D.}\ \bibnamefont
  {Buckle}}, \bibinfo {author} {\bibfnamefont {T.}~\bibnamefont {Ashley}},\
  and\ \bibinfo {author} {\bibfnamefont {R.~J.}\ \bibnamefont {Nicholas}},\
  }\href {https://doi.org/10.1103/PhysRevB.86.045404} {\bibfield  {journal}
  {\bibinfo  {journal} {Phys. Rev. B}\ }\textbf {\bibinfo {volume} {86}},\
  \bibinfo {pages} {045404} (\bibinfo {year} {2012})}\BibitemShut {NoStop}%
\bibitem [{\citenamefont {Dietrich}\ \emph {et~al.}(2012)\citenamefont
  {Dietrich}, \citenamefont {Byrnes}, \citenamefont {Vitkalov}, \citenamefont
  {Dmitriev},\ and\ \citenamefont {Bykov}}]{Dietrich2012}%
  \BibitemOpen
  \bibfield  {author} {\bibinfo {author} {\bibfnamefont {S.}~\bibnamefont
  {Dietrich}}, \bibinfo {author} {\bibfnamefont {S.}~\bibnamefont {Byrnes}},
  \bibinfo {author} {\bibfnamefont {S.}~\bibnamefont {Vitkalov}}, \bibinfo
  {author} {\bibfnamefont {D.~V.}\ \bibnamefont {Dmitriev}},\ and\ \bibinfo
  {author} {\bibfnamefont {A.~A.}\ \bibnamefont {Bykov}},\ }\href
  {https://doi.org/10.1103/PhysRevB.85.155307} {\bibfield  {journal} {\bibinfo
  {journal} {Phys. Rev. B}\ }\textbf {\bibinfo {volume} {85}},\ \bibinfo
  {pages} {155307} (\bibinfo {year} {2012})}\BibitemShut {NoStop}%
\bibitem [{\citenamefont {Bockhorn}\ \emph {et~al.}(2013)\citenamefont
  {Bockhorn}, \citenamefont {Hodaei}, \citenamefont {Schuh}, \citenamefont
  {Wegscheider},\ and\ \citenamefont {Haug}}]{bockhorn2013magnetoresistance}%
  \BibitemOpen
  \bibfield  {author} {\bibinfo {author} {\bibfnamefont {L.}~\bibnamefont
  {Bockhorn}}, \bibinfo {author} {\bibfnamefont {A.}~\bibnamefont {Hodaei}},
  \bibinfo {author} {\bibfnamefont {D.}~\bibnamefont {Schuh}}, \bibinfo
  {author} {\bibfnamefont {W.}~\bibnamefont {Wegscheider}},\ and\ \bibinfo
  {author} {\bibfnamefont {R.~J.}\ \bibnamefont {Haug}},\ }\href
  {https://doi.org/10.1088/1742-6596/456/1/012003} {\bibfield  {journal}
  {\bibinfo  {journal} {J. Phys. Conf. Ser.}\ }\textbf {\bibinfo {volume}
  {456}},\ \bibinfo {pages} {012003} (\bibinfo {year} {2013})}\BibitemShut
  {NoStop}%
\bibitem [{\citenamefont {Hatke}\ \emph {et~al.}(2012)\citenamefont {Hatke},
  \citenamefont {Zudov}, \citenamefont {Reno}, \citenamefont {Pfeiffer},\ and\
  \citenamefont {West}}]{hatke2012giant}%
  \BibitemOpen
  \bibfield  {author} {\bibinfo {author} {\bibfnamefont {A.}~\bibnamefont
  {Hatke}}, \bibinfo {author} {\bibfnamefont {M.}~\bibnamefont {Zudov}},
  \bibinfo {author} {\bibfnamefont {J.}~\bibnamefont {Reno}}, \bibinfo {author}
  {\bibfnamefont {L.}~\bibnamefont {Pfeiffer}},\ and\ \bibinfo {author}
  {\bibfnamefont {K.}~\bibnamefont {West}},\ }\href
  {https://doi.org/10.1103/PhysRevB.85.081304} {\bibfield  {journal} {\bibinfo
  {journal} {Phys. Rev. B}\ }\textbf {\bibinfo {volume} {85}},\ \bibinfo
  {pages} {081304} (\bibinfo {year} {2012})}\BibitemShut {NoStop}%
\bibitem [{\citenamefont {Mani}\ \emph {et~al.}(2013)\citenamefont {Mani},
  \citenamefont {Kriisa},\ and\ \citenamefont {Wegscheider}}]{mani2013size}%
  \BibitemOpen
  \bibfield  {author} {\bibinfo {author} {\bibfnamefont {R.~G.}\ \bibnamefont
  {Mani}}, \bibinfo {author} {\bibfnamefont {A.}~\bibnamefont {Kriisa}},\ and\
  \bibinfo {author} {\bibfnamefont {W.}~\bibnamefont {Wegscheider}},\ }\href
  {https://doi.org/10.1038/srep02747} {\bibfield  {journal} {\bibinfo
  {journal} {Sci. Rep.}\ }\textbf {\bibinfo {volume} {3}},\ \bibinfo {pages}
  {2747} (\bibinfo {year} {2013})}\BibitemShut {NoStop}%
\bibitem [{\citenamefont {Wang}\ \emph {et~al.}(2016)\citenamefont {Wang},
  \citenamefont {Samaraweera}, \citenamefont {Reichl}, \citenamefont
  {Wegscheider},\ and\ \citenamefont {Mani}}]{Wang2016}%
  \BibitemOpen
  \bibfield  {author} {\bibinfo {author} {\bibfnamefont {Z.}~\bibnamefont
  {Wang}}, \bibinfo {author} {\bibfnamefont {R.~L.}\ \bibnamefont
  {Samaraweera}}, \bibinfo {author} {\bibfnamefont {C.}~\bibnamefont {Reichl}},
  \bibinfo {author} {\bibfnamefont {W.}~\bibnamefont {Wegscheider}},\ and\
  \bibinfo {author} {\bibfnamefont {R.~G.}\ \bibnamefont {Mani}},\ }\href
  {https://doi.org/10.1038/srep38516} {\bibfield  {journal} {\bibinfo
  {journal} {Sci. Rep.}\ }\textbf {\bibinfo {volume} {6}},\ \bibinfo {pages}
  {38516} (\bibinfo {year} {2016})}\BibitemShut {NoStop}%
\bibitem [{\citenamefont {Samaraweera}\ \emph {et~al.}(2017)\citenamefont
  {Samaraweera}, \citenamefont {Liu}, \citenamefont {Wang}, \citenamefont
  {Reichl}, \citenamefont {Wegscheider},\ and\ \citenamefont
  {Mani}}]{Samaraweera2017}%
  \BibitemOpen
  \bibfield  {author} {\bibinfo {author} {\bibfnamefont {R.~L.}\ \bibnamefont
  {Samaraweera}}, \bibinfo {author} {\bibfnamefont {H.-C.}\ \bibnamefont
  {Liu}}, \bibinfo {author} {\bibfnamefont {Z.}~\bibnamefont {Wang}}, \bibinfo
  {author} {\bibfnamefont {C.}~\bibnamefont {Reichl}}, \bibinfo {author}
  {\bibfnamefont {W.}~\bibnamefont {Wegscheider}},\ and\ \bibinfo {author}
  {\bibfnamefont {R.~G.}\ \bibnamefont {Mani}},\ }\href
  {https://doi.org/10.1038/s41598-017-05351-8} {\bibfield  {journal} {\bibinfo
  {journal} {Sci. Rep.}\ }\textbf {\bibinfo {volume} {7}},\ \bibinfo {pages}
  {5074} (\bibinfo {year} {2017})}\BibitemShut {NoStop}%
\bibitem [{\citenamefont {Samaraweera}\ \emph {et~al.}(2018)\citenamefont
  {Samaraweera}, \citenamefont {Liu}, \citenamefont {Gunawardana},
  \citenamefont {Kriisa}, \citenamefont {Reichl}, \citenamefont {Wegscheider},\
  and\ \citenamefont {Mani}}]{Samaraweera2018}%
  \BibitemOpen
  \bibfield  {author} {\bibinfo {author} {\bibfnamefont {R.~L.}\ \bibnamefont
  {Samaraweera}}, \bibinfo {author} {\bibfnamefont {H.-C.}\ \bibnamefont
  {Liu}}, \bibinfo {author} {\bibfnamefont {B.}~\bibnamefont {Gunawardana}},
  \bibinfo {author} {\bibfnamefont {A.}~\bibnamefont {Kriisa}}, \bibinfo
  {author} {\bibfnamefont {C.}~\bibnamefont {Reichl}}, \bibinfo {author}
  {\bibfnamefont {W.}~\bibnamefont {Wegscheider}},\ and\ \bibinfo {author}
  {\bibfnamefont {R.~G.}\ \bibnamefont {Mani}},\ }\href
  {https://doi.org/10.1038/s41598-018-28359-0} {\bibfield  {journal} {\bibinfo
  {journal} {Sci. Rep.}\ }\textbf {\bibinfo {volume} {8}},\ \bibinfo {pages}
  {10061} (\bibinfo {year} {2018})}\BibitemShut {NoStop}%
\bibitem [{\citenamefont {Samaraweera}\ \emph {et~al.}(2020)\citenamefont
  {Samaraweera}, \citenamefont {Gunawardana}, \citenamefont {Nanayakkara},
  \citenamefont {Munasinghe}, \citenamefont {Kriisa}, \citenamefont {Reichl},
  \citenamefont {Wegscheider},\ and\ \citenamefont {Mani}}]{Samaraweera2020}%
  \BibitemOpen
  \bibfield  {author} {\bibinfo {author} {\bibfnamefont {R.~L.}\ \bibnamefont
  {Samaraweera}}, \bibinfo {author} {\bibfnamefont {B.}~\bibnamefont
  {Gunawardana}}, \bibinfo {author} {\bibfnamefont {T.~R.}\ \bibnamefont
  {Nanayakkara}}, \bibinfo {author} {\bibfnamefont {R.~C.}\ \bibnamefont
  {Munasinghe}}, \bibinfo {author} {\bibfnamefont {A.}~\bibnamefont {Kriisa}},
  \bibinfo {author} {\bibfnamefont {C.}~\bibnamefont {Reichl}}, \bibinfo
  {author} {\bibfnamefont {W.}~\bibnamefont {Wegscheider}},\ and\ \bibinfo
  {author} {\bibfnamefont {R.~G.}\ \bibnamefont {Mani}},\ }\href
  {https://doi.org/10.1038/s41598-019-57331-9} {\bibfield  {journal} {\bibinfo
  {journal} {Sci. Rep.}\ }\textbf {\bibinfo {volume} {10}},\ \bibinfo {pages}
  {781} (\bibinfo {year} {2020})}\BibitemShut {NoStop}%
\bibitem [{\citenamefont {Zudov}\ \emph {et~al.}(2001)\citenamefont {Zudov},
  \citenamefont {Ponomarev}, \citenamefont {Efros}, \citenamefont {Du},
  \citenamefont {Simmons},\ and\ \citenamefont {Reno}}]{Zudov2001}%
  \BibitemOpen
  \bibfield  {author} {\bibinfo {author} {\bibfnamefont {M.~A.}\ \bibnamefont
  {Zudov}}, \bibinfo {author} {\bibfnamefont {I.~V.}\ \bibnamefont
  {Ponomarev}}, \bibinfo {author} {\bibfnamefont {A.~L.}\ \bibnamefont
  {Efros}}, \bibinfo {author} {\bibfnamefont {R.~R.}\ \bibnamefont {Du}},
  \bibinfo {author} {\bibfnamefont {J.~A.}\ \bibnamefont {Simmons}},\ and\
  \bibinfo {author} {\bibfnamefont {J.~L.}\ \bibnamefont {Reno}},\ }\href
  {https://doi.org/10.1103/physrevlett.86.3614} {\bibfield  {journal} {\bibinfo
   {journal} {Phys. Rev. Lett.}\ }\textbf {\bibinfo {volume} {86}},\ \bibinfo
  {pages} {3614} (\bibinfo {year} {2001})}\BibitemShut {NoStop}%
\bibitem [{\citenamefont {Bockhorn}\ \emph {et~al.}(2011)\citenamefont
  {Bockhorn}, \citenamefont {Barthold}, \citenamefont {Schuh}, \citenamefont
  {Wegscheider},\ and\ \citenamefont {Haug}}]{Bockhorn2011}%
  \BibitemOpen
  \bibfield  {author} {\bibinfo {author} {\bibfnamefont {L.}~\bibnamefont
  {Bockhorn}}, \bibinfo {author} {\bibfnamefont {P.}~\bibnamefont {Barthold}},
  \bibinfo {author} {\bibfnamefont {D.}~\bibnamefont {Schuh}}, \bibinfo
  {author} {\bibfnamefont {W.}~\bibnamefont {Wegscheider}},\ and\ \bibinfo
  {author} {\bibfnamefont {R.~J.}\ \bibnamefont {Haug}},\ }\href
  {https://doi.org/10.1103/PhysRevB.83.113301} {\bibfield  {journal} {\bibinfo
  {journal} {Phys. Rev. B}\ }\textbf {\bibinfo {volume} {83}},\ \bibinfo
  {pages} {113301} (\bibinfo {year} {2011})}\BibitemShut {NoStop}%
\bibitem [{\citenamefont {Bockhorn}\ \emph {et~al.}(2014)\citenamefont
  {Bockhorn}, \citenamefont {Gornyi}, \citenamefont {Schuh}, \citenamefont
  {Reichl}, \citenamefont {Wegscheider},\ and\ \citenamefont
  {Haug}}]{Bockhorn2014}%
  \BibitemOpen
  \bibfield  {author} {\bibinfo {author} {\bibfnamefont {L.}~\bibnamefont
  {Bockhorn}}, \bibinfo {author} {\bibfnamefont {I.~V.}\ \bibnamefont
  {Gornyi}}, \bibinfo {author} {\bibfnamefont {D.}~\bibnamefont {Schuh}},
  \bibinfo {author} {\bibfnamefont {C.}~\bibnamefont {Reichl}}, \bibinfo
  {author} {\bibfnamefont {W.}~\bibnamefont {Wegscheider}},\ and\ \bibinfo
  {author} {\bibfnamefont {R.~J.}\ \bibnamefont {Haug}},\ }\href
  {https://doi.org/10.1103/PhysRevB.90.165434} {\bibfield  {journal} {\bibinfo
  {journal} {Phys. Rev. B}\ }\textbf {\bibinfo {volume} {90}},\ \bibinfo
  {pages} {165434} (\bibinfo {year} {2014})}\BibitemShut {NoStop}%
\bibitem [{\citenamefont {Choi}\ \emph {et~al.}(1986)\citenamefont {Choi},
  \citenamefont {Tsui},\ and\ \citenamefont {Palmateer}}]{Choi1986}%
  \BibitemOpen
  \bibfield  {author} {\bibinfo {author} {\bibfnamefont {K.~K.}\ \bibnamefont
  {Choi}}, \bibinfo {author} {\bibfnamefont {D.~C.}\ \bibnamefont {Tsui}},\
  and\ \bibinfo {author} {\bibfnamefont {S.~C.}\ \bibnamefont {Palmateer}},\
  }\href {https://doi.org/10.1103/PhysRevB.33.8216} {\bibfield  {journal}
  {\bibinfo  {journal} {Phys. Rev. B}\ }\textbf {\bibinfo {volume} {33}},\
  \bibinfo {pages} {8216} (\bibinfo {year} {1986})}\BibitemShut {NoStop}%
\bibitem [{\citenamefont {Thornton}\ \emph {et~al.}(1989)\citenamefont
  {Thornton}, \citenamefont {Roukes}, \citenamefont {Scherer},\ and\
  \citenamefont {Van~de Gaag}}]{thornton1989boundary}%
  \BibitemOpen
  \bibfield  {author} {\bibinfo {author} {\bibfnamefont {T.}~\bibnamefont
  {Thornton}}, \bibinfo {author} {\bibfnamefont {M.}~\bibnamefont {Roukes}},
  \bibinfo {author} {\bibfnamefont {A.}~\bibnamefont {Scherer}},\ and\ \bibinfo
  {author} {\bibfnamefont {B.}~\bibnamefont {Van~de Gaag}},\ }\href
  {https://doi.org/10.1103/PhysRevLett.63.2128} {\bibfield  {journal} {\bibinfo
   {journal} {Phys. Rev. Lett.}\ }\textbf {\bibinfo {volume} {63}},\ \bibinfo
  {pages} {2128} (\bibinfo {year} {1989})}\BibitemShut {NoStop}%
\bibitem [{\citenamefont {Van~Loosdrecht}\ \emph {et~al.}(1988)\citenamefont
  {Van~Loosdrecht}, \citenamefont {Beenakker}, \citenamefont {Van~Houten},
  \citenamefont {Williamson}, \citenamefont {Van~Wees}, \citenamefont {Mooij},
  \citenamefont {Foxon},\ and\ \citenamefont {Harris}}]{van1988aharonov}%
  \BibitemOpen
  \bibfield  {author} {\bibinfo {author} {\bibfnamefont {P.}~\bibnamefont
  {Van~Loosdrecht}}, \bibinfo {author} {\bibfnamefont {C.}~\bibnamefont
  {Beenakker}}, \bibinfo {author} {\bibfnamefont {H.}~\bibnamefont
  {Van~Houten}}, \bibinfo {author} {\bibfnamefont {J.}~\bibnamefont
  {Williamson}}, \bibinfo {author} {\bibfnamefont {B.}~\bibnamefont
  {Van~Wees}}, \bibinfo {author} {\bibfnamefont {J.}~\bibnamefont {Mooij}},
  \bibinfo {author} {\bibfnamefont {C.}~\bibnamefont {Foxon}},\ and\ \bibinfo
  {author} {\bibfnamefont {J.}~\bibnamefont {Harris}},\ }\href
  {https://doi.org/10.1103/PhysRevB.38.10162} {\bibfield  {journal} {\bibinfo
  {journal} {Phys. Rev. B}\ }\textbf {\bibinfo {volume} {38}},\ \bibinfo
  {pages} {10162} (\bibinfo {year} {1988})}\BibitemShut {NoStop}%
\bibitem [{\citenamefont {Van~Houten}\ \emph {et~al.}(1988)\citenamefont
  {Van~Houten}, \citenamefont {Beenakker}, \citenamefont {Van~Loosdrecht},
  \citenamefont {Thornton}, \citenamefont {Ahmed}, \citenamefont {Pepper},
  \citenamefont {Foxon},\ and\ \citenamefont {Harris}}]{van1988four}%
  \BibitemOpen
  \bibfield  {author} {\bibinfo {author} {\bibfnamefont {H.}~\bibnamefont
  {Van~Houten}}, \bibinfo {author} {\bibfnamefont {C.}~\bibnamefont
  {Beenakker}}, \bibinfo {author} {\bibfnamefont {P.}~\bibnamefont
  {Van~Loosdrecht}}, \bibinfo {author} {\bibfnamefont {T.}~\bibnamefont
  {Thornton}}, \bibinfo {author} {\bibfnamefont {H.}~\bibnamefont {Ahmed}},
  \bibinfo {author} {\bibfnamefont {M.}~\bibnamefont {Pepper}}, \bibinfo
  {author} {\bibfnamefont {C.}~\bibnamefont {Foxon}},\ and\ \bibinfo {author}
  {\bibfnamefont {J.}~\bibnamefont {Harris}},\ }\href
  {https://doi.org/10.1103/PhysRevB.37.8534} {\bibfield  {journal} {\bibinfo
  {journal} {Phys. Rev. B}\ }\textbf {\bibinfo {volume} {37}},\ \bibinfo
  {pages} {8534} (\bibinfo {year} {1988})}\BibitemShut {NoStop}%
\bibitem [{\citenamefont {Mani}\ \emph {et~al.}(1993)\citenamefont {Mani},
  \citenamefont {von Klitzing},\ and\ \citenamefont {Ploog}}]{Mani1993}%
  \BibitemOpen
  \bibfield  {author} {\bibinfo {author} {\bibfnamefont {R.~G.}\ \bibnamefont
  {Mani}}, \bibinfo {author} {\bibfnamefont {K.}~\bibnamefont {von Klitzing}},\
  and\ \bibinfo {author} {\bibfnamefont {K.}~\bibnamefont {Ploog}},\ }\href
  {https://doi.org/10.1103/PhysRevB.48.4571} {\bibfield  {journal} {\bibinfo
  {journal} {Phys. Rev. B}\ }\textbf {\bibinfo {volume} {48}},\ \bibinfo
  {pages} {4571} (\bibinfo {year} {1993})}\BibitemShut {NoStop}%
\bibitem [{\citenamefont {Paalanen}\ \emph {et~al.}(1984)\citenamefont
  {Paalanen}, \citenamefont {Tsui}, \citenamefont {Lin},\ and\ \citenamefont
  {Gossard}}]{Paalanen1984}%
  \BibitemOpen
  \bibfield  {author} {\bibinfo {author} {\bibfnamefont {M.}~\bibnamefont
  {Paalanen}}, \bibinfo {author} {\bibfnamefont {D.}~\bibnamefont {Tsui}},
  \bibinfo {author} {\bibfnamefont {B.}~\bibnamefont {Lin}},\ and\ \bibinfo
  {author} {\bibfnamefont {A.}~\bibnamefont {Gossard}},\ }\href
  {https://doi.org/https://doi.org/10.1016/0039-6028(84)90279-6} {\bibfield
  {journal} {\bibinfo  {journal} {Surf. Sci.}\ }\textbf {\bibinfo {volume}
  {142}},\ \bibinfo {pages} {29} (\bibinfo {year} {1984})}\BibitemShut
  {NoStop}%
\bibitem [{\citenamefont {Kramer}\ and\ \citenamefont
  {MacKinnon}(1993)}]{kramer1993localization}%
  \BibitemOpen
  \bibfield  {author} {\bibinfo {author} {\bibfnamefont {B.}~\bibnamefont
  {Kramer}}\ and\ \bibinfo {author} {\bibfnamefont {A.}~\bibnamefont
  {MacKinnon}},\ }\href
  {https://doi.org/https://doi.org/10.1088/0034-4885/56/12/001} {\bibfield
  {journal} {\bibinfo  {journal} {Rep. Prog. Phys.}\ }\textbf {\bibinfo
  {volume} {56}},\ \bibinfo {pages} {1469} (\bibinfo {year}
  {1993})}\BibitemShut {NoStop}%
\bibitem [{\citenamefont {Glazman}\ and\ \citenamefont
  {Khaetskii}(1989)}]{glazman1989quantum}%
  \BibitemOpen
  \bibfield  {author} {\bibinfo {author} {\bibfnamefont {L.}~\bibnamefont
  {Glazman}}\ and\ \bibinfo {author} {\bibfnamefont {A.}~\bibnamefont
  {Khaetskii}},\ }\href {https://doi.org/10.1088/0953-8984/1/30/015} {\bibfield
   {journal} {\bibinfo  {journal} {J. Phys. Condens. Matter}\ }\textbf
  {\bibinfo {volume} {1}},\ \bibinfo {pages} {5005} (\bibinfo {year}
  {1989})}\BibitemShut {NoStop}%
\bibitem [{\citenamefont {Masubuchi}\ \emph {et~al.}(2012)\citenamefont
  {Masubuchi}, \citenamefont {Iguchi}, \citenamefont {Yamaguchi}, \citenamefont
  {Onuki}, \citenamefont {Arai}, \citenamefont {Watanabe}, \citenamefont
  {Taniguchi},\ and\ \citenamefont {Machida}}]{Masubuchi2012}%
  \BibitemOpen
  \bibfield  {author} {\bibinfo {author} {\bibfnamefont {S.}~\bibnamefont
  {Masubuchi}}, \bibinfo {author} {\bibfnamefont {K.}~\bibnamefont {Iguchi}},
  \bibinfo {author} {\bibfnamefont {T.}~\bibnamefont {Yamaguchi}}, \bibinfo
  {author} {\bibfnamefont {M.}~\bibnamefont {Onuki}}, \bibinfo {author}
  {\bibfnamefont {M.}~\bibnamefont {Arai}}, \bibinfo {author} {\bibfnamefont
  {K.}~\bibnamefont {Watanabe}}, \bibinfo {author} {\bibfnamefont
  {T.}~\bibnamefont {Taniguchi}},\ and\ \bibinfo {author} {\bibfnamefont
  {T.}~\bibnamefont {Machida}},\ }\href
  {https://doi.org/10.1103/PhysRevLett.109.036601} {\bibfield  {journal}
  {\bibinfo  {journal} {Phys. Rev. Lett.}\ }\textbf {\bibinfo {volume} {109}},\
  \bibinfo {pages} {036601} (\bibinfo {year} {2012})}\BibitemShut {NoStop}%
\bibitem [{\citenamefont {Ditlefsen}\ and\ \citenamefont
  {Lothe}(1966)}]{Ditlefsen1966}%
  \BibitemOpen
  \bibfield  {author} {\bibinfo {author} {\bibfnamefont {E.}~\bibnamefont
  {Ditlefsen}}\ and\ \bibinfo {author} {\bibfnamefont {J.}~\bibnamefont
  {Lothe}},\ }\href {https://doi.org/10.1080/14786436608211970} {\bibfield
  {journal} {\bibinfo  {journal} {Philos. Mag.}\ }\textbf {\bibinfo {volume}
  {14}},\ \bibinfo {pages} {759} (\bibinfo {year} {1966})}\BibitemShut
  {NoStop}%
\bibitem [{\citenamefont {Hikami}\ \emph {et~al.}(1980)\citenamefont {Hikami},
  \citenamefont {Larkin},\ and\ \citenamefont {Nagaoka}}]{hikami1980spin}%
  \BibitemOpen
  \bibfield  {author} {\bibinfo {author} {\bibfnamefont {S.}~\bibnamefont
  {Hikami}}, \bibinfo {author} {\bibfnamefont {A.~I.}\ \bibnamefont {Larkin}},\
  and\ \bibinfo {author} {\bibfnamefont {Y.}~\bibnamefont {Nagaoka}},\ }\href
  {https://doi.org/https://doi.org/10.1143/PTP.63.707} {\bibfield  {journal}
  {\bibinfo  {journal} {Prog. Theor. Phys.}\ }\textbf {\bibinfo {volume}
  {63}},\ \bibinfo {pages} {707} (\bibinfo {year} {1980})}\BibitemShut
  {NoStop}%
\bibitem [{\citenamefont {Miller}\ \emph {et~al.}(2003)\citenamefont {Miller},
  \citenamefont {Zumb\"uhl}, \citenamefont {Marcus}, \citenamefont
  {Lyanda-Geller}, \citenamefont {Goldhaber-Gordon}, \citenamefont {Campman},\
  and\ \citenamefont {Gossard}}]{Miller2003}%
  \BibitemOpen
  \bibfield  {author} {\bibinfo {author} {\bibfnamefont {J.~B.}\ \bibnamefont
  {Miller}}, \bibinfo {author} {\bibfnamefont {D.~M.}\ \bibnamefont
  {Zumb\"uhl}}, \bibinfo {author} {\bibfnamefont {C.~M.}\ \bibnamefont
  {Marcus}}, \bibinfo {author} {\bibfnamefont {Y.~B.}\ \bibnamefont
  {Lyanda-Geller}}, \bibinfo {author} {\bibfnamefont {D.}~\bibnamefont
  {Goldhaber-Gordon}}, \bibinfo {author} {\bibfnamefont {K.}~\bibnamefont
  {Campman}},\ and\ \bibinfo {author} {\bibfnamefont {A.~C.}\ \bibnamefont
  {Gossard}},\ }\href {https://doi.org/10.1103/PhysRevLett.90.076807}
  {\bibfield  {journal} {\bibinfo  {journal} {Phys. Rev. Lett.}\ }\textbf
  {\bibinfo {volume} {90}},\ \bibinfo {pages} {076807} (\bibinfo {year}
  {2003})}\BibitemShut {NoStop}%
\bibitem [{\citenamefont {Yu}\ and\ \citenamefont {Cardona}(2005)}]{Yu2005}%
  \BibitemOpen
  \bibfield  {author} {\bibinfo {author} {\bibfnamefont {P.~Y.}\ \bibnamefont
  {Yu}}\ and\ \bibinfo {author} {\bibfnamefont {M.}~\bibnamefont {Cardona}},\
  }\href {https://doi.org/10.1007/b137661} {\emph {\bibinfo {title}
  {Fundamentals of Semiconductors}}}\ (\bibinfo  {publisher} {Springer Berlin
  Heidelberg},\ \bibinfo {year} {2005})\ p.\ \bibinfo {pages} {207}\BibitemShut
  {NoStop}%
\bibitem [{\citenamefont {Chaplik}(1971)}]{chaplik}%
  \BibitemOpen
  \bibfield  {author} {\bibinfo {author} {\bibfnamefont {A.~V.}\ \bibnamefont
  {Chaplik}},\ }\href@noop {} {\bibfield  {journal} {\bibinfo  {journal} {Sov.
  Phys. JETP}\ }\textbf {\bibinfo {volume} {33}},\ \bibinfo {pages} {997}
  (\bibinfo {year} {1971})}\BibitemShut {NoStop}%
\bibitem [{\citenamefont {Yacoby}\ \emph {et~al.}(1991)\citenamefont {Yacoby},
  \citenamefont {Sivan}, \citenamefont {Umbach},\ and\ \citenamefont
  {Hong}}]{yacobytauee}%
  \BibitemOpen
  \bibfield  {author} {\bibinfo {author} {\bibfnamefont {A.}~\bibnamefont
  {Yacoby}}, \bibinfo {author} {\bibfnamefont {U.}~\bibnamefont {Sivan}},
  \bibinfo {author} {\bibfnamefont {C.~P.}\ \bibnamefont {Umbach}},\ and\
  \bibinfo {author} {\bibfnamefont {J.~M.}\ \bibnamefont {Hong}},\ }\href
  {https://doi.org/10.1103/PhysRevLett.66.1938} {\bibfield  {journal} {\bibinfo
   {journal} {Phys. Rev. Lett.}\ }\textbf {\bibinfo {volume} {66}},\ \bibinfo
  {pages} {1938} (\bibinfo {year} {1991})}\BibitemShut {NoStop}%
\bibitem [{yac()}]{yacobydelta}%
  \BibitemOpen
  \href@noop {} {}\bibinfo {note} {Fitting the empirical parabolic dependence
  reflected in Eq. \eqref{eq:rhoxxbg} with Eq. \eqref{eq:ee}, we find
  $\Delta$=0.88$eV_{DC}$, which is close to $\Delta$=0.82$eV_{DC}$ obtained by
  Yacoby \textit{et al.} in Ref. \cite{yacobytauee}.}\BibitemShut {Stop}%
\bibitem [{\citenamefont {Krishna~Kumar}\ \emph {et~al.}(2017)\citenamefont
  {Krishna~Kumar}, \citenamefont {Bandurin}, \citenamefont {Pellegrino},
  \citenamefont {Cao}, \citenamefont {Principi}, \citenamefont {Guo},
  \citenamefont {Auton}, \citenamefont {Ben~Shalom}, \citenamefont
  {Ponomarenko}, \citenamefont {Falkovich}, \citenamefont {Watanabe},
  \citenamefont {Taniguchi}, \citenamefont {Grigorieva}, \citenamefont
  {Levitov}, \citenamefont {Polini},\ and\ \citenamefont
  {Geim}}]{krishna2017superballistic}%
  \BibitemOpen
  \bibfield  {author} {\bibinfo {author} {\bibfnamefont {R.}~\bibnamefont
  {Krishna~Kumar}}, \bibinfo {author} {\bibfnamefont {D.}~\bibnamefont
  {Bandurin}}, \bibinfo {author} {\bibfnamefont {F.}~\bibnamefont
  {Pellegrino}}, \bibinfo {author} {\bibfnamefont {Y.}~\bibnamefont {Cao}},
  \bibinfo {author} {\bibfnamefont {A.}~\bibnamefont {Principi}}, \bibinfo
  {author} {\bibfnamefont {H.}~\bibnamefont {Guo}}, \bibinfo {author}
  {\bibfnamefont {G.}~\bibnamefont {Auton}}, \bibinfo {author} {\bibfnamefont
  {M.}~\bibnamefont {Ben~Shalom}}, \bibinfo {author} {\bibfnamefont {L.~A.}\
  \bibnamefont {Ponomarenko}}, \bibinfo {author} {\bibfnamefont
  {G.}~\bibnamefont {Falkovich}}, \bibinfo {author} {\bibfnamefont
  {K.}~\bibnamefont {Watanabe}}, \bibinfo {author} {\bibfnamefont
  {T.}~\bibnamefont {Taniguchi}}, \bibinfo {author} {\bibfnamefont {I.~V.}\
  \bibnamefont {Grigorieva}}, \bibinfo {author} {\bibfnamefont {L.~S.}\
  \bibnamefont {Levitov}}, \bibinfo {author} {\bibfnamefont {M.}~\bibnamefont
  {Polini}},\ and\ \bibinfo {author} {\bibfnamefont {A.~K.}\ \bibnamefont
  {Geim}},\ }\href {https://doi.org/https://doi.org/10.1038/nphys4240}
  {\bibfield  {journal} {\bibinfo  {journal} {Nat. Phys.}\ }\textbf {\bibinfo
  {volume} {13}},\ \bibinfo {pages} {1182} (\bibinfo {year}
  {2017})}\BibitemShut {NoStop}%
\bibitem [{\citenamefont {Blaikie}\ \emph {et~al.}(1995)\citenamefont
  {Blaikie}, \citenamefont {Cumming}, \citenamefont {Cleaver}, \citenamefont
  {Ahmed},\ and\ \citenamefont {Nakazato}}]{blaikie1995}%
  \BibitemOpen
  \bibfield  {author} {\bibinfo {author} {\bibfnamefont {R.~J.}\ \bibnamefont
  {Blaikie}}, \bibinfo {author} {\bibfnamefont {D.~R.~S.}\ \bibnamefont
  {Cumming}}, \bibinfo {author} {\bibfnamefont {J.~R.~A.}\ \bibnamefont
  {Cleaver}}, \bibinfo {author} {\bibfnamefont {H.}~\bibnamefont {Ahmed}},\
  and\ \bibinfo {author} {\bibfnamefont {K.}~\bibnamefont {Nakazato}},\ }\href
  {https://doi.org/10.1063/1.360680} {\bibfield  {journal} {\bibinfo  {journal}
  {J. Appl. Phys.}\ }\textbf {\bibinfo {volume} {78}},\ \bibinfo {pages} {330}
  (\bibinfo {year} {1995})}\BibitemShut {NoStop}%
\bibitem [{\citenamefont {Vavilov}\ and\ \citenamefont
  {Aleiner}(2004)}]{vavilov2004}%
  \BibitemOpen
  \bibfield  {author} {\bibinfo {author} {\bibfnamefont {M.~G.}\ \bibnamefont
  {Vavilov}}\ and\ \bibinfo {author} {\bibfnamefont {I.~L.}\ \bibnamefont
  {Aleiner}},\ }\href {https://doi.org/10.1103/PhysRevB.69.035303} {\bibfield
  {journal} {\bibinfo  {journal} {Phys. Rev. B}\ }\textbf {\bibinfo {volume}
  {69}},\ \bibinfo {pages} {035303} (\bibinfo {year} {2004})}\BibitemShut
  {NoStop}%
\bibitem [{\citenamefont {Dmitriev}\ \emph {et~al.}(2005)\citenamefont
  {Dmitriev}, \citenamefont {Vavilov}, \citenamefont {Aleiner}, \citenamefont
  {Mirlin},\ and\ \citenamefont {Polyakov}}]{Dmitriev2005}%
  \BibitemOpen
  \bibfield  {author} {\bibinfo {author} {\bibfnamefont {I.~A.}\ \bibnamefont
  {Dmitriev}}, \bibinfo {author} {\bibfnamefont {M.~G.}\ \bibnamefont
  {Vavilov}}, \bibinfo {author} {\bibfnamefont {I.~L.}\ \bibnamefont
  {Aleiner}}, \bibinfo {author} {\bibfnamefont {A.~D.}\ \bibnamefont
  {Mirlin}},\ and\ \bibinfo {author} {\bibfnamefont {D.~G.}\ \bibnamefont
  {Polyakov}},\ }\href {https://doi.org/10.1103/PhysRevB.71.115316} {\bibfield
  {journal} {\bibinfo  {journal} {Phys. Rev. B}\ }\textbf {\bibinfo {volume}
  {71}},\ \bibinfo {pages} {115316} (\bibinfo {year} {2005})}\BibitemShut
  {NoStop}%
\bibitem [{\citenamefont {Vavilov}\ \emph {et~al.}(2007)\citenamefont
  {Vavilov}, \citenamefont {Aleiner},\ and\ \citenamefont
  {Glazman}}]{vavilov2007nonlinear}%
  \BibitemOpen
  \bibfield  {author} {\bibinfo {author} {\bibfnamefont {M.~G.}\ \bibnamefont
  {Vavilov}}, \bibinfo {author} {\bibfnamefont {I.~L.}\ \bibnamefont
  {Aleiner}},\ and\ \bibinfo {author} {\bibfnamefont {L.~I.}\ \bibnamefont
  {Glazman}},\ }\href {https://doi.org/10.1103/PhysRevB.76.115331} {\bibfield
  {journal} {\bibinfo  {journal} {Phys. Rev. B}\ }\textbf {\bibinfo {volume}
  {76}},\ \bibinfo {pages} {115331} (\bibinfo {year} {2007})}\BibitemShut
  {NoStop}%
\bibitem [{\citenamefont {Khodas}\ and\ \citenamefont
  {Vavilov}(2008)}]{khodas2008}%
  \BibitemOpen
  \bibfield  {author} {\bibinfo {author} {\bibfnamefont {M.}~\bibnamefont
  {Khodas}}\ and\ \bibinfo {author} {\bibfnamefont {M.~G.}\ \bibnamefont
  {Vavilov}},\ }\href {https://doi.org/10.1103/PhysRevB.78.245319} {\bibfield
  {journal} {\bibinfo  {journal} {Phys. Rev. B}\ }\textbf {\bibinfo {volume}
  {78}},\ \bibinfo {pages} {245319} (\bibinfo {year} {2008})}\BibitemShut
  {NoStop}%
\bibitem [{\citenamefont {Studenikin}\ \emph {et~al.}(2005)\citenamefont
  {Studenikin}, \citenamefont {Potemski}, \citenamefont {Sachrajda},
  \citenamefont {Hilke}, \citenamefont {Pfeiffer},\ and\ \citenamefont
  {West}}]{Studenikin2005}%
  \BibitemOpen
  \bibfield  {author} {\bibinfo {author} {\bibfnamefont {S.~A.}\ \bibnamefont
  {Studenikin}}, \bibinfo {author} {\bibfnamefont {M.}~\bibnamefont
  {Potemski}}, \bibinfo {author} {\bibfnamefont {A.}~\bibnamefont {Sachrajda}},
  \bibinfo {author} {\bibfnamefont {M.}~\bibnamefont {Hilke}}, \bibinfo
  {author} {\bibfnamefont {L.~N.}\ \bibnamefont {Pfeiffer}},\ and\ \bibinfo
  {author} {\bibfnamefont {K.~W.}\ \bibnamefont {West}},\ }\href
  {https://doi.org/10.1103/PhysRevB.71.245313} {\bibfield  {journal} {\bibinfo
  {journal} {Phys. Rev. B}\ }\textbf {\bibinfo {volume} {71}},\ \bibinfo
  {pages} {245313} (\bibinfo {year} {2005})}\BibitemShut {NoStop}%
\bibitem [{\citenamefont {Hatke}\ \emph {et~al.}(2008)\citenamefont {Hatke},
  \citenamefont {Chiang}, \citenamefont {Zudov}, \citenamefont {Pfeiffer},\
  and\ \citenamefont {West}}]{Hatke2008}%
  \BibitemOpen
  \bibfield  {author} {\bibinfo {author} {\bibfnamefont {A.~T.}\ \bibnamefont
  {Hatke}}, \bibinfo {author} {\bibfnamefont {H.-S.}\ \bibnamefont {Chiang}},
  \bibinfo {author} {\bibfnamefont {M.~A.}\ \bibnamefont {Zudov}}, \bibinfo
  {author} {\bibfnamefont {L.~N.}\ \bibnamefont {Pfeiffer}},\ and\ \bibinfo
  {author} {\bibfnamefont {K.~W.}\ \bibnamefont {West}},\ }\href
  {https://doi.org/10.1103/PhysRevLett.101.246811} {\bibfield  {journal}
  {\bibinfo  {journal} {Phys. Rev. Lett.}\ }\textbf {\bibinfo {volume} {101}},\
  \bibinfo {pages} {246811} (\bibinfo {year} {2008})}\BibitemShut {NoStop}%
\end{thebibliography}
\end{document}